\documentclass[12pt,preprint]{aastex}
\usepackage{epsfig}
\usepackage{natbib}
\usepackage{graphicx}
\usepackage{multirow}
\usepackage{lscape}
\usepackage{mathrsfs,amssymb}
\usepackage{amsmath}
\usepackage{color}

\newcommand	\beq	{\begin{equation}}	
\newcommand	\eeq	{\end{equation}}	

\newcommand       \cm           {\,{\rm cm}}

\newcommand       \km           {\,{\rm km}}
\newcommand       \erg          {\,{\rm erg}}

\newcommand       \g            {\,{\rm g}}
\newcommand       \K            {\,{\rm K}}

\newcommand       \Mpc          {\,{\rm Mpc}}
\newcommand       \s            {\,{\rm s}}

\newcommand       \simlt        {\lesssim}
\newcommand       \simgt        {\gtrsim}
\newcommand       \gtsim        {\gtrsim}

\newcommand       \um           {\mu{\rm m}}
\newcommand       \mum          {\,{\rm \mu m}}
\newcommand       \ppm          {\,{\rm ppm}}

\newcommand       \simali       {\sim\,}

\newcommand       \Msil           {M_{\rm sil}}
\newcommand       \Mcarb          {M_{\rm carb}}

\newcommand	  \xtot         {\left[{\rm X/H}\right]_{\rm tot}}
\newcommand	  \csun         {\left[{\rm C/H}\right]_{\odot}}

\newcommand	  \fesun        {\left[{\rm Fe/H}\right]_{\odot}}
\newcommand	  \mgsun        {\left[{\rm Mg/H}\right]_{\odot}}
\newcommand	  \sisun        {\left[{\rm Si/H}\right]_{\odot}}

\newcommand	  \ctot         {\left[{\rm C/H}\right]_{\rm tot}}

\newcommand	  \xdust        {\left[{\rm X/H}\right]_{\rm dust}}

\newcommand	  \fedust       {\left[{\rm Fe/H}\right]_{\rm dust}}
\newcommand	  \mgdust       {\left[{\rm Mg/H}\right]_{\rm dust}}
\newcommand	  \sidust       {\left[{\rm Si/H}\right]_{\rm dust}}
\newcommand	  \xgas         {\left[{\rm X/H}\right]_{\rm gas}}
\newcommand	  \cgas        {\left[{\rm C/H}\right]_{\rm gas}}

\newcommand	  \mux         {\mu_{\rm X}}
\newcommand	  \muc         {\mu_{\rm C}}

\newcommand	  \musi        {\mu_{\rm Si}}
\newcommand	  \mufe        {\mu_{\rm Fe}}
\newcommand	  \mumg        {\mu_{\rm Mg}}

\newcommand{\spitzerirs}{{\em Spitzer}/IRS\ }

\newcommand{\etal}{\textrm{et al.\ }}

\newcommand{\sd}{SDSS~J0808+3948}
\newcommand \Tw           {T_{\rm w}}
\newcommand \Tc           {T_{\rm c}}
\newcommand \Twi          {T_{{\rm w},i}}
\newcommand \Tci          {T_{{\rm c},i}}
\newcommand \Mwi          {M_{{\rm w},i}}
\newcommand \Mci          {M_{{\rm c},i}}
\newcommand \knui         {\kappa_{{\rm abs},i}(\nu)}
\newcommand \Twsil        {T_{\rm w}^{\rm S}}
\newcommand \Tcsil        {T_{\rm c}^{\rm S}}
\newcommand \Mwsil        {M_{\rm w}^{\rm S}}
\newcommand \Mcsil        {M_{\rm c}^{\rm S}}
\newcommand \Twcarb        {T_{\rm w}^{\rm C}}
\newcommand \Tccarb        {T_{\rm c}^{\rm C}}
\newcommand \Mwcarb        {M_{\rm w}^{\rm C}}
\newcommand \Mccarb        {M_{\rm c}^{\rm C}}
\countdef\decade=200
\advance\decade by \year
\countdef\hours=201
\advance\hours by \time
\divide\hours by 60
\countdef\mins=202
\advance\mins by \hours
\multiply\mins by 60
\multiply\hours by 100
\countdef\miltime=203
\advance\miltime by \hours
\advance\miltime by \time
\advance\miltime by -\mins
\def\today{\number\decade.\number\month.\number\day.\number\miltime}
\shorttitle{A Tale of Three Galaxies}
\title{
A Tale of Three Galaxies: 
Deciphering the Infrared Emission of 
the Spectroscopically Anomalous Galaxies 
IRAS~F10398+1455, 
IRAS~F21013-0739 and 
SDSS~J0808+3948
\\{\small DRAFT: \today ~~}
}
\author{
Yanxia~Xie\altaffilmark{1,2},
Aigen~Li\altaffilmark{2}, 
Lei~Hao\altaffilmark{1}, and 
Robert~Nikutta\altaffilmark{3}
}
\altaffiltext{1}{Shanghai Astronomical Observatory,
                       Chinese Academy of Sciences,
                       80 Nandan Road, Shanghai 200030, China
                       }
\altaffiltext{2}{Department of Physics and Astronomy,
                       University of Missouri,
                       Columbia, MO 65211, USA
                       }
\altaffiltext{3}{Instituto de Astrofisica,
                      Facultad de Fisica,
                      Pontificia Universidad Cat\'olica de Chile,
                      306, Santiago 22, Chile
                      }
\begin{document}

\begin{abstract}
The \textit{Spitzer}/Infrared Spectrograph spectra of 
three spectroscopically anomalous galaxies
(IRAS~F10398+1455, IRAS~F21013-0739 
and SDSS~J0808+3948)
are modeled in terms of 
a mixture of 
warm and cold silicate dust, and
warm and cold carbon dust.
Their unique infrared (IR) emission spectra
are characterized by
a steep $\simali$5--8$\mum$ emission continuum, 
strong emission bands from 
polycyclic aromatic hydrocarbon (PAH) molecules,
and prominent silicate emission. 
The steep $\simali$5--8$\mum$ emission continuum 
and strong PAH emission features suggest 
the dominance of starbursts,
while the silicate emission is indicative of
significant heating from active galactic nuclei (AGNs).
%
%
With warm and cold silicate dust 
of various compositions 
(``astronomical silicate,''  
amorphous olivine, or amorphous pyroxene) 
combined with warm and cold carbon dust 
(amorphous carbon, or graphite), 
we are able to closely reproduce 
the observed IR emission of these 
galaxies.
We find that the dust temperature is 
the primary cause in regulating 
the steep $\sim$5--8$\mum$ continuum 
and silicate emission,
insensitive to the exact silicate 
or carbon dust mineralogy
and grain size $a$
as long as $a\simlt1\mum$.
More specifically, the temperature of 
the $\simali$5--8$\mum$ continuum emitter
(which is essentially carbon dust) of
these galaxies is $\sim$250--400$\K$,
much lower than that of typical quasars
which is $\sim$640$\K$.
Moreover, it appears that larger 
dust grains are 
preferred in quasars.
The lower dust temperature and 
smaller grain sizes 
inferred for these three galaxies 
compared with that of quasars
could be due to the fact that they
may harbor a young/weak AGN 
which is not maturely developed yet. 
\end{abstract}


\section{Introduction\label{sec:intro}}
%
Dust is a ubiquitous feature of the cosmos. 
It is present in a wide variety of astrophysical
systems, including active galactic nuclei (AGNs) 
and starburst galaxies.
Amorphous silicates and some form of carbonaceous 
materials (e.g., graphite, amorphous carbon,
and hydrogenated amorphous carbon) are thought 
to be the two dominant cosmic dust species.
Silicate dust reveals its presence in AGN 
and starbursts through the 9.7 and 18$\mum$
spectral features which respectively arise from 
the Si--O stretching and O--Si--O bending 
vibrational modes (Henning 2010).
The detection of the 3.4$\mum$ C--H stretching 
absorption feature in AGN indicates the presence 
of aliphatic, chain-like hydrocarbon dust in AGN
(e.g., see Imanishi et al.\ 1997, Mason et al.\ 2004). 
Starbursts lack the 3.4$\mum$ absorption feature.
This may be related to the fact that starbursts,
  particularly in their nuclear regions, 
  are often heavily obscured by dust and ice.
  In the Milky Way, the 3.4$\mum$ absorption feature 
  is seen in the diffuse interstellar medium (ISM), 
  but not in dense clouds where the 3.1$\mum$ H$_2$O
  ice absorption feature is strong
  (see Pendleton \& Allamandola 2002). 
  Starbursts often exhibit strong H$_2$O ice absorption
  at 3.1 and 6.0$\mum$ (e.g., see Spoon et al.\ 2004),
  while these ice features are not seen in AGN
  where it is too hot for ice to survive 
  against sublimation (see Li 2007).
Furthermore, starbursts also emit prominently 
at 3.3, 6.2, 7.7, 8.6 and 11.3$\mum$ 
which are often collectively ascribed
to polycyclic aromatic hydrocarbon (PAH) molecules
(L\'eger \& Puget 1984, Allamandola et al.\ 1985),  
however, these PAH emission bands are not detected in
AGN (see Li 2007). The absence of PAHs in AGN 
is commonly attributed to the destruction of PAHs 
by extreme UV and soft X-ray photons in AGN 
(Roche et al.\ 1991; Voit 1991, 1992; 
Siebenmorgen et al.\ 2004). 

In starbursts, the 9.7 and 18$\mum$ silicate features 
are always seen in \textit{absorption}. 
In AGN, in contrast, the silicate features can be seen 
either in \textit{absorption} or in \textit{emission}, 
or can not be made out at all in the continuum.
AGN are thought to be surrounded by an optically thick 
dust torus which would block the photons from 
the broad line region and accretion disk 
if they are 
viewed edge-on, through the torus (``type 2''), 
or would allow the detection of 
broad emission lines if they are viewed 
face-on (``type 1'').
The anisotropic torus is invoked 
by the unification theory of AGN to 
account for the observational dichotomy 
(Antonucci 1993; Urry \& Padovani 1995).
In type 1 AGN, silicates are expected to 
\textit{emit} at 9.7 and 18$\mum$ 
where hot dust can be directly detected, 
while in type 2 AGN, silicates are expected
to be seen in \textit{absorption} 
because of torus obscuration 
(see Li 2007, Mason 2015).\footnote{%
   There are some exceptions:
   some type 2 AGN exhibit silicate features
   in \textit{emission} instead of \textit{absorption}
   (Sturm \etal 2006; Mason \etal 2009; Nikutta \etal 2009).
   }
Moreover, starbursts show a \textit{steeply rising} emission
continuum at $\simali$5--8$\mum$ before the onset of
the 9.7$\mum$ silicate feature, 
while the $\simali$5--8$\mum$ emission continuum 
is \textit{flat} in AGN.

Very recently, Xie \etal (2014) 
studied the $\simali$5--40$\mum$ infrared (IR) 
spectra of IRAS~F10398+1455, IRAS~F21013-0739, 
and SDSS~J0808+3948 obtained with 
the {\it Infrared Spectrograph} (IRS) 
on board the {\it Spitzer Space Telescope} 
(Houck \etal 2004). 
They found that the IR spectra of these galaxies
are \textit{anomalous}.
Spectroscopically, on one hand they resemble
that of AGN in the sense that the silicate features 
in these galaxies are seen in \textit{emission};
however, they also exhibit strong PAH emission
features which are absent in AGN.
On the other hand, they are like starbursts
in the sense that the $\simali$5--8$\mum$ 
emission continua of both these galaxies and 
starbursts all steeply rise with wavelength $\lambda$
and they all show strong PAH emission features.
However, the silicate features seen in \textit{emission} 
in these galaxies are often seen in \textit{absorption} 
in starbursts. Furthermore, the \textit{steep} 
$\simali$5--8$\mum$ emission continuum
seen in these galaxies is much \textit{flatter} in AGN 
(see Figure~1 of Xie et al.\ 2014).
Let $d\ln F_\nu/d\ln\lambda$
be the slope of the $\simali$5--8$\mum$ emission continuum,
where $F_\nu$ is the observed flux at frequency $\nu$, 
$\lambda=c/\nu$ is wavelength, 
and $c$ is the speed of light.
On average, $d\ln F_\nu/d\ln\lambda\approx0.8$ for AGN,
while for IRAS~F10398+1455, IRAS~F21013-0739, 
and SDSS~J0808+3948,
$d\ln F_\nu/d\ln\lambda$\,$\approx$\,4.1, 4.2, and 4.6,
respectively.
A detailed description of the spectral properties and 
a comparison of the {\it Spitzer}/IRS spectra of 
these three galaxies 
with that of typical starubursts and quasars 
can be found in Xie \etal (2014).

The focus of this paper is to explore
the unique properties of the dust in terms
of composition, size and temperature in these
three spectroscopically anomalous galaxies.
This paper is organized as follows.
We describe the data in \S\ref{sec:data}.  
In \S\ref{sec:model} we describe 
the dust model and calculate the mass 
absorption coefficients of various silicate
and carbon dust species in the IR.
The model-fitting to the observed IR emission
is presented in \S\ref{sec:results}.
In \S\ref{sec:discussion} we compare
the PAHFIT approach with the spline approach,
and also compare the IR emission of 
our three galaxies
with that of quasars and IRAS~FSC10214+4724,
an ultraluminous IR galaxy (ULIRG). 
The major conclusions are summarized
in \S\ref{sec:summary}.
Throughout the paper, we assume a 
cosmological model with 
$H_{0} = 70\,h_{70}\km\s^{-1}\Mpc^{-1}$, 
$\Omega_{m} = 0.3$ and $\Omega_{\land} = 0.7$. 
The $\rm L_{\odot}$ represents solar luminosity 
of $\rm 3.826\times10^{33}\,ergs\,s^{-1}$ 
and $\rm M_{\odot}$ represents solar mass 
of $\rm 1.989\times10^{33}\,g$.

\section{Observations and Data\label{sec:data}}
These three galaxies were found when cross-matching 
the \textit{Sloan Digital Sky Survey} (SDSS) 
and {\it Spitzer}/IRS low resolution spectra 
(L.~Hao \etal 2015, in preparation).
The \textit{Spitzer}/IRS spectra of these galaxies
are taken from 
the {\it Cornell AtlaS of Spitzer/IRS Sources} (CASSIS) 
which includes $\simali$13,000 low resolution 
spectra of $>$11,000 distinct sources observed 
in the standard staring mode 
and provides publishable quality spectra
(Lebouteiller \etal 2011; 
L.~Hao \etal 2015, in preparation).
We tabulate their basic properties
in Table~\ref{tab:basicpara}.

%

\begin{deluxetable}{lccccccc}   
\tablecolumns{9}
\tablewidth{0pc}
\tablecaption{Basic Parameters for
              the Three Spectroscopically 
              Anomalous Galaxies 
              IRAS~F10398+1455, 
              IRAS~F21013-0739 and 
              SDSS~J0808+3948
              \label{tab:basicpara}}
\tablehead{
\colhead{Sources} & \colhead{R.A.} & \colhead{Decl.} & \colhead{Redshift} & 
\colhead{Classification\tablenotemark{a}} & 
\colhead{Program ID\tablenotemark{b}} & 
\colhead{$\rm M_{\ast}\tablenotemark{c}$} & 
\colhead{$L_{\rm IR}$\tablenotemark{d}}  \\
\colhead{} & \colhead{} & \colhead{}  & 
\colhead{}  &  \colhead{} &  \colhead{} & \colhead{($\rm M_{\odot}$)} & 
\colhead{($\rm L_{\odot}$)} 
}
\rotate
\startdata
IRAS F10398\,+\,1455 & 10h42m33.32s & +14d39m54.1s  & 0.099 & AGN+SB & 40991 & 10$^{10.7}$  & 10$^{10.50}$ \\
IRAS F21013\,--\,0739 & 21h03m58.75s & -07d28m02.5s  & 0.136 & AGN+SB  &  
40444 & 10$^{10.9}$ & 10$^{10.53}$ \\
SDSS J0808\,+\,3948  & 08h08m44.27s & 39d48m52.36s  & 0.091 & AGN+SB & 
40444 & 10$^{9.8}$ & 10$^{10.84}$ \\
\enddata
\tablenotetext{a}{The optical classification was based on 
                  the Baldwin, Phillips \& Terlevich (1981; BPT) 
                  diagram which plots the emission line ratio of 
                  [$\rm O~{III}$]/H$\beta$ against [$\rm N~{II}$]/H$\alpha$ 
                  (Y. Xie \etal 2015, in preparation). 
                  These three galaxies are classified 
                  as a composite of type 2 AGN and starburst.
                  } 
\tablenotetext{b}{The program ID from which the {\it Spitzer}/IRS 
                  spectra were taken.
                  }
\tablenotetext{c}{Stellar mass (taken from 
                  the MPA-JHU SDSS DR7 catalog).
                  }
\tablenotetext{d}{Infrared luminosity.}
\end{deluxetable}

\begin{figure}
\begin{center}
\resizebox{0.75\hsize}{!}{\includegraphics{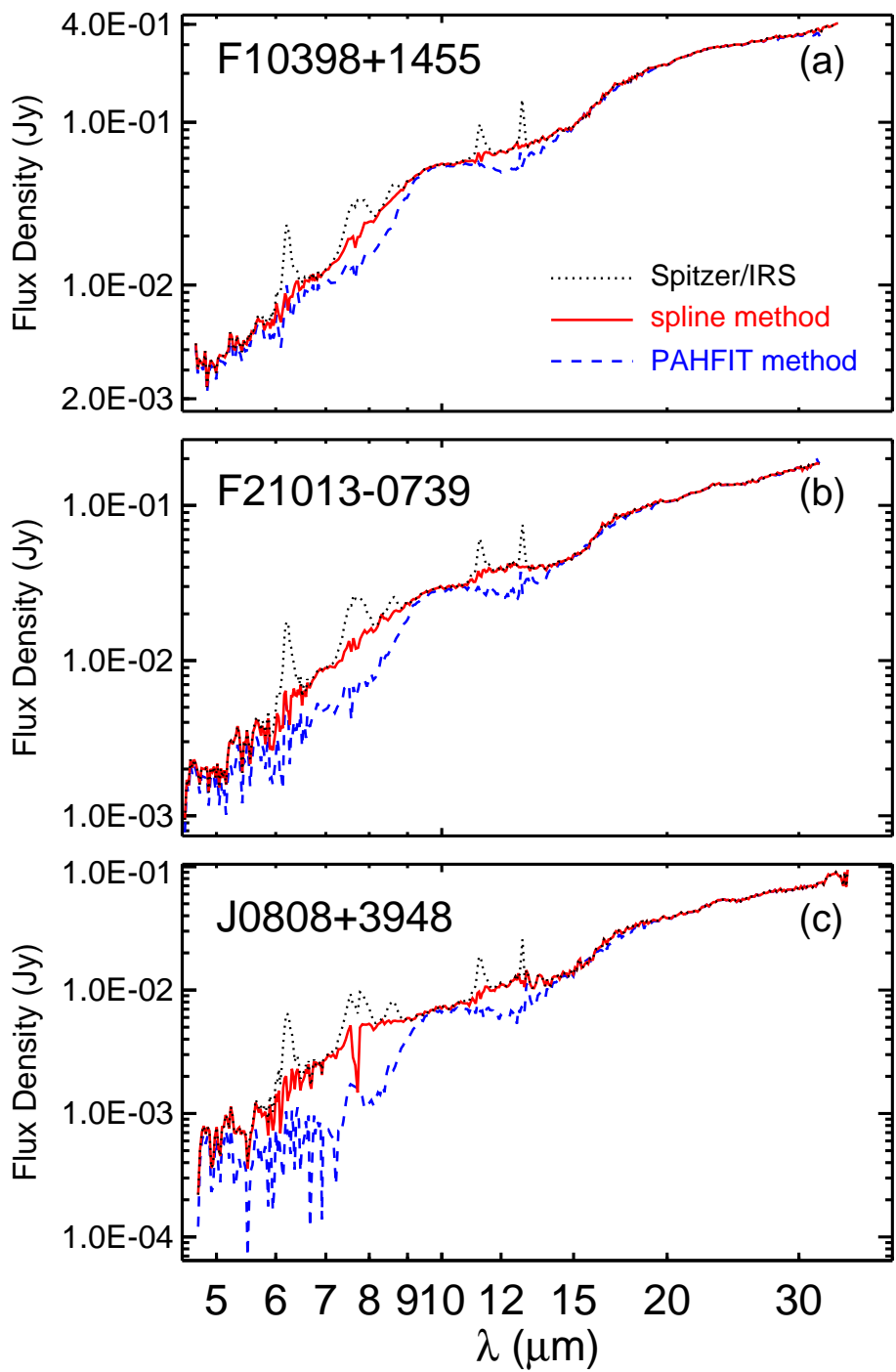}}
\caption{\footnotesize
         \label{fig:sil_pahfit_spline}
         The ``residual'' dust emission
         of IRAS~F10398+1455 (a), IRAS~F21013-0739 (b) 
         and \sd (c) obtained
         by subtracting from 
         the \textit{Spitzer}/IRS spectrum (black dotted line) 
         the PAH and ionic emission lines 
         determined from 
         the spline approach (red solid line)
         or from the PAHFIT approach (blue dashed line).
         In the spline approach, 
         the continuum underneath the PAH emission
         features is fitted with a spline function.
         In the PAHFIT approach,
         the continuum underneath the PAH emission
         features is approximated as a sum of starlight 
         and blackbodies of different temperatures 
         derived from the PAHFIT software 
         of Smith et al.\ (2007).
         }
\end{center}
\end{figure}

%
While the $\simali$5--8$\mum$ continuum
and the 9.7 and 18$\mum$ silicate emission features
result from dust grains which, 
heated by the AGN central engine, 
attain an equilibrium temperature (see Li 2007), 
the PAH emission features arise from PAH molecules 
which are stochastically heated by single stellar photons
from their host galaxies (see Draine \& Li 2001). 
Due to the differences in their emission mechanisms, 
carriers, and emitting regions,
to facilitate a detailed examination of the dust emission, 
we first remove the PAH emission features
and the ionic emission lines 
from the \textit{Spitzer}/IRS spectra of these galaxies.
This requires an estimation of the dust emission continuum
underneath the PAH and ionic emission lines.
We take two approaches 
in defining the continuum. 
First, we approximate the continuum as
a sum of starlight and blackbodies of 
different temperatures using a modified version
of the PAHFIT software (Smith et al.\ 2007).
The continuum determined in this way is hereafter 
referred to as ``the PAHFIT continuum.'' 
%
We also define a ``spline continuum'' 
which is obtained by selecting anchor points 
at 5--7, 14.5--15.0 and 29--30$\mum$
of each \textit{Spitzer}/IRS spectrum 
to form a underlying continuum and then 
fitting these points with a spline function
(L. Hao et al.\ 2015 in preparation).
%
In both approaches, 
the PAH emission features are fitted with a number 
of Drude profiles which are expected for 
classical damped harmonic oscillators (see Li 2009),
while the ionic lines are fitted with Gaussian profiles.

%

We are mostly interested in 
the ``residual'' dust emission 
obtained by subtracting 
the PAH and ionic emission lines
from the observed \textit{Spitzer}/IRS spectra.
%
%
For illustration, we show in Figure~\ref{fig:sil_pahfit_spline}
the ``residual'' dust emission of 
IRAS~F10398+1455, IRAS~F21013-0739 and \sd \,%
obtained by subtracting 
the PAH and ionic emission lines 
determined from the PAHFIT approach.
We also show in Figure~\ref{fig:sil_pahfit_spline}
the ``residual'' dust emission 
of three galaxies obtained with the
spline approach.  
%
For F10398+1455, the overall ``residual'' 
dust emission profiles resulting from 
these two approaches are in close agreement with each other, 
except that the 9.7$\mum$ silicate emission feature 
determined from the PAHFIT method
is somewhat narrower than that from the spline method.
In specific, the fluxes in the blue and red wings of 
the 9.7$\mum$ silicate emission feature 
resulting from the PAHFIT method
are lower by $\simlt$5\% than that 
from the spline method.
The flux differences between that derived from
the spline method and that from PAHFIT 
are more pronounced 
in F21013-0739 and \sd\ than F10398+1455, 
with the spline-based 9.7$\mum$ emission
feature being broader and less prominent.\footnote{%
   We note that, although the spline-based 
   9.7$\mum$ silicate emission feature of \sd\ 
   is almost too weak to be noticeable,
   the 18$\mum$ O--Si--O silicate bending
   feature is clearly seen
   (see Figure~2 of Xie et al.\ 2014). 
   }
With $d\ln F_\nu/d\ln\lambda$\,$\approx$\,2.7 
and 2.6 for F21013-0739 and \sd\ respectively, 
the PAHFIT method results in a less steep
$\simali$5--8$\mum$ continuum, but still much
steeper than that of AGN
($d\ln F_\nu/d\ln\lambda$\,$\approx$\,0.8).
The relatively high flux density drawn from 
the spline approach could be due to an incomplete 
subtraction of the PAH wings. 
We will discuss in \S\ref{sec:spline}
the differences in our modeling results 
derived from the two residual spectra.
%
%

%

\section{Dust Model\label{sec:model}}
To model the dust emission, we need to specify 
the dust chemical composition, size, and temperature. 
For dust composition, we will consider amorphous silicate 
and graphite or amorphous carbon, although other dust species
(e.g., SiC, oxides) may be present in AGN tori
(e.g., see Laor \& Draine 1993,
Markwick-Kemper et al.\ 2007, 
K\"oher \& Li 2010). 
%

In the Milky Way diffuse ISM the dust grain sizes range 
from a few Angstroms to a few tenth micrometers 
(see Li 2004a). In AGN the dust is expected to 
be skewed toward a larger average grain size
due to the preferential destruction of small dust in 
the harsh environments of AGN (see Li 2007), 
and probably also because of grain growth 
in the dense circumnuclear regions
(see Maiolino et al.\ 2001).
For simplicity, 
we will not consider size distributions,
instead, we will only consider seven discrete sizes:
$a$\,=\,0.1, 0.5, 1.0, 1.5, 2.0, 5.0 and 10.0$\mum$
where $a$ is the radius of the dust grains 
(assumed spherical).
The dust sizes considered here are broad enough 
to cover both sub-$\mu$m-sized dust typical of
the Galactic diffuse ISM (see Li 2004a)
and large, $\mu$m-sized dust typical of AGN 
(e.g., see Li et al.\ 2008, Smith \etal 2010, 
Lyu et al.\ 2014, Z.~Shao et al.\ 2015, in preparation). 

The dust in AGN tori is expected to 
have a range of temperatures. 
In the inner ``wall'' region of the torus,
the dust temperature reaches $\simgt$1500$\K$, 
the sublimation temperature of silicate and 
graphite materials. 
The dust temperature decreases when moving away
from the inner ``wall.'' 
To fully account for the temperature distribution,
we need to perform radiative transfer calculations
for the dust in AGN tori 
heated by the central engine.
For simplicity, we will only consider two temperatures
--- a warm component and a cold component --- to 
represent the temperature distribution.

We assume that the observed IR emission 
results from silicate dust and carbon dust 
(i.e., graphite or amorphous carbon). 
For each dust species, we consider 
a warm component with temperature $\Tw$ 
and a cold component with temperature $\Tc$. 
Assuming the torus is optically thin in the IR, 
we model the dust IR emission as
\begin{equation}
F_{\nu} = \frac{1}{d^2} \times 
          \sum_{i} \left\{
          B_\nu(\Twi) 
          \times \knui
          \times \Mwi 
        +  B_\nu(\Tci) 
          \times \knui
          \times \Mci\right\} ~~,
\end{equation}
where the sum is over the two dust species
(silicate and graphite or amorphous carbon),
$d$ is the luminosity distance of the object,
$\knui$ is the mass absorption coefficient
or opacity (in unit of $\cm^{2}\g^{-1}$)
of dust of type $i$,
$B_\nu(T)$ is the Planck function
of temperature $T$ at frequency $\nu$,
$\Twi$ and $\Tci$ are respectively
the temperatures of the warm and cold components
of dust of type $i$, and
$\Mwi$ and $\Mci$ are respectively
the masses of the warm and cold components
of dust of type $i$.
For a given composition and size, 
we calculate the mass absorption coefficient
$\kappa_{\rm abs}(\nu)$ from Mie theory 
(Bohren \& Huffman 1983)
using the refractive index of 
the corresponding dust material. 

To summarize, there are in total eight parameters 
in fitting the observed IR emission:
the temperature ($\Twsil$) and mass ($\Mwsil$) 
for the warm silicate component,
the temperature ($\Tcsil$) and mass ($\Mcsil$) 
for the cold silicate component,
the temperature ($\Twcarb$) and mass ($\Mwcarb$) 
for the warm carbon dust component, and
the temperature ($\Tccarb$) and mass ($\Mccarb$) 
for the cold carbon dust component.
We obtain the best fit for each galaxy 
using the MPFIT code, 
an IDL $\chi^{2}$-minimization routine 
based on the Levenberg--Marquardt 
algorithm (Markwardt 2009).

We require the dust temperatures not to exceed
the sublimation temperature 
($T_{\rm subl}$\,$\simali$1500$\K$) 
of silicate and graphite materials.
For the dust mass, we could apply 
the elemental abundances to constrain
the mass ratio of the silicate component 
to the carbon dust component. 
Let $\xtot$ be the total abundance
of element X relative to H,
$\xgas$ be the amount of X in the gas phase,
and $\xdust$ be the amount of X contained in dust.
As an element will be in the gas-phase 
and/or locked up in the solid-phase,  
one naturally obtains $\xdust$\,=\,$\xtot-\xgas$.
Let $\mux$ be the atomic weight of X
($\mux$\,$\approx$\,12, 16, 56, 24, and 28
for carbon, oxygen, iron, magnesium, and silicon).
If we assume a stoichiometric composition of
Mg$_{\rm 2x}$Fe$_{\rm 2(1-x)}$SiO$_4$
for silicate (i.e., each silicon atom
corresponds to four oxygen atoms), 
the mass ratio of the carbon dust component
to the silicate component is
\begin{equation}\label{eq:Msil2MH}
\Mcarb/\Msil = \frac{\muc\left\{\ctot-\cgas\right\}} 
               {\mufe\fedust + \mumg\mgdust
                + \musi\sidust + 4\times \mu_{\rm O}\sidust} ~~.
\end{equation}
If we assume that 
(i) these galaxies have
the solar C, Fe, Mg and Si abundances,\footnote{%
  $\csun\approx269\pm31\ppm$,
  $\fesun\approx31.6\pm2.9\ppm$,
  $\mgsun\approx39.8\pm3.7\ppm$, and
  $\sisun\approx32.4\pm2.2\ppm$ 
  (Asplund et al.\ 2009).
  }
(ii) all Fe, Mg and Si elements 
are depleted in silicate dust, and 
(iii) $\simali$37\% of the C is in the gas-phase 
(i.e., $\cgas\approx100\ppm$, Sofia et al.\ 2011),
we derive $\Mcarb/\Msil\approx0.36$. 
With C atoms all depleted in dust
(i.e., $\cgas=0\ppm$), 
one obtains an upper limit of 
$\Mcarb/\Msil\approx0.57$.
A higher $\Mcarb/\Msil$ ratio could be achieved
if one assumes an enstatite (MgSiO$_3$) composition
for the silicate component:
$\Mcarb/\Msil\approx0.59$ 
for $\cgas = 100\ppm$, or
$\Mcarb/\Msil\approx0.95$ 
for $\cgas = 0\ppm$.
As we have no information about
the elemental abundances and depletion
for these three galaxies,
in the modeling we will allow the mass ratio
to vary between $0.2<\Mcarb/\Msil<2$.
    
We estimate the uncertainties
for the eight model parameters 
by performing Monte-Carlo simulations.
For the {\it Spitzer}/IRS spectrum of each source, 
we assume that the flux density error  
statistically follows a normal distribution. 
The dispersion is characterized by 
the observed 1$\sigma$ error,
composed of the statistical 
and systematic errors. 
The latter arises from 
the flux differences between the two nods of 
the {\it Spitzer}/IRS spectra, 
the sky background contamination,
and the {\it Spitzer}/IRS pointing
and flux calibration errors
(Lebouteiller \etal 2011).\footnote{%
  We note that the signal-to-noise ratio (SNR) 
  of the {\it Spitzer}/IRS spectrum 
  in the $\simali$5--14.5$\mum$ wavelength interval 
  is lower than that in the interval of 
  $\simali$14.5--38$\mum$.
  This was caused by the different
  observational modules
  (i.e., the 5--14.5$\mum$ {\it Short Low} IRS module
  with a slit width of $3\farcs6$, 
  and the 14.5--38$\mum$ {\it Long Low} IRS module
  with a slit width of $11\farcs2$).
  To fit the 5--8$\mum$ continuum emission
  and the 9.7$\mum$ silicate emission feature,
  we arbitrarily increase the weights 
  for the data points at 5--8 and 8--14.5$\mum$ 
  by ten and two times, respectively,
  compared to that at 14.5--38$\mum$. 
  }
We generate a new ``observational'' spectrum 
through randomly sampling a point at each wavelength 
from the normal distribution.
We then model the new spectrum 
and derive a set of model parameters. 
We conduct 100 simulations for each source
as the parameters derived from 10,000 simulations
only slightly differ from that derived from 100 simulations.
The final model spectrum is calculated 
from the median values of the model parameters.
The error of each parameter 
is derived from the standard deviation 
of 100 simulations. 
%

\begin{figure}
\vspace{-2mm}
\begin{center}
$
\begin{array}{cc} 
\resizebox{0.42\hsize}{!}{\includegraphics
{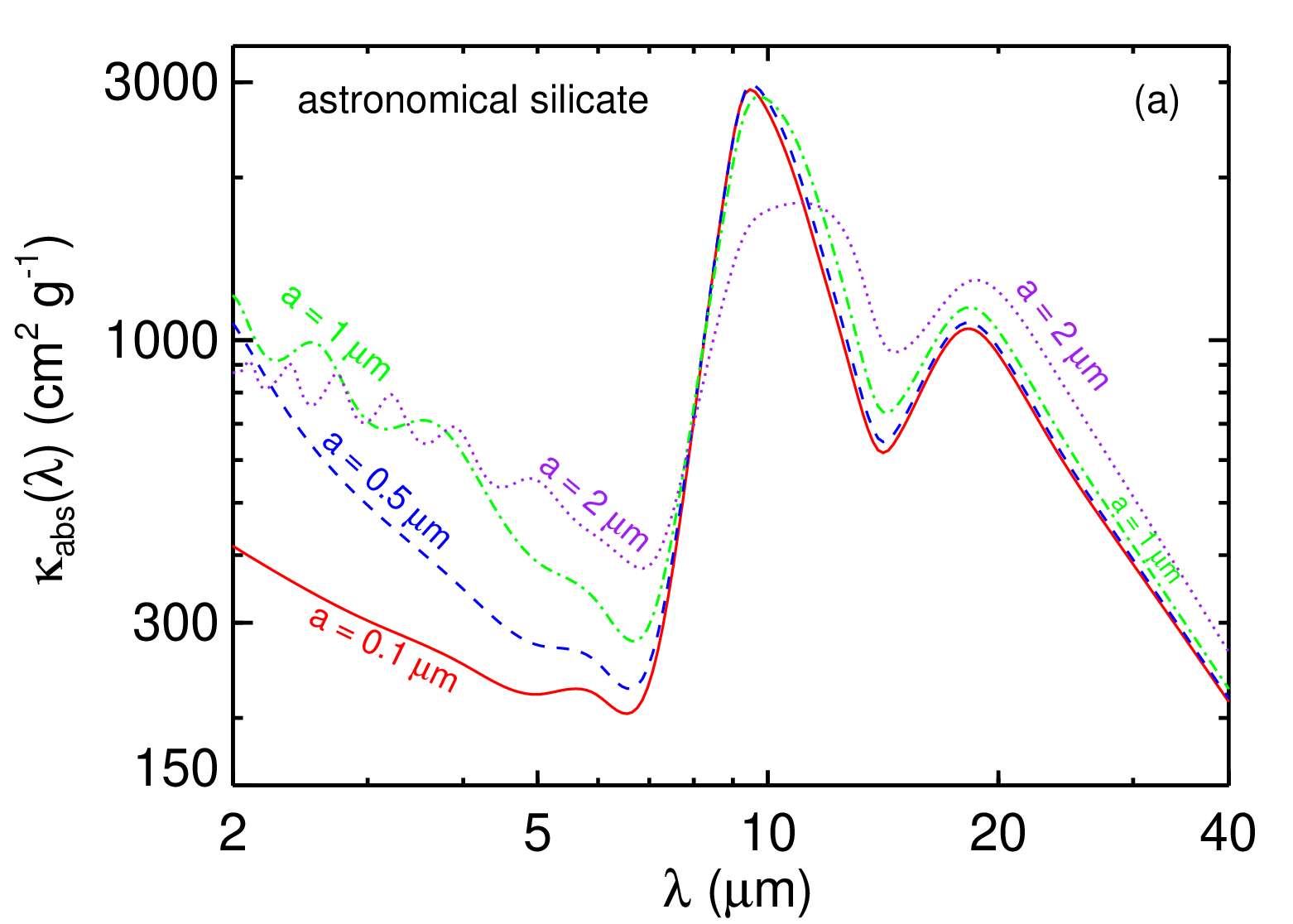}}
&
\resizebox{0.42\hsize}{!}{\includegraphics
{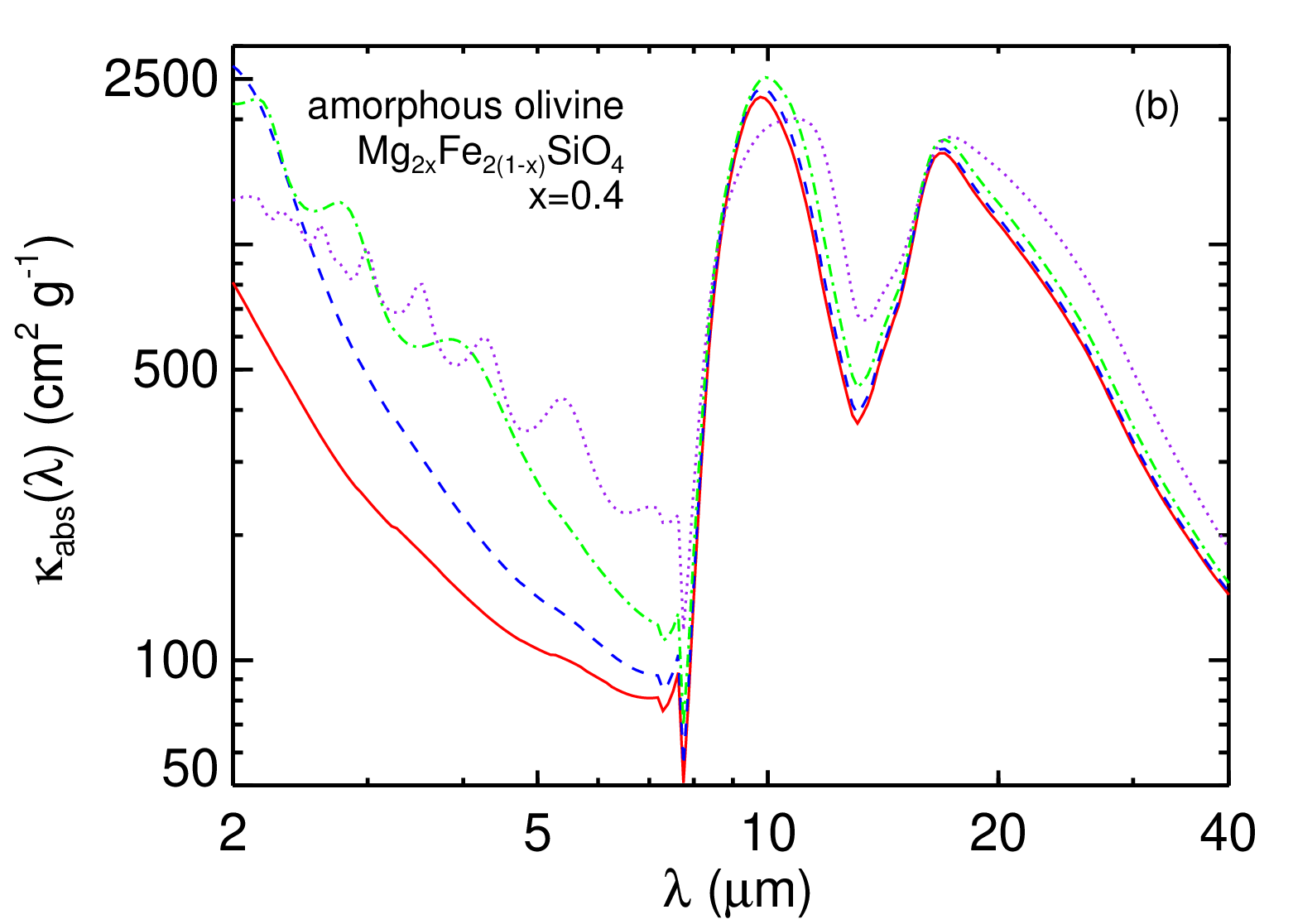}} \\
\resizebox{0.42\hsize}{!}{\includegraphics
{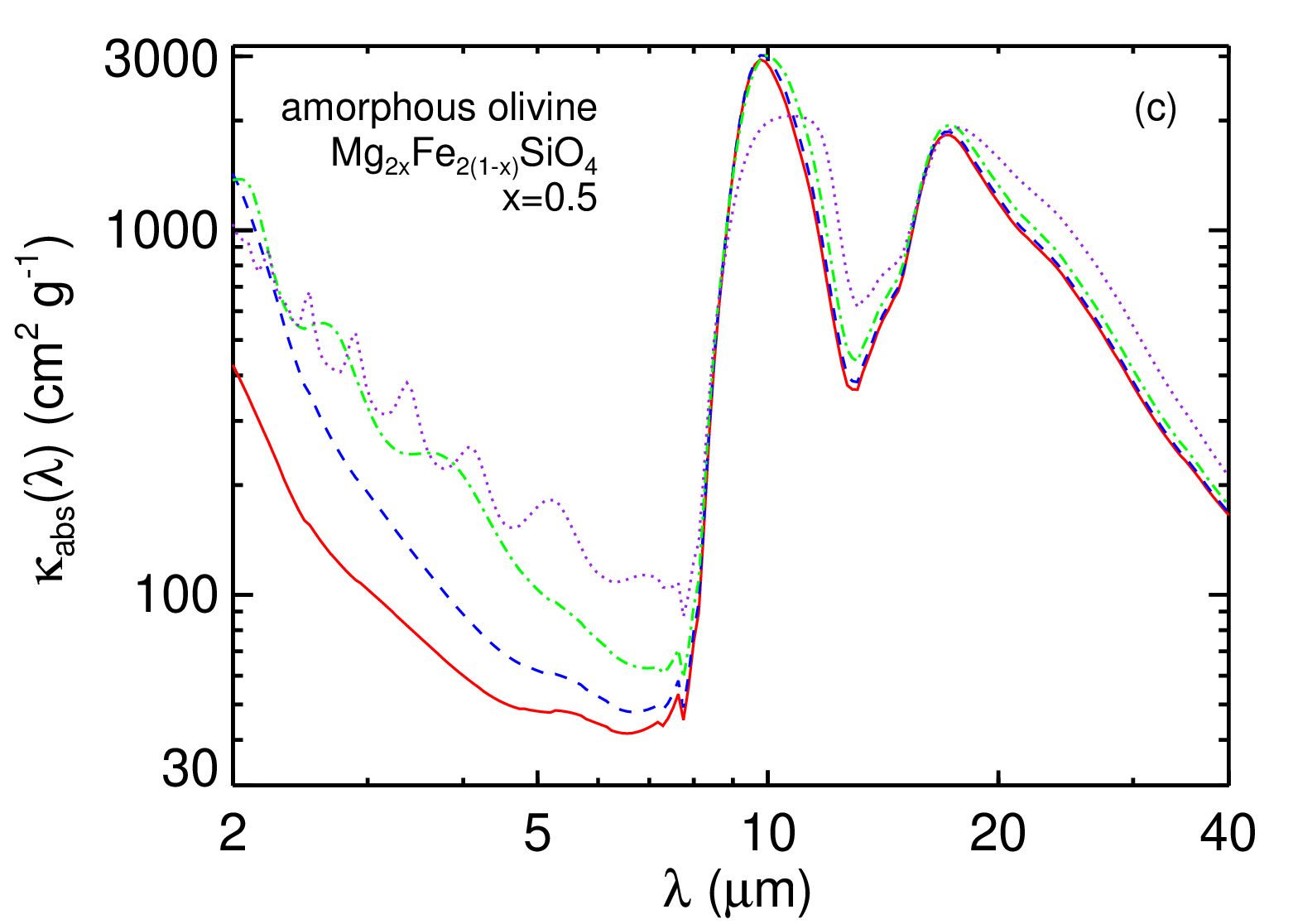}}
&
\resizebox{0.42\hsize}{!}{\includegraphics
{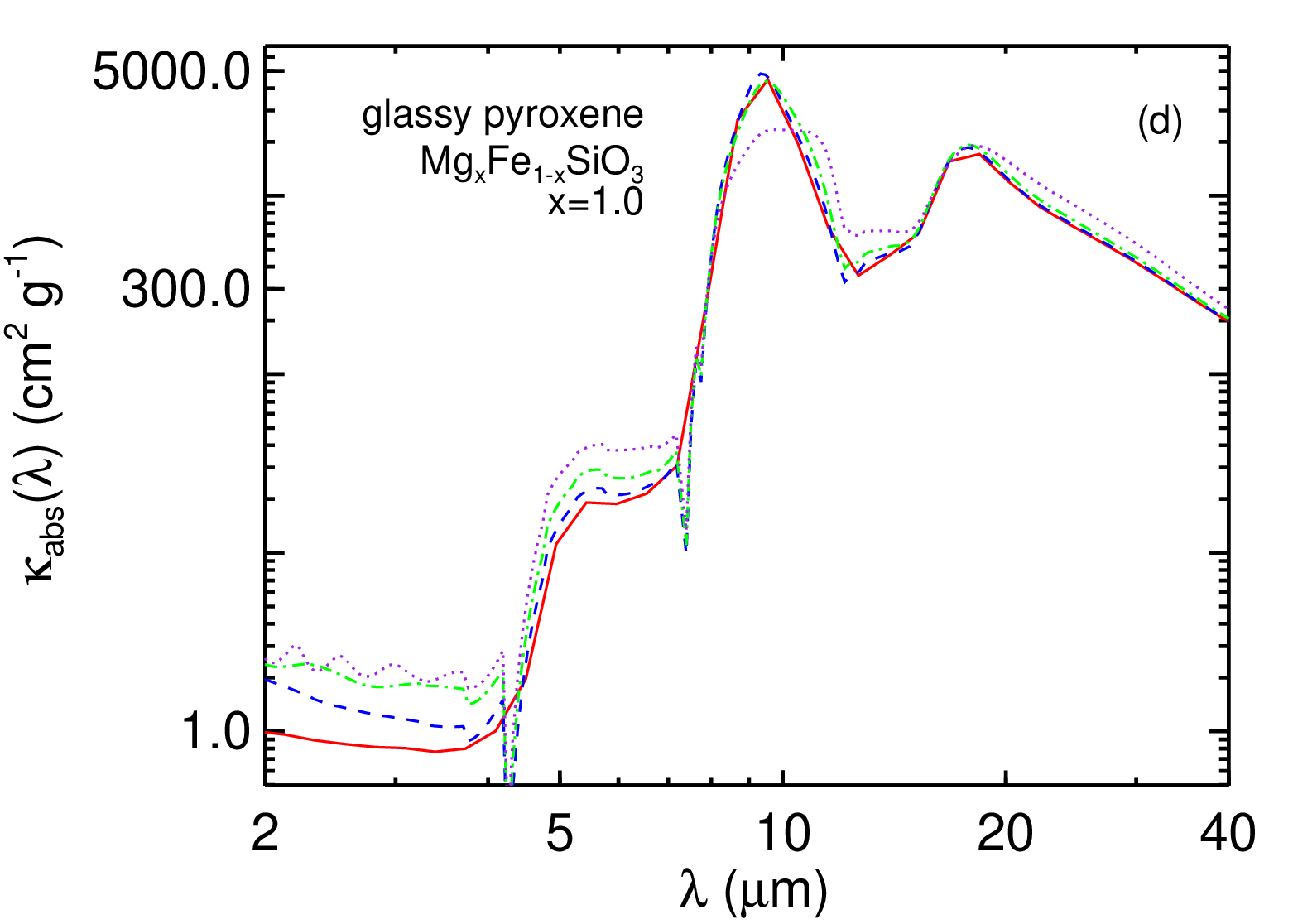}} \\
\resizebox{0.42\hsize}{!}{\includegraphics
{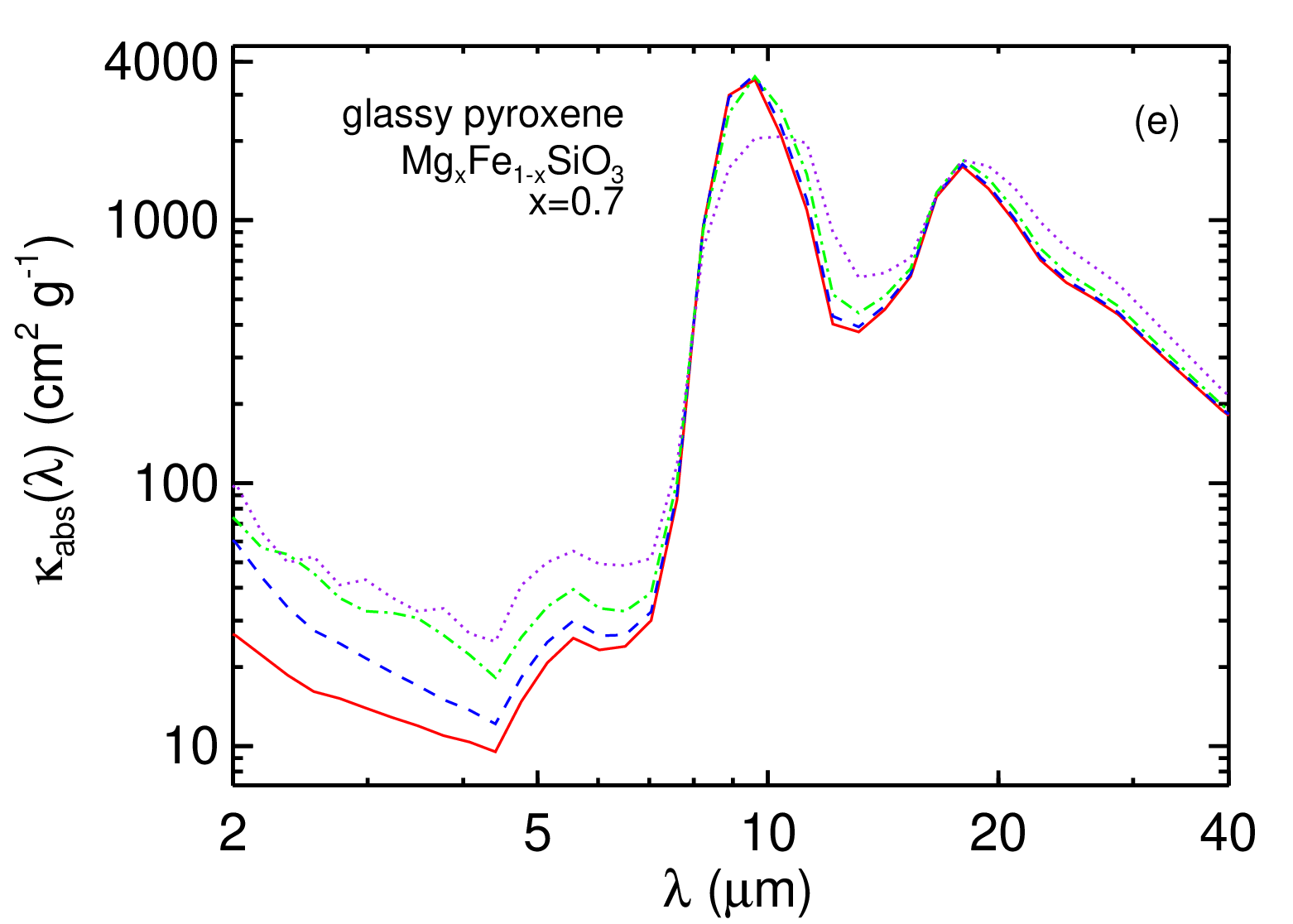}} 
&
\resizebox{0.42\hsize}{!}{\includegraphics
{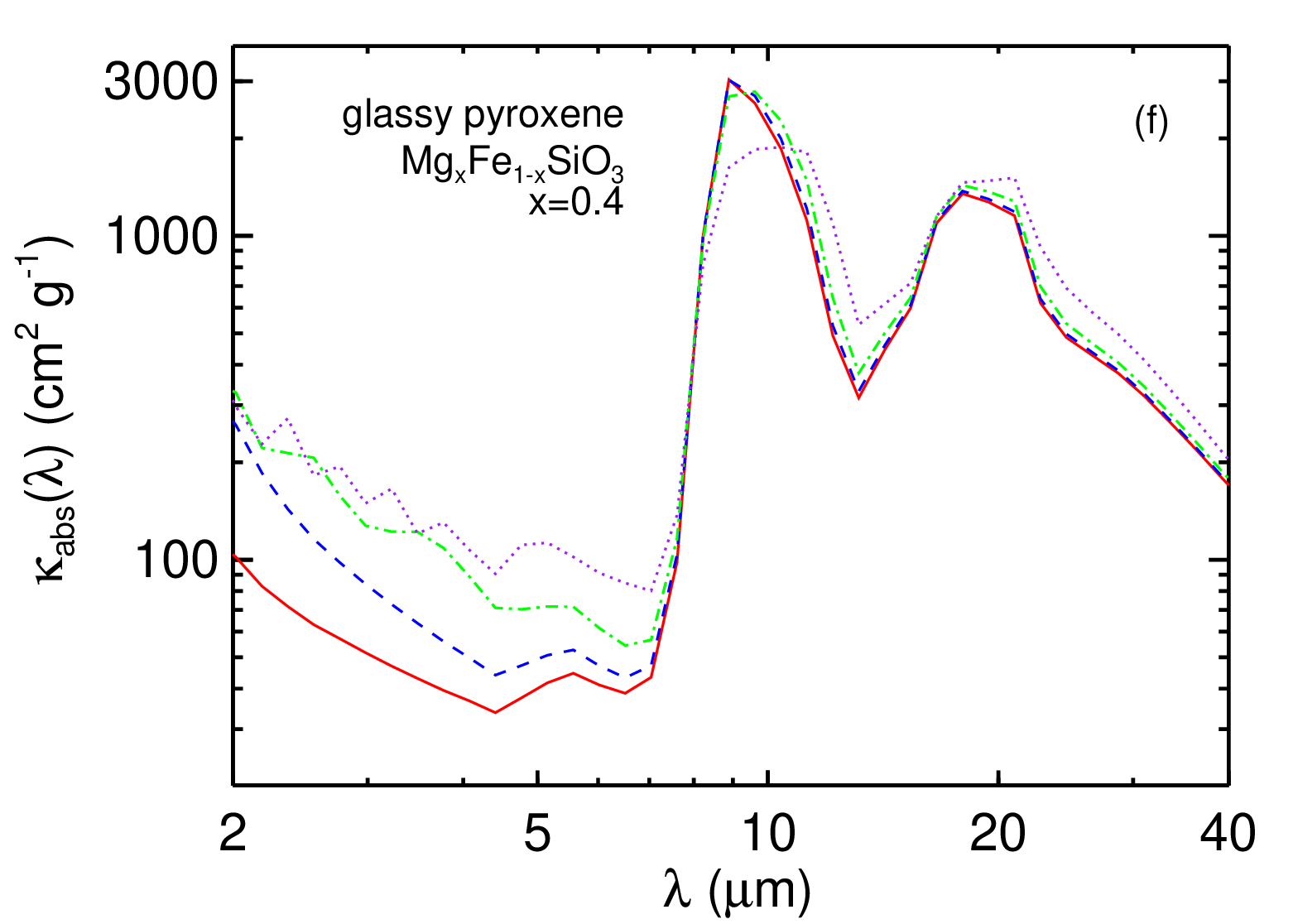}}  \\
\resizebox{0.42\hsize}{!}{\includegraphics
{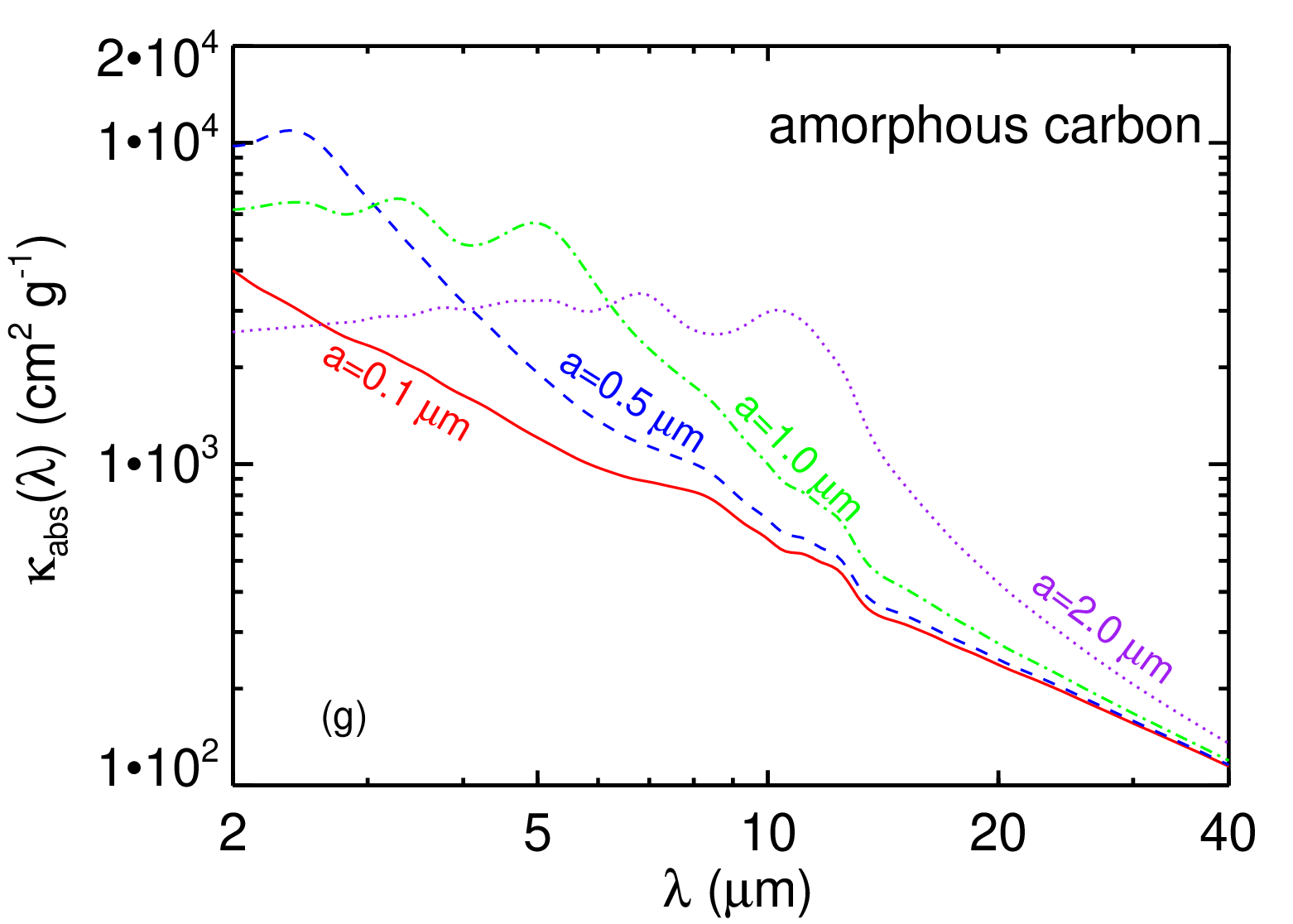}}
&
\resizebox{0.42\hsize}{!}{\includegraphics
{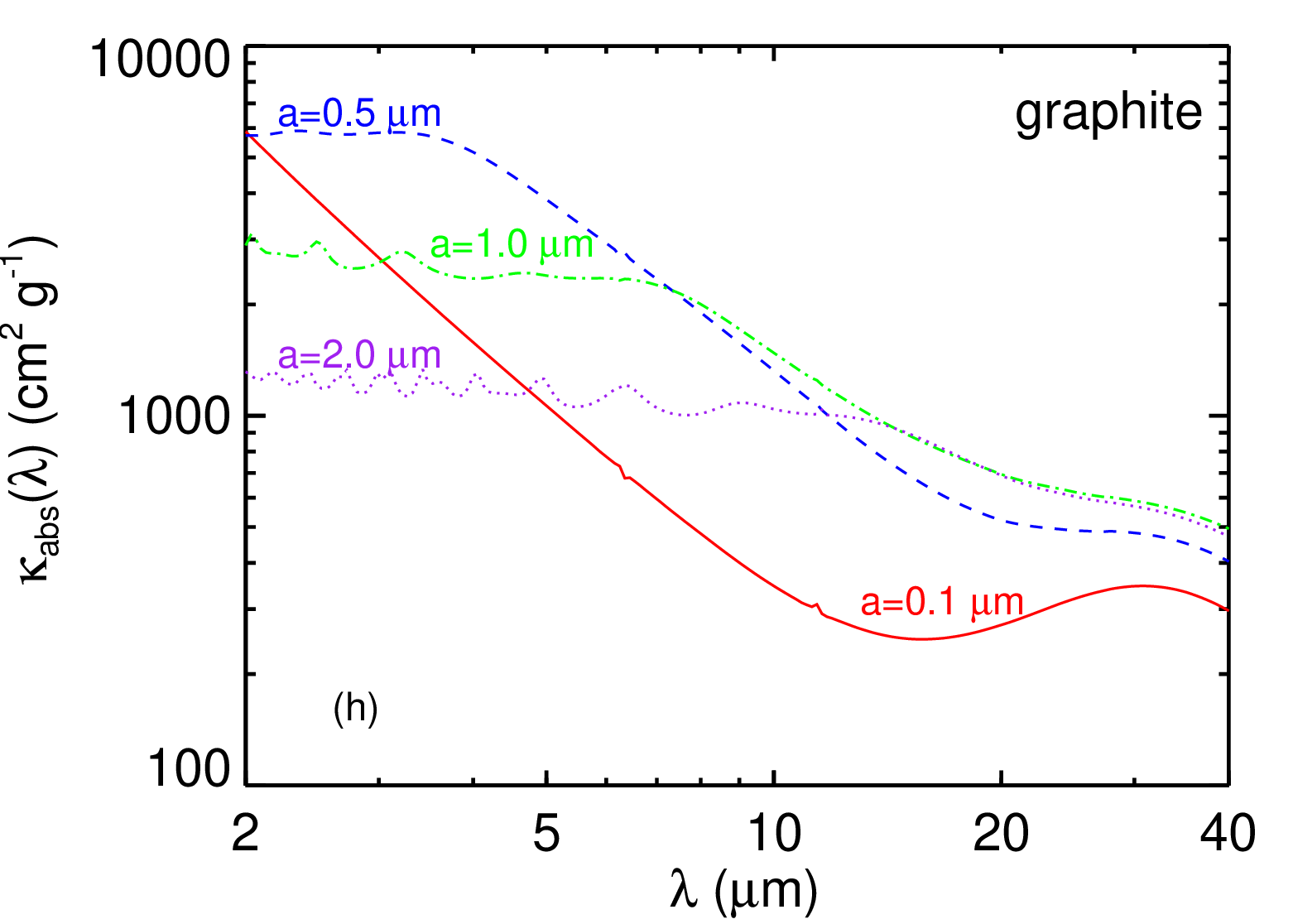}}
\end{array}
$
\caption{\footnotesize 
         \label{fig:kappa1} 
         Mass absorption coefficients or 
         opacities $\kappa_{\rm abs}(\lambda)$ for
         (a) ``astronomical silicates'' 
         (Draine \& Lee 1984),
         (b) amorphous olivine Mg$_{0.8}$Fe$_{1.2}$SiO$_4$ 
             (Dorschner et al.\ 1995),
         (c) amorphous olivine MgFeSiO$_4$ 
             (Dorschner et al.\ 1995),
         (d) glassy pyroxene MgSiO$_{3}$
             (Dorschner et al. 1995),
         (e) glassy pyroxene Mg$_{0.7}$Fe$_{0.3}$SiO$_{3}$
             (Dorschner et al. 1995),
         (f) glassy pyroxene Mg$_{0.4}$Fe$_{0.6}$SiO$_{3}$
             (Dorschner et al. 1995),
         (g) amorphous carbon (Rouleau \& Martin 1991), and
         (h) graphite (Draine \& Lee 1984).
           All dust species are assumed spherical
          with a range of radii:
          $a$\,=\,0.1$\mum$ (red solid),
          $a$\,=\,0.5$\mum$ (blue dashed),
          $a$\,=\,1$\mum$ (green dotted--dashed), and
          $a$\,=\,2$\mum$ (purple dotted).
          }
\end{center}
\end{figure}

The mass absorption coefficient or opacity
$\kappa_{\rm abs}(\nu)$ depends on
the dust size, shape, and composition
through the dielectric function or
index of refraction of the dust material.
For graphite, we take the index of refraction 
of Draine \& Lee (1984). 
For amorphous carbon, we take the index of 
refraction of Rouleau \& Martin (1991).
For silicate, we consider a range of compositions:
(i) ``astronomical silicates'' of Draine \& Lee (1984), 
(ii) pyroxene Mg$_{\rm x}$Fe$_{\rm 1-x}$SiO$_3$ 
     with ${\rm x}$\,=\,0.4, 0.7, 1.0 
     (Dorschner \etal 1995), and 
(iii) olivine Mg$_{\rm 2x}$Fe$_{\rm 2\left(1-x\right)}$SiO$_4$ 
      with ${\rm x}$\,=\,0.4, 0.5 (Dorschner \etal 1995).  
In Figure~\ref{fig:kappa1}, we show the mass absorption 
coefficient $\kappa_{\rm abs}(\lambda)$ of each dust species 
of different sizes spanning the wavelength range of 2--40$\mum$. 

For all silicate species, 
the mass absorption coefficients 
$\kappa_{\rm abs}(\lambda)$ 
in the 2--40$\mum$ wavelength range 
are featured by a relatively smooth continuum 
at $\simali$2--8$\mum$ and two prominent bumps 
peaking around 9.7 and 18$\mum$. 
The $\simali$2--8$\mum$ continuum increases
with the dust size. 
The 9.7 and 18$\mum$ features also vary with 
the dust size: they substantially broaden 
and shift to longer wavelengths 
for $a\gtsim1.5\mum$.\footnote{%
  For $a\simlt0.5\mum$ 
  (and even up to $a\simlt1\mum$), 
  the 9.7 and 18$\mum$ features are 
  insensitive to the dust size
  since grains with $a\simlt0.5\mum$ satisfy
  the Rayleigh scattering condition of 
  $2\pi a/\lambda\ll 1$ at these spectral bands
  (Bohren \& Huffman 1983).
  }
For amorphous carbon and graphite,
the $\simali$2--8$\mum$ continuum 
mass absorption coefficients  
flatten off with increasing dust size.

\begin{figure}
\begin{center}
\resizebox{0.8\hsize}{!}{
\includegraphics{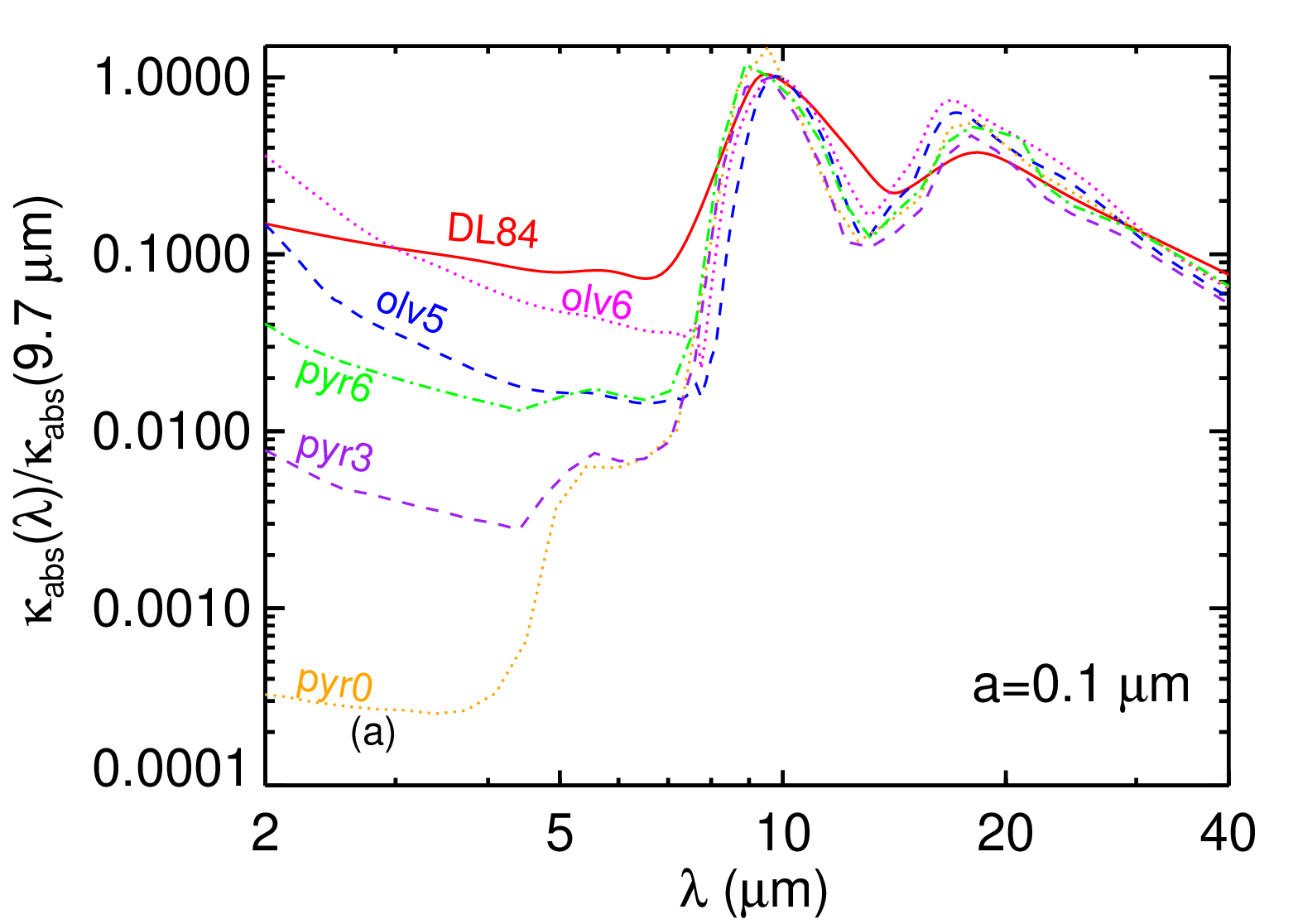}} \\
\resizebox{0.8\hsize}{!}{
\includegraphics{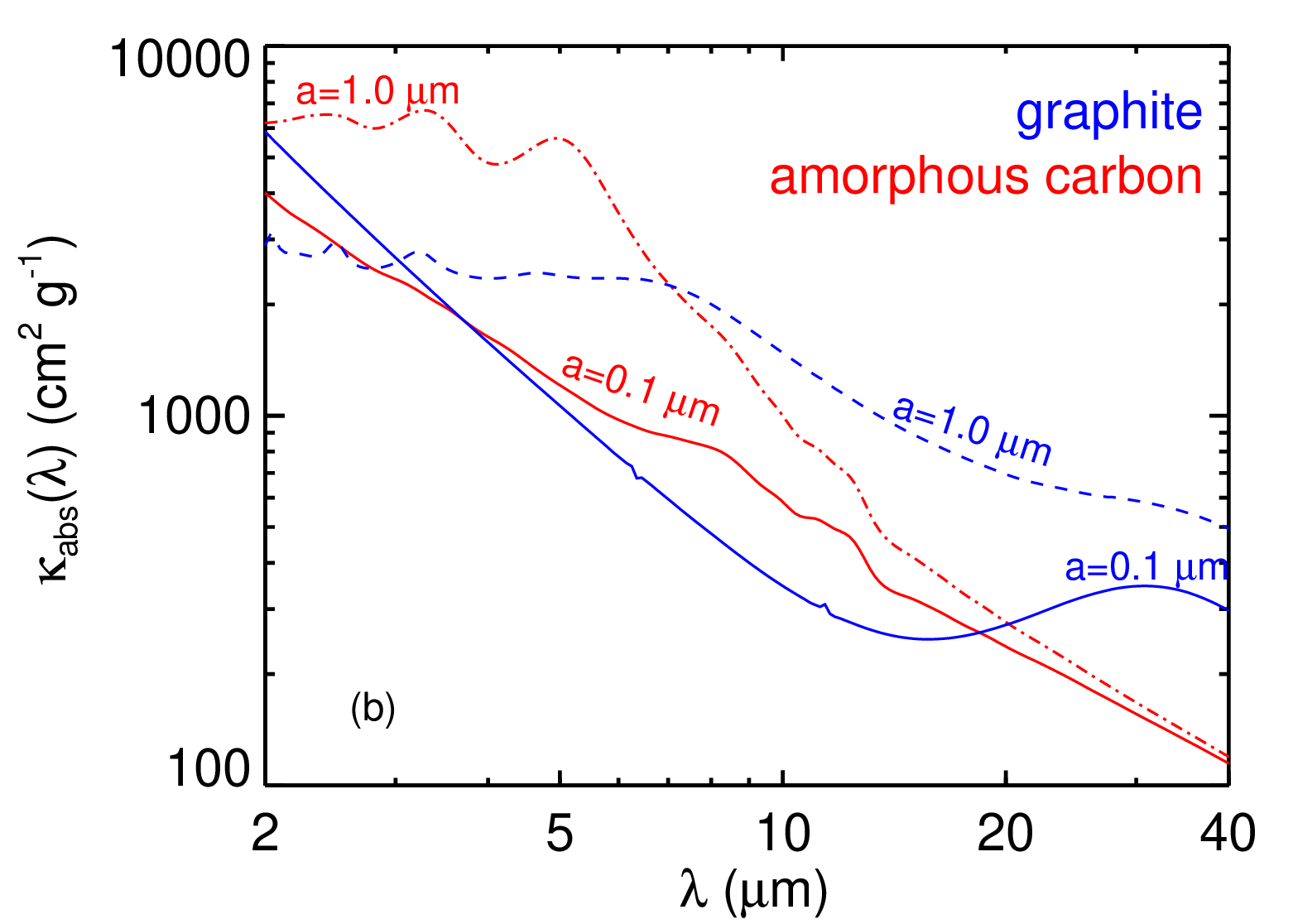}}
\caption{\footnotesize
         \label{fig:kappa2}
         Top panel (a): comparison of the 2--40$\mum$
         opacity of the Draine \& Lee (1984) 
         ``astronomical silicate'' 
         (``DL84,'' solid red) 
         with that of pyroxene MgSiO$_{3}$ 
         (``pyr0,'' dotted orange, Dorschner et al.\ 1995), 
         pyroxene Mg$_{0.7}$Fe$_{0.3}$SiO$_{3}$ 
         (``pyr3,'' dashed purple, Dorschner et al.\ 1995), 
         pyroxene Mg$_{0.4}$Fe$_{0.6}$SiO$_{3}$ 
         (``pyr6,'' dotted-dashed green, Dorschner et al.\ 1995), 
         olivine MgFeSiO$_{4}$ 
         (``olv5,'' dashed blue, Dorschner et al.\ 1995), and
         olivine Mg$_{0.8}$Fe$_{1.2}$SiO$_{4}$ 
         (``olv6,'' dotted magenta, Dorschner et al.\ 1995).
         All silicate dust species are taken to be spherical. 
         All dust species are assumed spherical
         with a radius of $a=0.1\mum$.
         Bottom panel (b): comparison of the 2--40$\mum$
         opacities of graphite (Draine \& Lee 1984)
         of $a=0.1\mum$ (solid blue)
         and $a=1\mum$ (dashed blue) with that of 
         amorphous carbon (Rouleau \& Martin 1991)
         of $a=0.1\mum$ (solid red)
         and $a=1\mum$ (dotted--dashed red).
         }
\end{center}
\end{figure}

The $\simali$2--8$\mum$ continuum 
mass absorption coefficients are
sensitive to the silicate composition.
As illustrated in Figure~\ref{fig:kappa2}a,
for a given dust size,
the $\simali$2--8$\mum$ continuum opacity
increases significantly with the iron 
fraction of the dust.
In general, amorphous olivine has
a higher $\simali$2--8$\mum$ continuum 
opacity than pyroxene.
We note that the dielectric functions of the 
Draine \& Lee (1984) ``astronomical'' silicate 
were synthesized to be ``dirty'' so that it is 
absorptive in the near-UV and optical wavelength 
ranges and also has a high
$\simali$2--8$\mum$ continuum opacity.
In Figure~\ref{fig:kappa2}b, we compare
the 2--40$\mum$ opacities of graphite
and amorphous carbon.
In the $\simali$5--8$\mum$ wavelength range
of interest here, amorphous carbon has
a higher opacity than graphite.

\begin{deluxetable} {lccrrrllllll}  
\tabletypesize\tiny
\tablecolumns{12}
\tablewidth{0pc}
\tablecaption{\footnotesize
              \label{tab:modpara}
              Model Parameters for 
              Three Spectroscopically 
              Anomalous Galaxies (IRAS~F10398+1455, 
              IRAS~F21013-0739, and SDSS~J0808+3948)
              and Two Comparison Sources
              (Quasar Average Spectrum, 
               and ULIRG IRAS~FSC\,10214+4724)
               } 
\tablehead
{ 
\multicolumn{1}{l}{Object} & 
\multicolumn{1}{c}{Composition} & 
\multicolumn{1}{c}{Dust Size} & 
\multicolumn{1}{c}{$\Twsil$} & 
\multicolumn{1}{c}{$\Twcarb$} & 
\multicolumn{1}{c}{$\Tcsil$} &
\multicolumn{1}{c}{$\Tccarb$} &
\multicolumn{1}{c}{$\Mwsil$} & 
\multicolumn{1}{c}{$\Mwcarb$} & 
\multicolumn{1}{c}{$\Mcsil$} &
\multicolumn{1}{c}{$\Mccarb$} & 
\multicolumn{1}{c}{$\rm\chi^{2}/{dof}$}  \\
 & 
 & 
\multicolumn{1}{c}{($\mu$m)} & 
\multicolumn{1}{c}{(K)} &
\multicolumn{1}{c}{(K)} & 
\multicolumn{1}{c}{(K)} &
\multicolumn{1}{c}{(K)} &
\multicolumn{1}{c}{(${10^{34}}$ g)} &
\multicolumn{1}{c}{(${10^{34}}$ g)} &
\multicolumn{1}{c}{(${10^{37}}$ g)} &  
\multicolumn{1}{c}{(${10^{37}}$ g)} & 
}
\rotate
\startdata
F10398+1455$^{a}$ & DL84\,+\,ac & 0.1 & 271$\pm$1.6 & 389$\pm$2.8 & 92$\pm$0.2 & 117$\pm$0.1 & 14.0$\pm$0.5 & 3.1$\pm$0.2 & 5.9$\pm$0.03 & 11.9$\pm$0.07 & 38\\
F10398+1455$^{b}$ & DL84\,+\,ac & 0.1 & 248$\pm$4.9 & 295$\pm$1.3 & 84$\pm$0.5 & 115$\pm$0.3 & 15.4$\pm$2.1 & 27.0$\pm$4.3 & 7.4$\pm$0.08 & 14.8$\pm$0.2  & 42 \\ 
F10398+1455$^{a}$ & olv6\,+\,ac & 0.1 & 303$\pm$2.0 & 323$\pm$2.2 & 76$\pm$0.3 & 113$\pm$0.1 & 8.8$\pm$0.3 & 10.9$\pm$0.5 & 8.7$\pm$0.04 & 17.4$\pm$0.09 & 25  \\
F10398+1455$^{a}$ & pyr3\,+\,ac & 1.0 & 248$\pm$3.0 & 259$\pm$1.5 & 74$\pm$0.8 & 109$\pm$0.3 & 16.0$\pm$1.1 & 20.6$\pm$1.9 & 9.1$\pm$0.2 & 18.3$\pm$0.4 & 24 \\
F10398+1455$^{a}$ & DL84\,+\,gra & 0.1 & 291$\pm$4.2 & 319$\pm$2.6 & 120$\pm$0.6 & 141$\pm$0.7 & 7.6$\pm$0.5 & 15.2$\pm$1.1 & 1.7$\pm$0.05  & 2.7$\pm$0.1 & 36 \\
F10398+1455$^{a}$ & DL84\,+\,ac & 1.0 & 350$\pm$30.0 & 284$\pm$3.9 & 93$\pm$0.4 & 113$\pm$1.1 & 3.7$\pm$2.9 & 7.4$\pm$6.0 & 6.1$\pm$0.1 & 12.2$\pm$0.2 & 59\\
F10398+1455$^{a}$ & DL84\,+\,gra & 1.0 & 395$\pm$13.8 & 385$\pm$5.7 & 112$\pm$0.1 & 121$\pm$0.4 & 2.4$\pm$0.3 & 1.1$\pm$0.3 & 1.8$\pm$0.02 & 3.1$\pm$0.08 & 30 \\
F10398+1455$^{a}$ & olv6\,+\,ac & 1.0 & 346$\pm$4.0 & 285$\pm$1.6 & 81$\pm$0.4 & 111$\pm$0.3 & 4.5$\pm$0.2 & 9.0$\pm$0.4 & 8.0$\pm$0.1 & 16.0$\pm$0.2 & 30 \\
F10398+1455$^{a}$ & DL84\,+\,ac & 0.1 & 244$\pm$3.3 & 357$\pm$3.2 & 92$\pm$1.0 & 115$\pm$0 & 23.7$\pm$2.1 & 5.4$\pm$0.6 & 6.2$\pm$0.2 & 12.5$\pm$0.3 & 42 \\ 
F10398+1455$^{a}$ & DL84\,+\,ac & 1.5 & 410$\pm$4.5 & 374$\pm$4.1 & 93$\pm$0.2 & 109$\pm$0.1 & 2.4$\pm$0.1 & 0.6$\pm$0.05 & 6.2$\pm$0.03 & 12.4$\pm$0.05 & 99\\
F10398+1455$^{a}$ & olv6\,+\,ac & 1.5 & 358$\pm$13.4 & 354$\pm$4.5 & 84$\pm$1.1 & 108$\pm$0.2 & 4.0$\pm$0.5 & 1.3$\pm$0.3 & 7.6$\pm$0.3 & 15.0$\pm$0.5 & 51 \\
F10398+1455$^{a}$ & pyr3\,+\,ac & 1.5 & 331$\pm$3.5 & 385$\pm$3.0 & 78$\pm$1.6 & 109$\pm$0.4 & 4.5$\pm$0.2 & 0.9$\pm$0.04 & 8.3$\pm$0.3 & 16.6$\pm$0.7 & 27 \\
F10398+1455$^{a}$ & DL84\,+\,ac & 2.0 & 419$\pm$3.1 & 334$\pm$3.0 & 99$\pm$0.1 &  99$\pm$0.1 & 2.8$\pm$0.06 & 1.2$\pm$0.1 & 6.0$\pm$0.02 & 12.0$\pm$0.04 & 201\\
F10398+1455$^{a}$ & olv6\,+\,ac & 2.0 & 464$\pm$50.0 & 324$\pm$14.0 & 89$\pm$0.2 & 99$\pm$0.5 & 1.5$\pm$1.3 & 2.6$\pm$2.4 & 7.9$\pm$0.1 & 15.8$\pm$0.2 & 168 \\
F10398+1455$^{a}$ & pyr3\,+\,ac & 2.0 & 641$\pm$8.4 & 382$\pm$2.4 & 89$\pm$0.1 & 102$\pm$0.1 & 0.6$\pm$0.02 & 1.1$\pm$0.04 & 7.5$\pm$0.02 & 15.0$\pm$0.05 & 101 \\
\hline
F21013-0731$^{a}$ & DL84\,+\,ac & 0.1 & 249$\pm$4.0 & 361$\pm$5.0 & 61$\pm$0.4 & 110$\pm$0.5 & 24.6$\pm$2.1 &  4.9$\pm$0.4 &  15.0$\pm$0.8  & 19.8$\pm$1.8 & 18 \\
F21013-0731$^{b}$ & DL84\,+\,ac & 0.1 & 240$\pm$0.0 & 301$\pm$1.3 & 74$\pm$0.8 & 120$\pm$3.6 & 16.2$\pm$0.5 & 32.5$\pm$1.0 & 20.7$\pm$4.4  & 10.3$\pm$4.0 & 21 \\
F21013-0731$^{a}$ & olv6\,+\,ac & 0.1 & 334$\pm$5.0 & 318$\pm$4.6 & 58$\pm$0.4 & 113$\pm$0.6 & 6.3$\pm$0.4 & 12.0$\pm$0.1 & 49.0$\pm$3.9  & 16.7$\pm$2.5 & 17\\
F21013-0731$^{a}$ & pyr3\,+\,ac & 1.0 & 251$\pm$9.2 & 255$\pm$3.2 & 66$\pm$3.7 & 110$\pm$1.1 & 16.4$\pm$4.5 & 21.7$\pm$6.9 & 26.0$\pm$9.4  & 16.3$\pm$19.3 & 17\\
F21013-0731$^{a}$ & DL84\,+\,gra & 0.1 & 267$\pm$0.5 & 327$\pm$5.5 & 111$\pm$0.4 & 144$\pm$0.6 & 14.4$\pm$0.02 & 12.2$\pm$1.4 & 1.7$\pm$0.003  & 2.6$\pm$0.03 & 19\\
F21013-0731$^{a}$ & DL84\,+\,ac & 0.1 & 270$\pm$4.3 & 374$\pm$5.3 & 74$\pm$2.0 & 115$\pm$0.0 & 15.0$\pm$1.4 &  3.7$\pm$0.6 &  9.7$\pm$1.1  & 15.0$\pm$2.6 & 20 \\
\hline
J0808+3948$^{a}$ & DL84\,+\,ac & 0.1 & 265$\pm$55.0 & 363$\pm$50.0 & 84$\pm$11.7 & 110$\pm$2.0 & 1.7$\pm$0.9 & 0.3$\pm$0.2 & 1.7$\pm$0.9  & 2.4$\pm$1.4 & 0.3 \\ 
J0808+3948$^{b}$ & DL84\,+\,ac & 0.1 & 200$\pm$0.0 & 285$\pm$3.5 & 78$\pm$11.6 & 106$\pm$2.0 & 3.2$\pm$0.2 & 6.3$\pm$0.5 & 2.1$\pm$0.4  & 3.0$\pm$0.3 & 0.7 \\
J0808+3948$^{a}$ & olv6\,+\,ac & 0.1 & 297$\pm$58 & 383$\pm$53.0 & 65$\pm$5.9 & 108$\pm$3.1 & 1.3$\pm$1.0 & 0.3$\pm$0.3 & 2.3$\pm$1.4 & 3.7$\pm$2.6 & 0.2 \\
J0808+3948$^{a}$ & pyr3\,+\,ac & 1.0 & 239$\pm$24.5 & 294$\pm$26.0 & 67$\pm$5.9 & 105$\pm$2.4 & 2.5$\pm$1.5 & 0.5$\pm$0.4 & 2.1$\pm$0.7 & 4.1$\pm$1.5 & 0.8 \\
J0808+3948$^{a}$ & DL84\,+\,gra & 0.1 & 265$\pm$56.0 & 379$\pm$66.0 & 106$\pm$6.5 & 129$\pm$7.4 & 1.7$\pm$1.3 & 0.3$\pm$0.3 & 0.5$\pm$0.2 & 0.5$\pm$0.3 & 0.5 \\
J0808+3948$^{a}$ & DL84\,+\,ac & 0.1 & 285$\pm$34.0 & 384$\pm$35.0 & 87$\pm$3.6 & 115$\pm$0.0 & 1.1$\pm$0.5 & 0.2$\pm$0.1 & 1.7$\pm$0.3  & 1.8$\pm$0.4 & 0.3 \\
\hline
Quasar & DL84\,+\,gra & 1.5 & 350$\pm$32.0 & 644$\pm$5.8 & 40$\pm$6.0 & 149$\pm$1.3 & -- & -- & -- & -- & 0.42 \\
\hline
FSC\,10214+4724 & DL84\,+\,gra & 0.5 & 394$\pm$78 & 377$\pm$132 & 72$\pm$35 &  138$\pm$11 & $\rm 5.2\times10^3$ & $\rm 2.4\times10^3$ & $\rm 1.9\times10^4$ & $\rm 5.6\times10^3$ & 5.8
\enddata
\tablenotetext{a}{Model parameters derived from 
                  the PAHFIT-based IR emission.}
\tablenotetext{b}{Model parameters derived from 
                  the spline-based IR emission.}
\end{deluxetable}



\section{Results\label{sec:results}}
A first glance of the mass absorption coefficient 
$\kappa_{\rm abs}(\lambda)$ profiles 
shown in Figures~\ref{fig:kappa1},~\ref{fig:kappa2}
would lead one to speculate that the steep $\simali$5--8$\mum$ 
emission continuum observed in the three spectroscopically
anomalous galaxies could be caused by silicate dust with 
an iron-poor pyroxene composition; and
for a given composition, small, sub-$\mu$m-sized dust 
may be preferred over $\mu$m-sized dust
(see Xie et al.\ 2014).
In this section we apply the dust model elaborated
in \S\ref{sec:model} to fit the observed \textit{Spitzer}/IRS
spectra of these galaxies by extensively exploring
the parameter space of eight model parameters
($\Twsil$, $\Tcsil$, $\Twcarb$, $\Tccarb$,
 $\Mwsil$, $\Mcsil$, $\Mwcarb$, and $\Mccarb$)
and different compositions
(``astronomical silicate'' [DL84],
pyroxene MgSiO$_{3}$ [pyr0],
pyroxene Mg$_{0.7}$Fe$_{0.3}$SiO$_{3}$ [pyr3],
pyroxene Mg$_{0.4}$Fe$_{0.6}$SiO$_{3}$ [pyr6], 
olivine MgFeSiO$_{4}$ [olv5],
olivine Mg$_{0.8}$Fe$_{1.2}$SiO$_{4}$ [olv6],
amorphous carbon [ac],
and graphite [gra]). 
%
%
We will focus on fitting the ``observed'' spectra
obtained by subtracting from the \textit{Spitzer}/IRS
spectra the PAH and ionic emission lines
determined with the PAHFIT method.
The model fits to the ``observed'' spectra
obtained 
with the spline method will 
be discussed in \S\ref{sec:spline}.


\begin{figure*}
\begin{center}
$
\begin{array}{c}
\resizebox{0.7\hsize}{!}{
\includegraphics{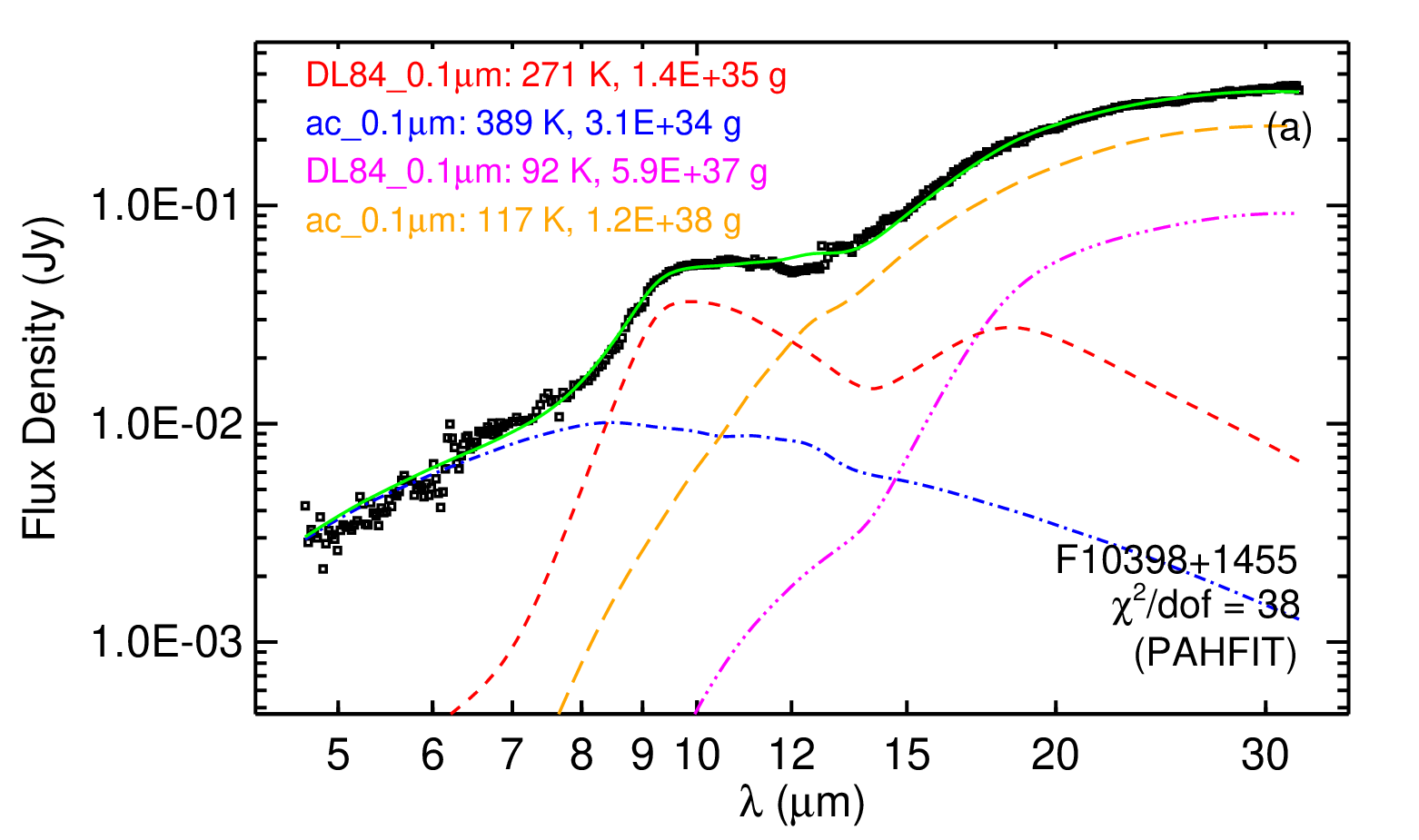}} \\
\resizebox{0.7\hsize}{!}{
\includegraphics{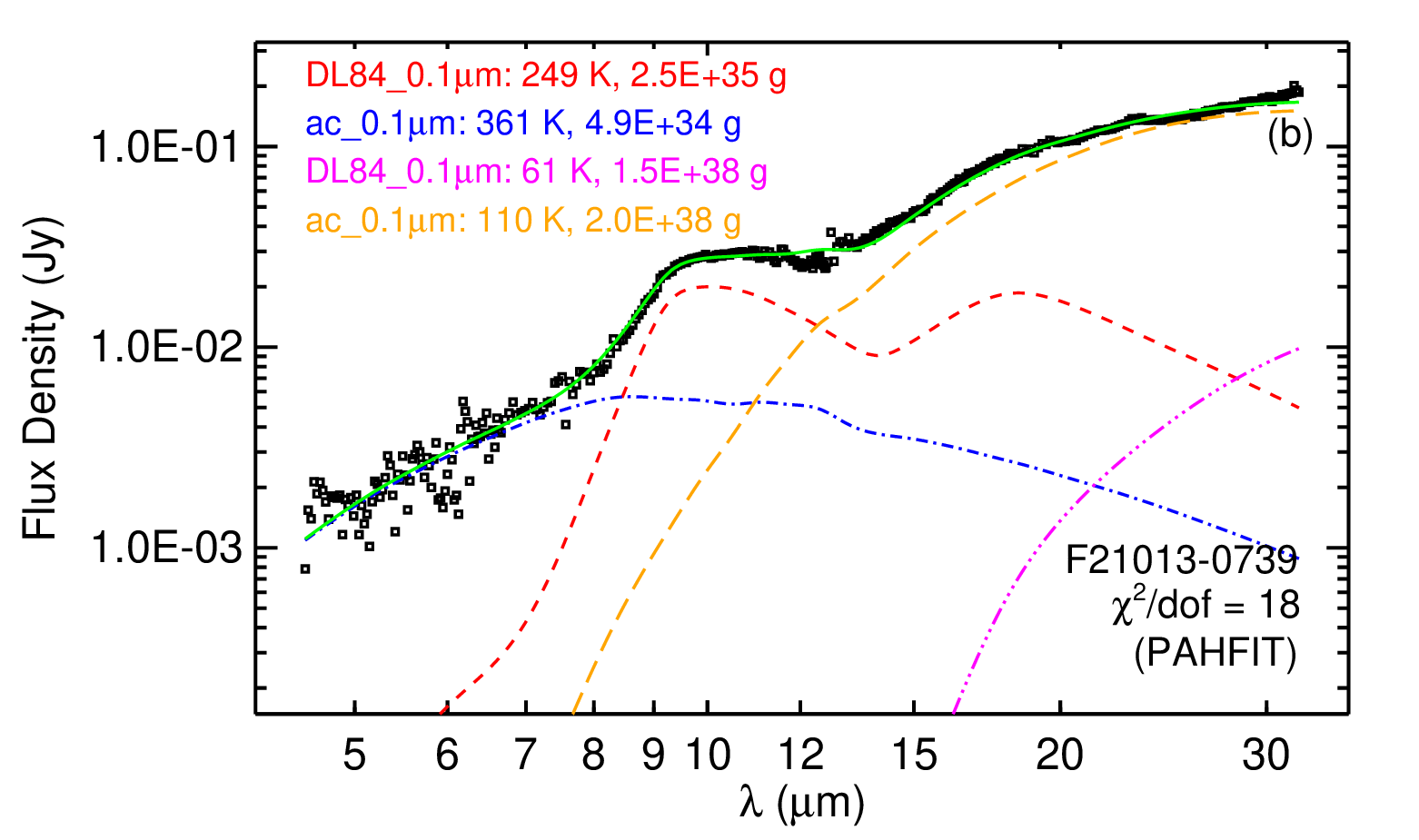}} \\
\resizebox{0.7\hsize}{!}{
\includegraphics{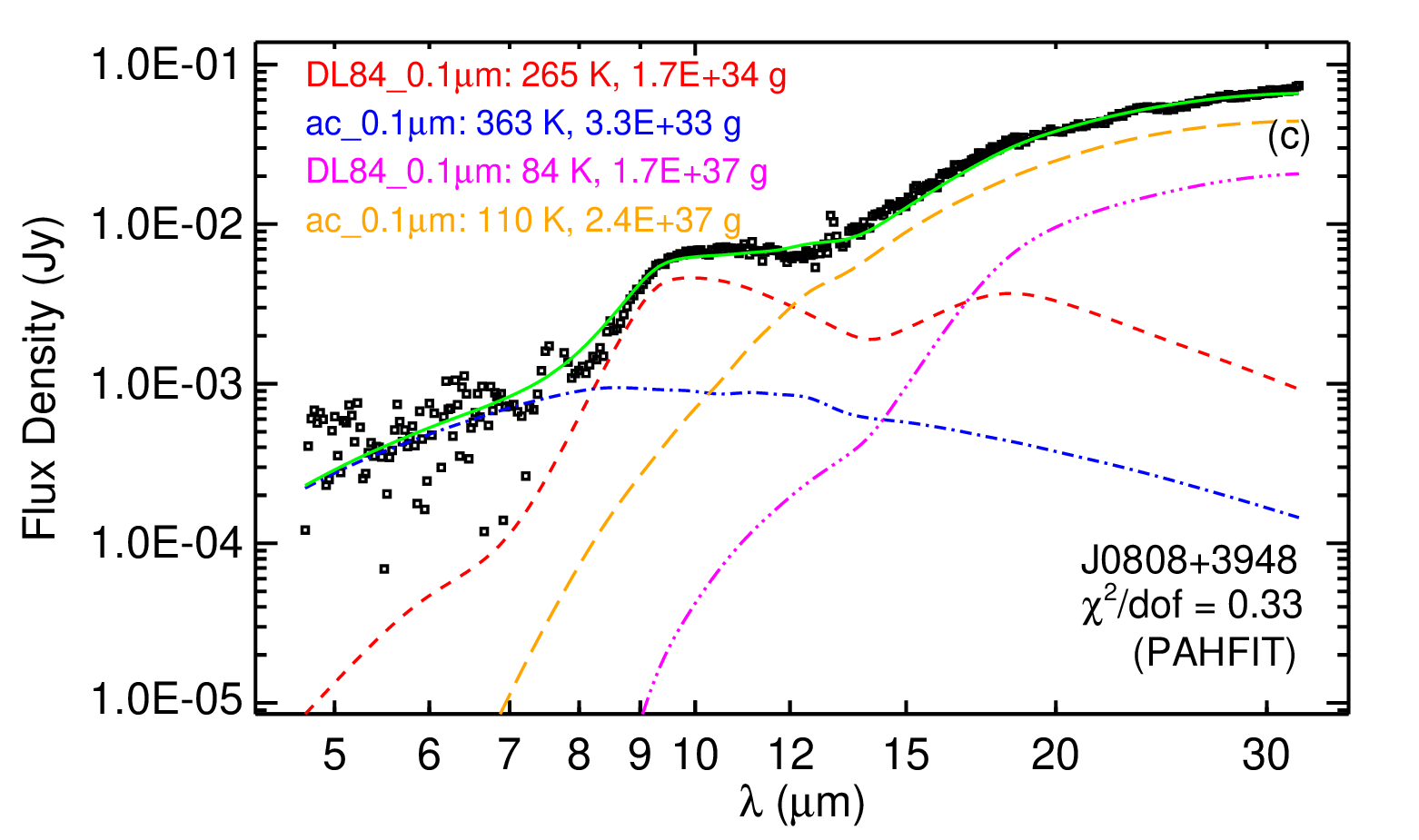}}
\end{array}
$ 
\caption{\footnotesize
         \label{fig:dl84ac} 
         Comparison of the model emission (solid green line)
         of IRAS~F10398+1455 (a), IRAS~F21013-0739 (b), and 
         SDSS~J0808+3948 (c)
         with the ``observed'' spectra (black open squares)
         obtained by subtracting from 
         the \textit{Spitzer}/IRS spectra 
         the PAH and ionic emission lines 
         determined with the PAHFIT method. 
         The model consists of four components:
         warm silicate (red dashed line),
         cold silicate (magenta dot-long dashed line),
         warm amorphous carbon (blue dotted--dashed line), and
         cold amorphous carbon (orange long dashed line).
         All dust components are taken to be 0.1$\mum$ in radii. 
         The DL84 ``astronomical'' silicate 
         is adopted for the silicate component.
         }
\end{center}
\end{figure*}

\subsection{Dust Composition\label{sec:composition}}
We first consider the DL84
``astronomical'' silicate 
and amorphous carbon,
and take all dust components 
to have the same grain size of 0.1$\mum$ in radii.
The effects of dust size will be discussed 
in \S\ref{sec:size}.
In Figure~\ref{fig:dl84ac} we show 
the model fits to the ``observed'' spectra 
of IRAS~F10398+1455, IRAS~F21013-0739 
and SDSS~J0808+3948.
The overall fits are very good
except for SDSS~J0808+3948; 
the model 9.7$\mum$ silicate profile
is a little bit too broad and its 
peak wavelength is a little bit too short.
Also, for IRAS~F10398+1455 and IRAS~F21013-0739,
the model slightly overpredicts
the red wing of the 9.7$\mum$ feature. 
For all three sources, the $\simali$5--8$\mum$
emission continuum is dominated by 
warm amorphous carbon, and the emission at
$\lambda>14\mum$ is dominated by 
cold amorphous carbon.

%

\begin{figure*}
\begin{center}
$
\begin{array}{c}
\resizebox{0.7\hsize}{!}{
\includegraphics{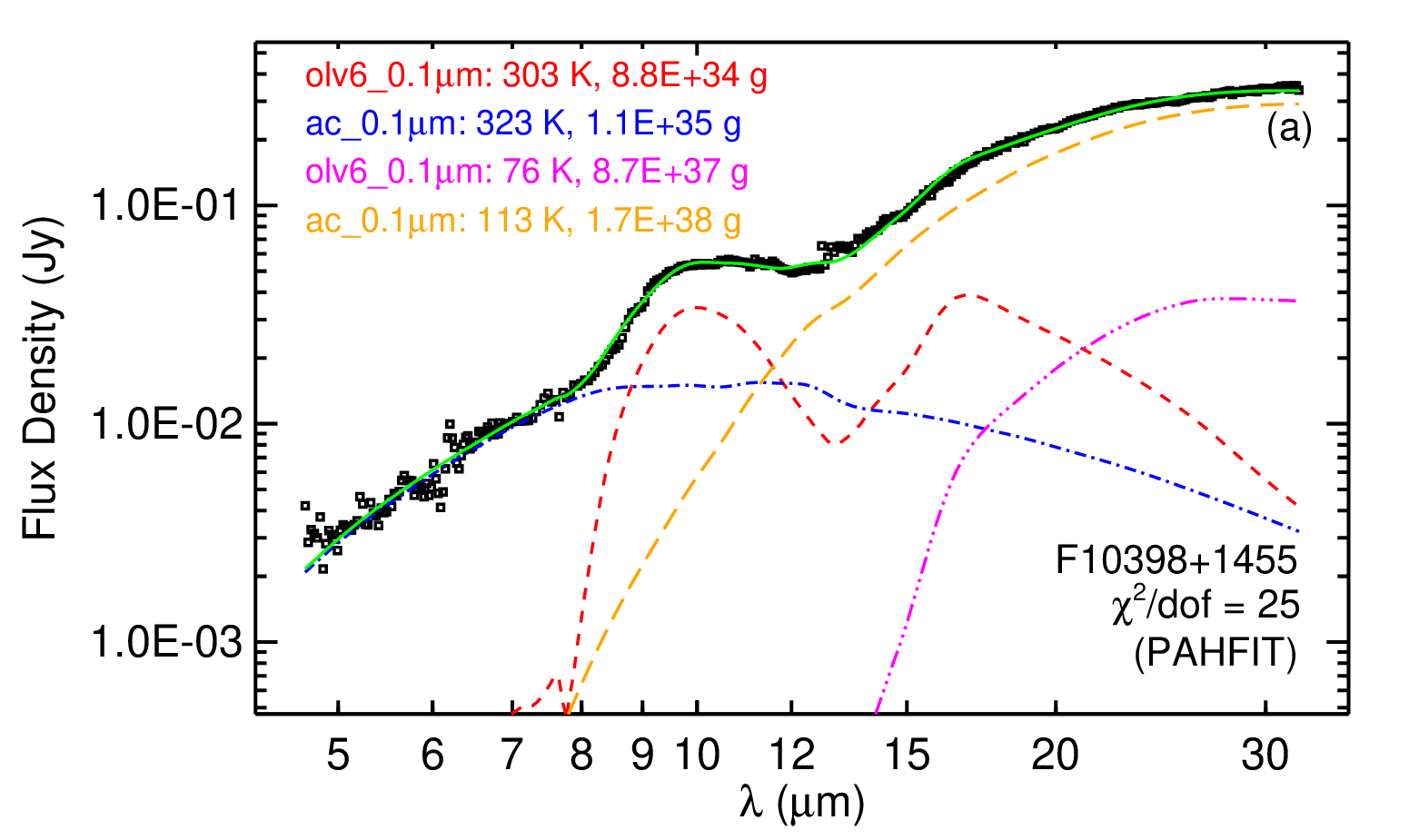} }\\
\resizebox{0.7\hsize}{!}{
\includegraphics{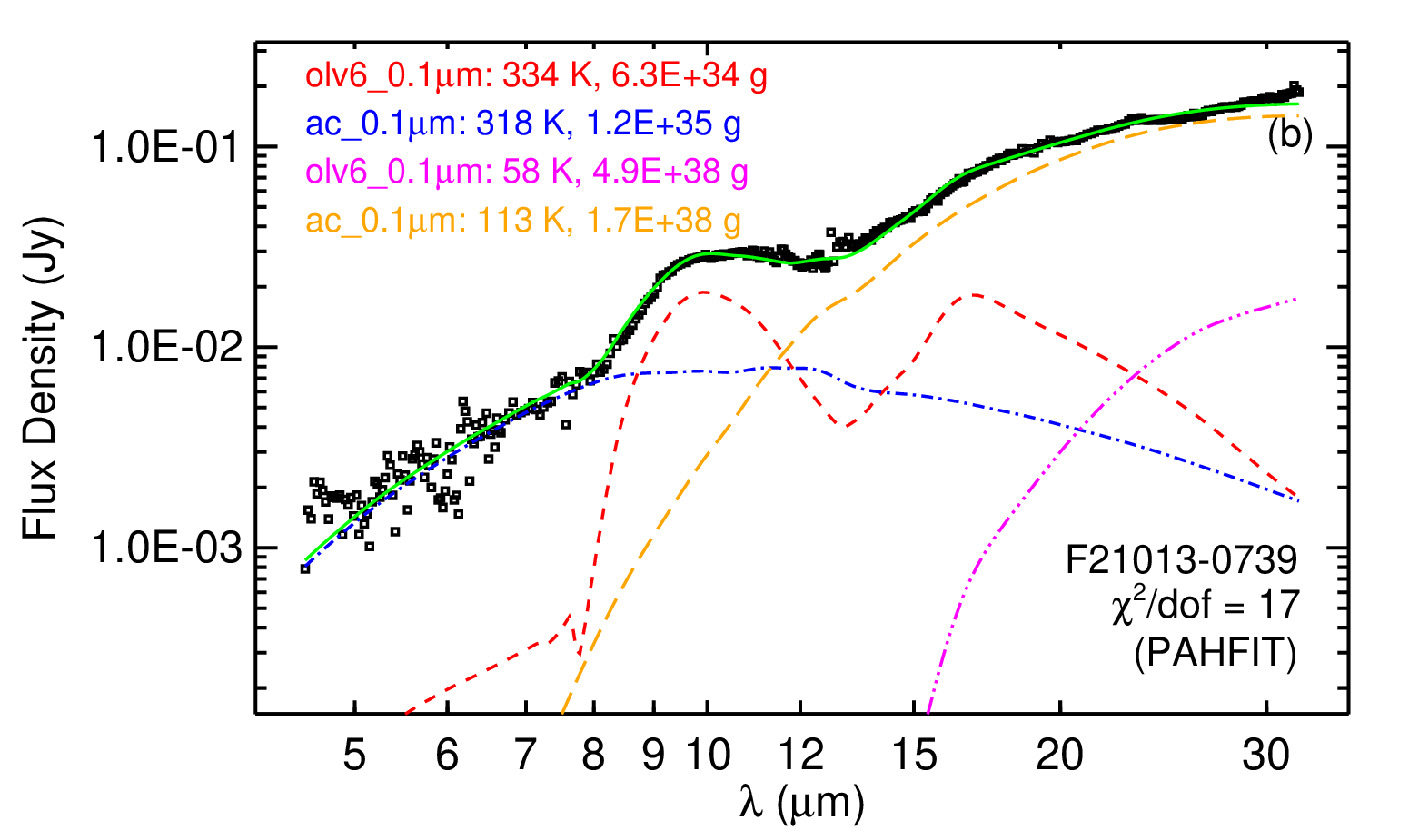} }\\
\resizebox{0.7\hsize}{!}{
\includegraphics{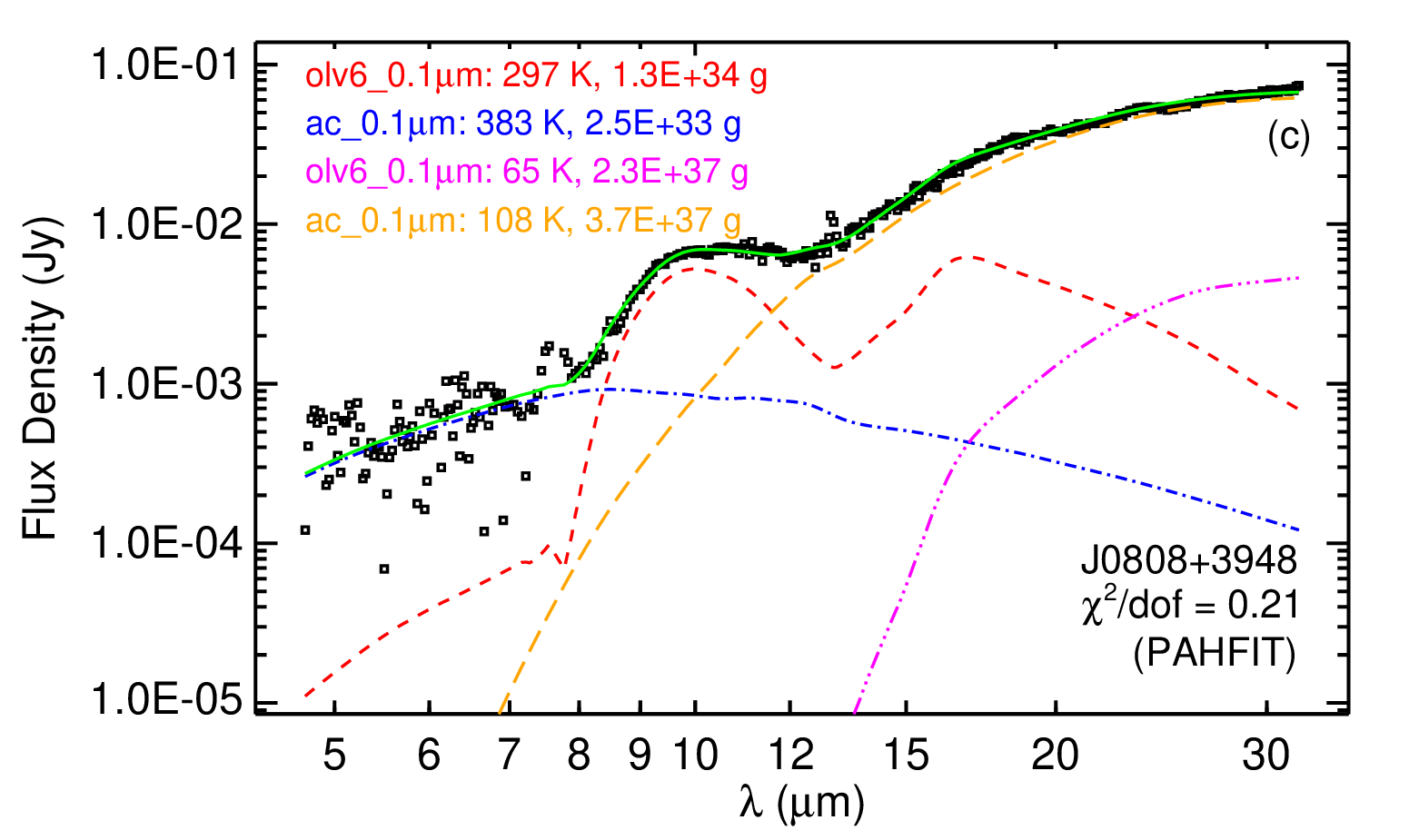} }
\end{array}
$
\caption{\footnotesize
         \label{fig:olivine_ac}
         Same as Figure~\ref{fig:dl84ac}
         but with the DL84 ``astronomical silicate''
         replaced by amorphous olivine 
         Mg$_{0.8}$Fe$_{1.2}$SiO$_{4}$ (``olv6''). 
         }
\end{center}
\end{figure*}

\begin{figure*}
\begin{center}
$
\begin{array}{c}
\resizebox{0.7\hsize}{!}{
\includegraphics{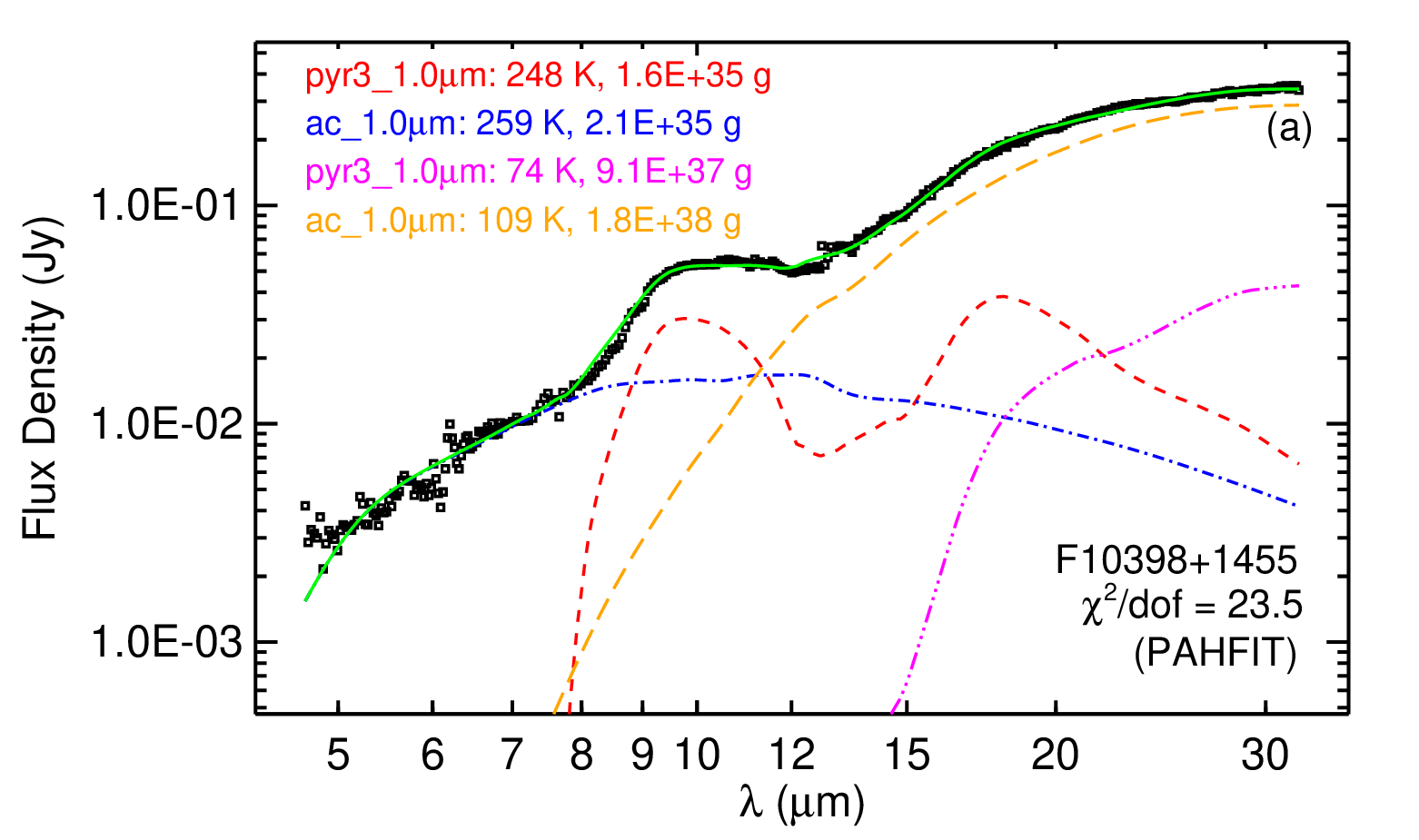} } \\
\resizebox{0.7\hsize}{!}{
\includegraphics{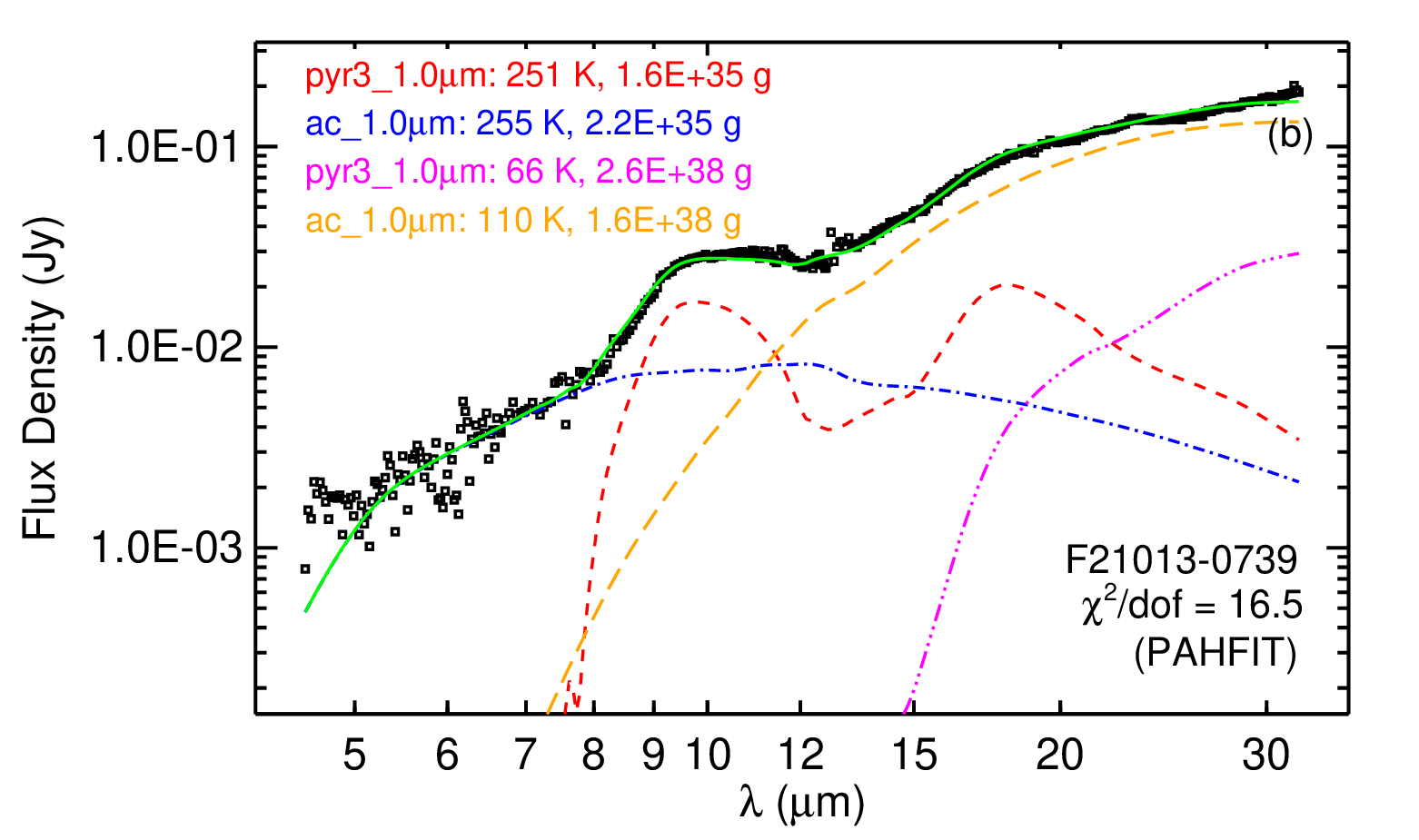} } \\
\resizebox{0.7\hsize}{!}{
\includegraphics{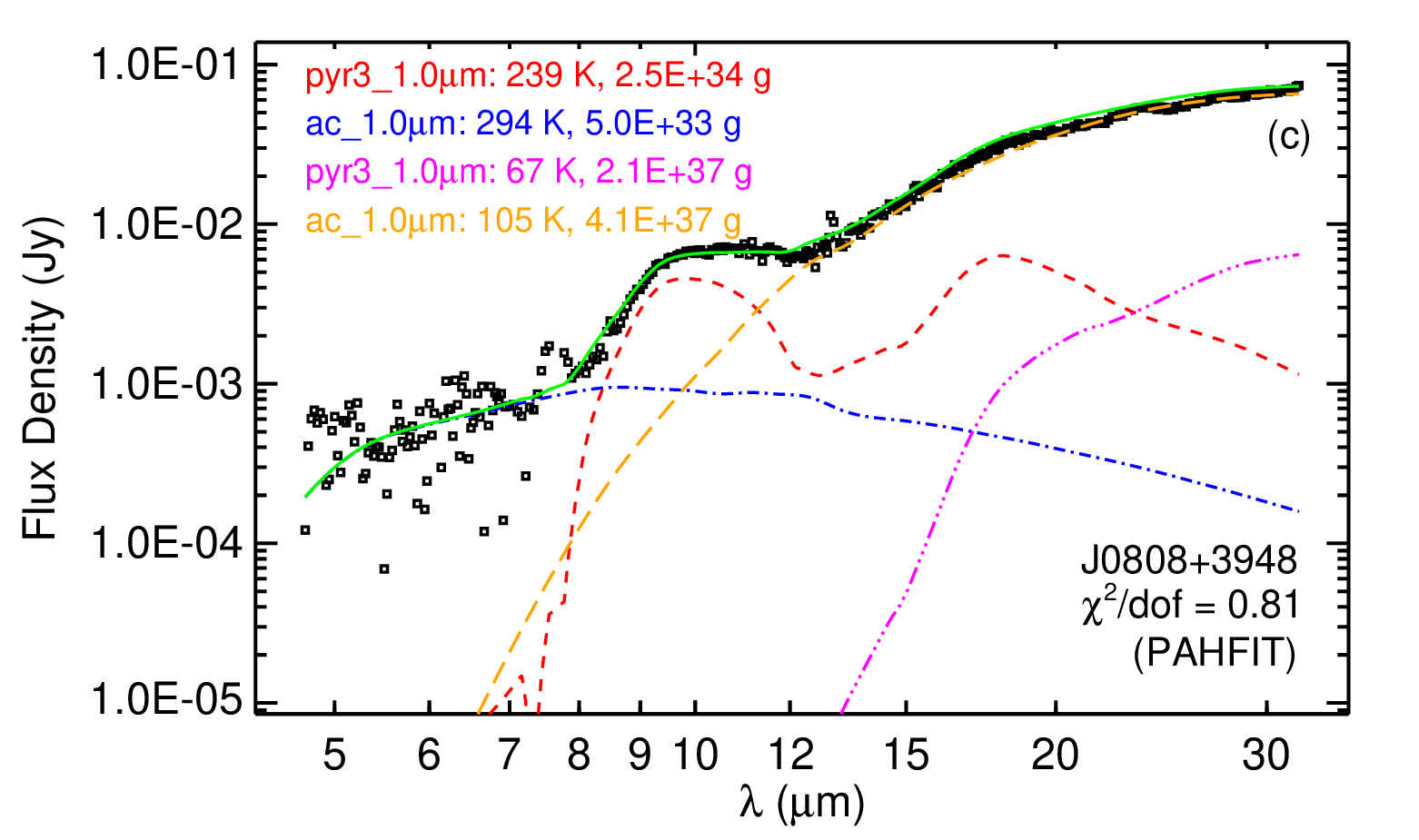} }
\end{array}
$
\caption{\footnotesize
         \label{fig:pyroxene_ac}
         Same as Figure~\ref{fig:dl84ac}
         but with the DL84 ``astronomical silicate''
         replaced by amorphous pyroxene
         Mg$_{0.7}$Fe$_{0.3}$SiO$_{3}$ (``pyr3'')
         and all dust components taken to
         have a size of $a=1.0\mum$. 
         }
\end{center}
\end{figure*}

We then consider models consisting of 
amorphous olivine and amorphous carbon.
Assuming a size of $a=0.1\mum$ for 
all dust components, we find that
Mg$_{0.8}$Fe$_{1.2}$SiO$_{4}$ 
(together with amorphous carbon)
provides an excellent fit to 
all three sources 
(see Figure~\ref{fig:olivine_ac}).
Similarly, we have also considered
models consisting of amorphous pyroxene 
and amorphous carbon.
As shown in Figure~\ref{fig:pyroxene_ac},
excellent fits are achieved with 
Mg$_{0.7}$Fe$_{0.3}$SiO$_{3}$ of $a=1.0\mum$. 
This is because the 9.7$\mum$ feature
of pyroxene dust is appreciably narrower 
than that of amorphous olivine 
and ``astronomical'' silicate
(see Figure~\ref{fig:kappa2}). 
Therefore, models consisting of $a=0.1\mum$ 
pyroxene and amorphous carbon produce too narrow 
a 9.7$\mum$ feature.
On the other hand, if the dust is too large
(say, $a=2.0\mum$), the model 9.7$\mum$ feature
will be too broad to be consistent with 
the observed spectra (see \S\ref{sec:size}).

\begin{figure*}
\begin{center}
$
\begin{array}{c}
\resizebox{0.7\hsize}{!}{
\includegraphics{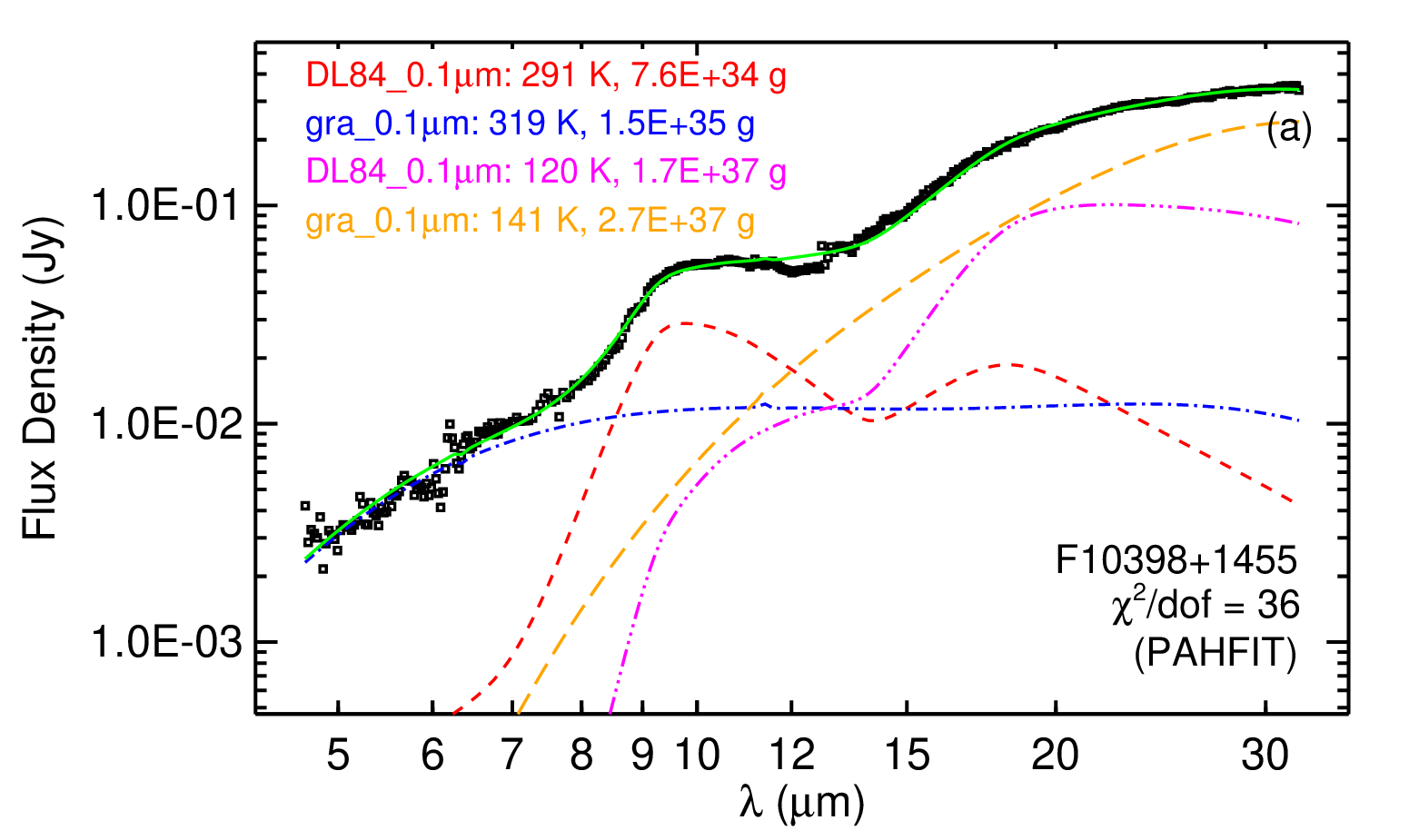} } \\
\resizebox{0.7\hsize}{!}{
\includegraphics{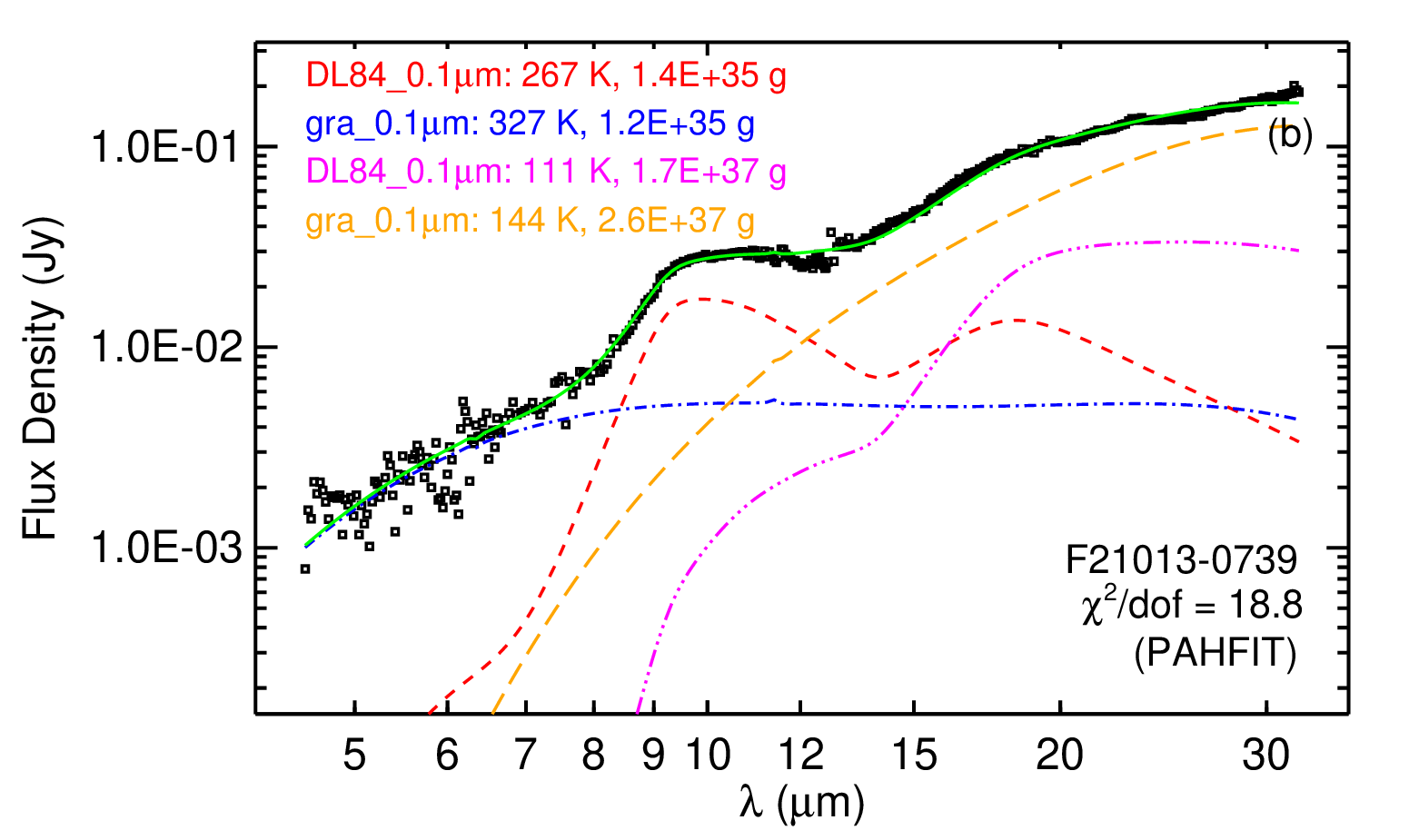} } \\
\resizebox{0.7\hsize}{!}{
\includegraphics{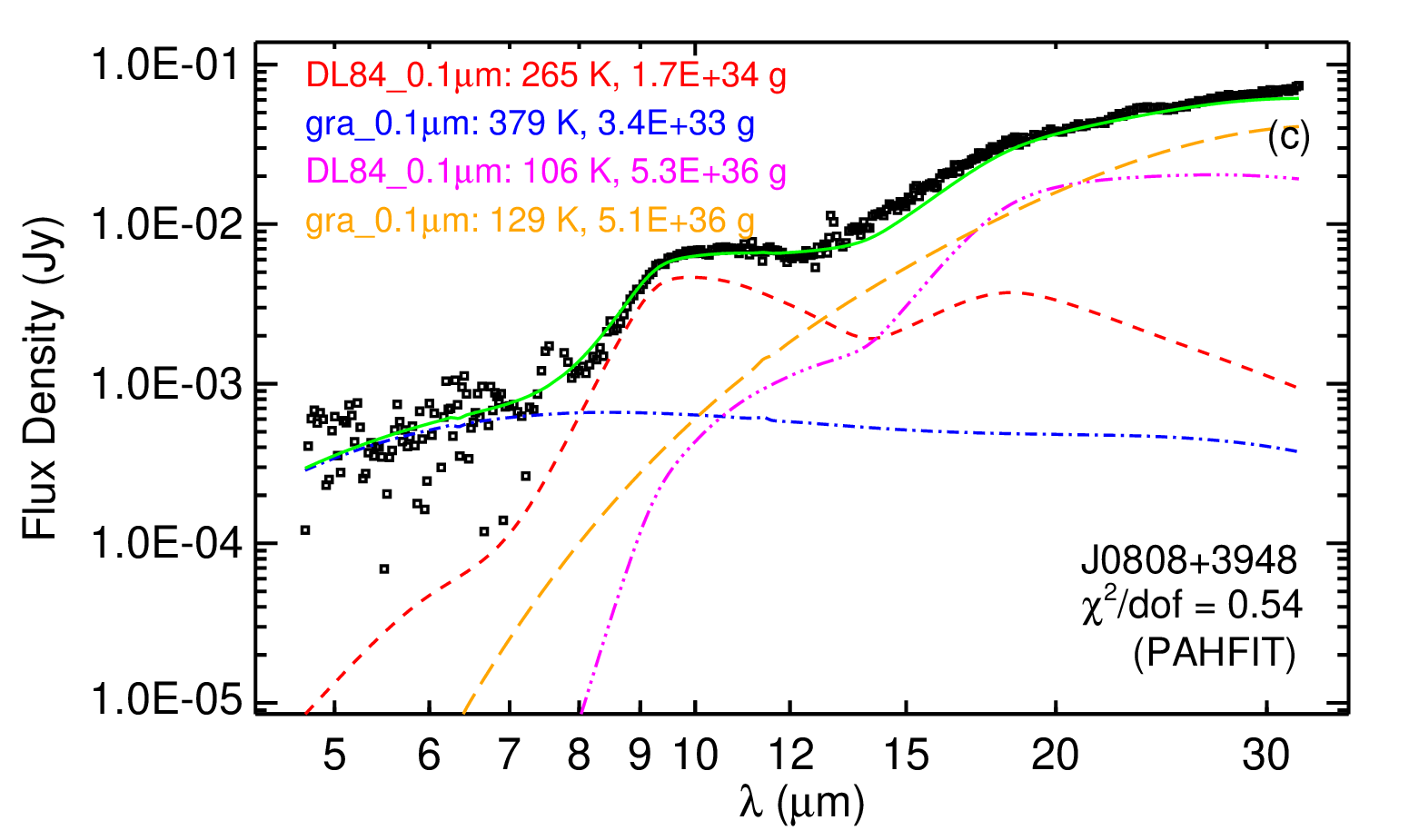} }
\end{array}
$
\caption{\footnotesize
         \label{fig:dl84gra}
         Same as Figure~\ref{fig:dl84ac}
         but with the amorphous carbon component
         replaced by graphite.
         }    
\end{center}
\end{figure*}

We have also considered graphite.
As shown in Figure~\ref{fig:dl84gra},
models consisting of graphite and
``astronomical silicate'' of $a=0.1\mum$ 
closely fit the observed IR emission
of all three galaxies, 
with the fitting $\rm \chi^2$ 
at a similar level as 
that of the ``astronomical silicate''/amorphous 
carbon model (see Figure~\ref{fig:dl84ac}).

\begin{figure*}
\begin{center}
$
\begin{array}{c}
\resizebox{0.7\hsize}{!}{
\includegraphics{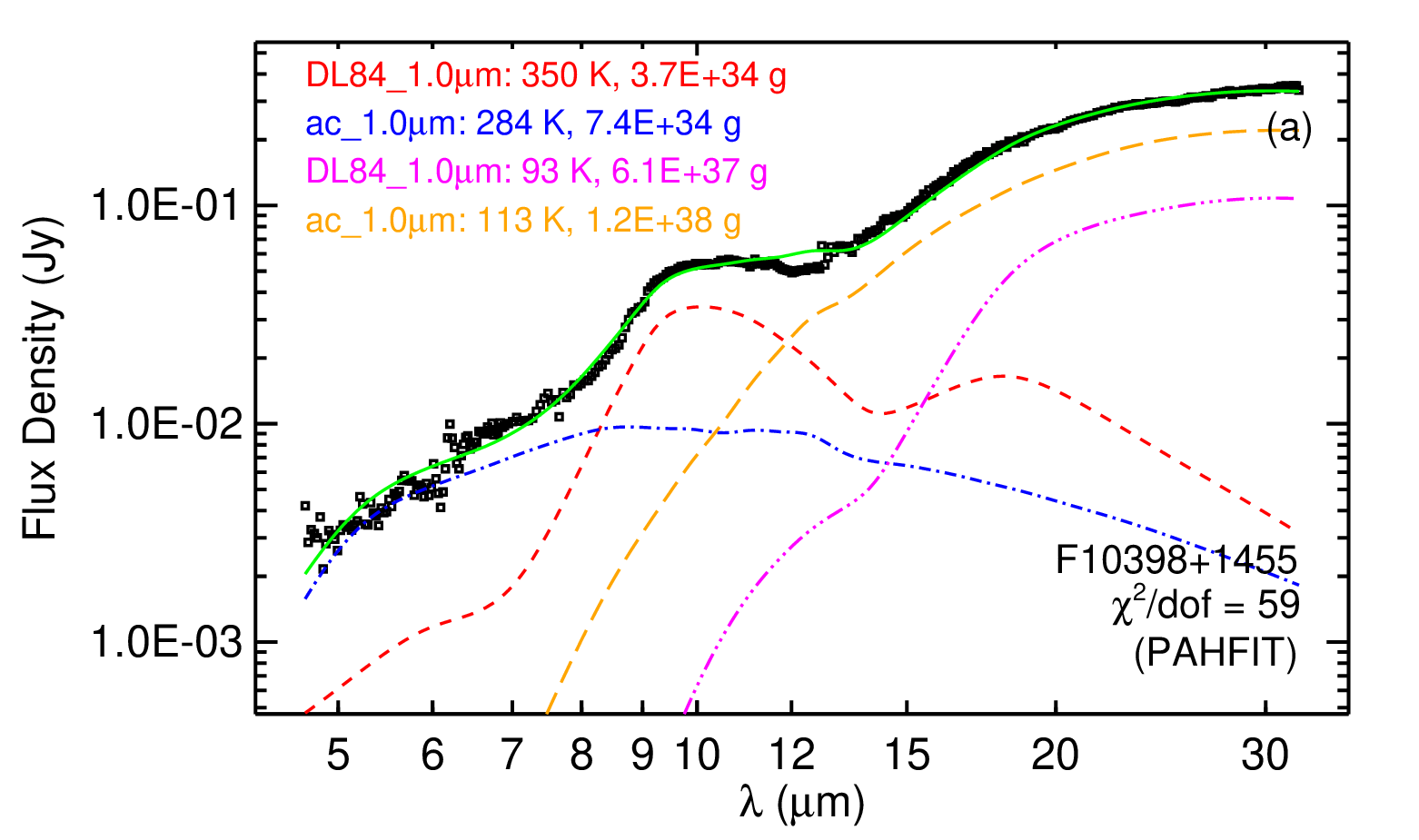}} \\
\resizebox{0.7\hsize}{!}{
\includegraphics{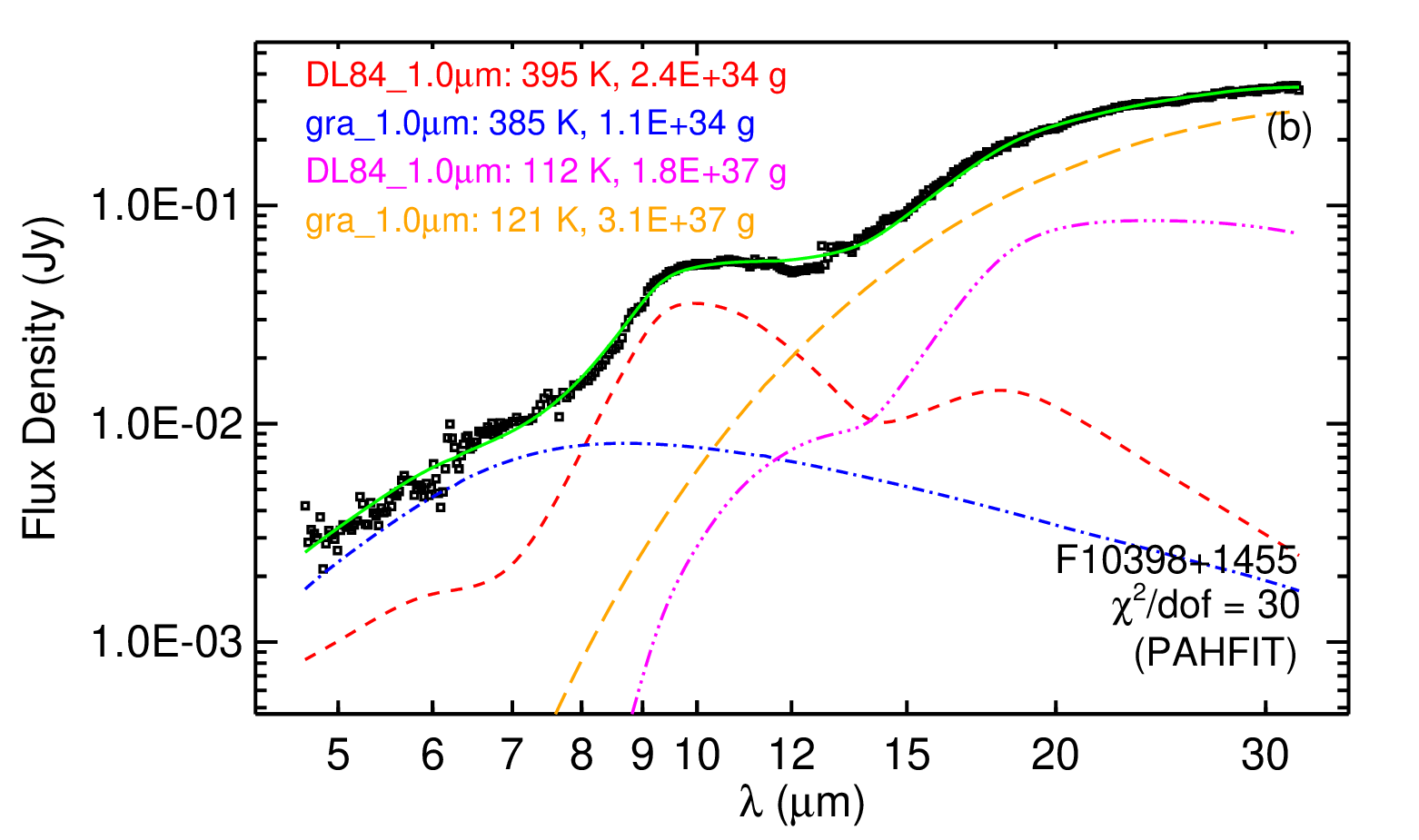}} \\
\resizebox{0.7\hsize}{!}{
\includegraphics{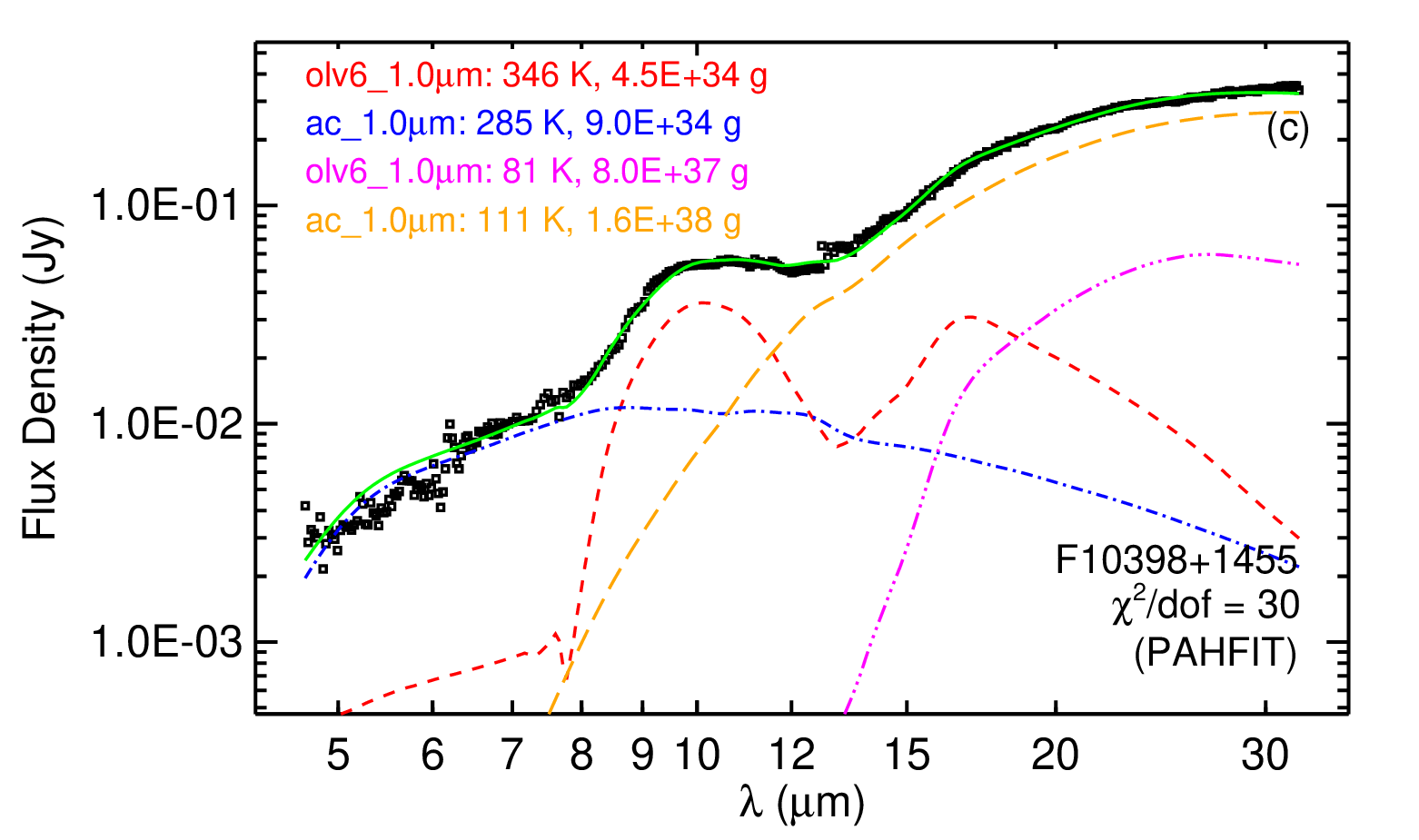}}
\end{array}
$
\caption{\footnotesize
         \label{fig:a1um} 
         Upper panel (a): Same as 
         Figure~\ref{fig:dl84ac}(a)
         but with $a=1.0\mum$ for all dust components.
         Middle panel (b): Same as 
         Figure~\ref{fig:dl84gra}(a)
         but with $a=1.0\mum$ for all dust components.
         Bottom panel (c): Same as 
         Figure~\ref{fig:olivine_ac}(a)
         but with $a=1.0\mum$ for all dust components.
         }
\end{center}
\end{figure*}

\begin{figure*}
\begin{center}
$
\begin{array}{c}
\resizebox{0.7\hsize}{!}{
\includegraphics{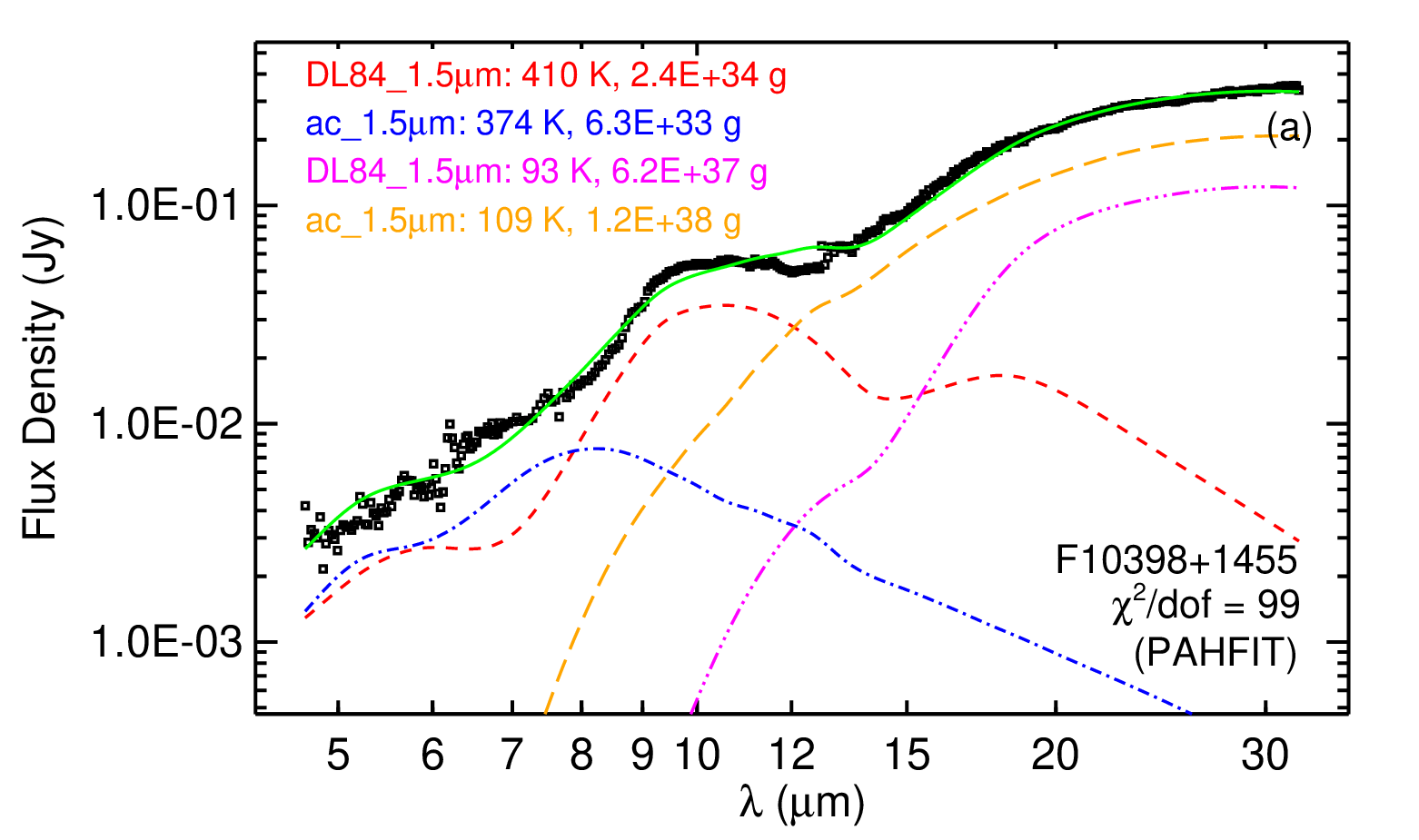}} \\
\resizebox{0.7\hsize}{!}{
\includegraphics{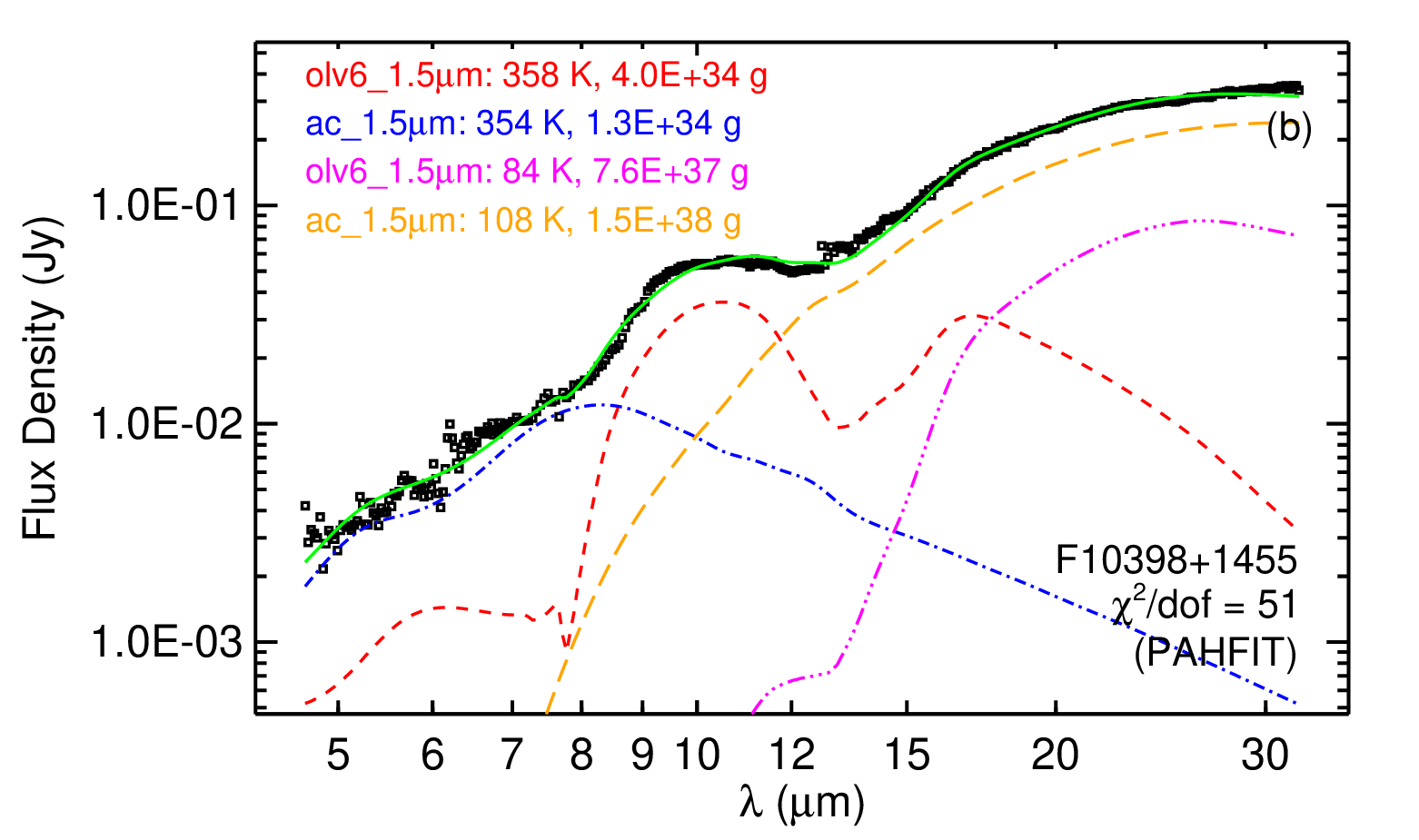}} \\
\resizebox{0.7\hsize}{!}{
\includegraphics{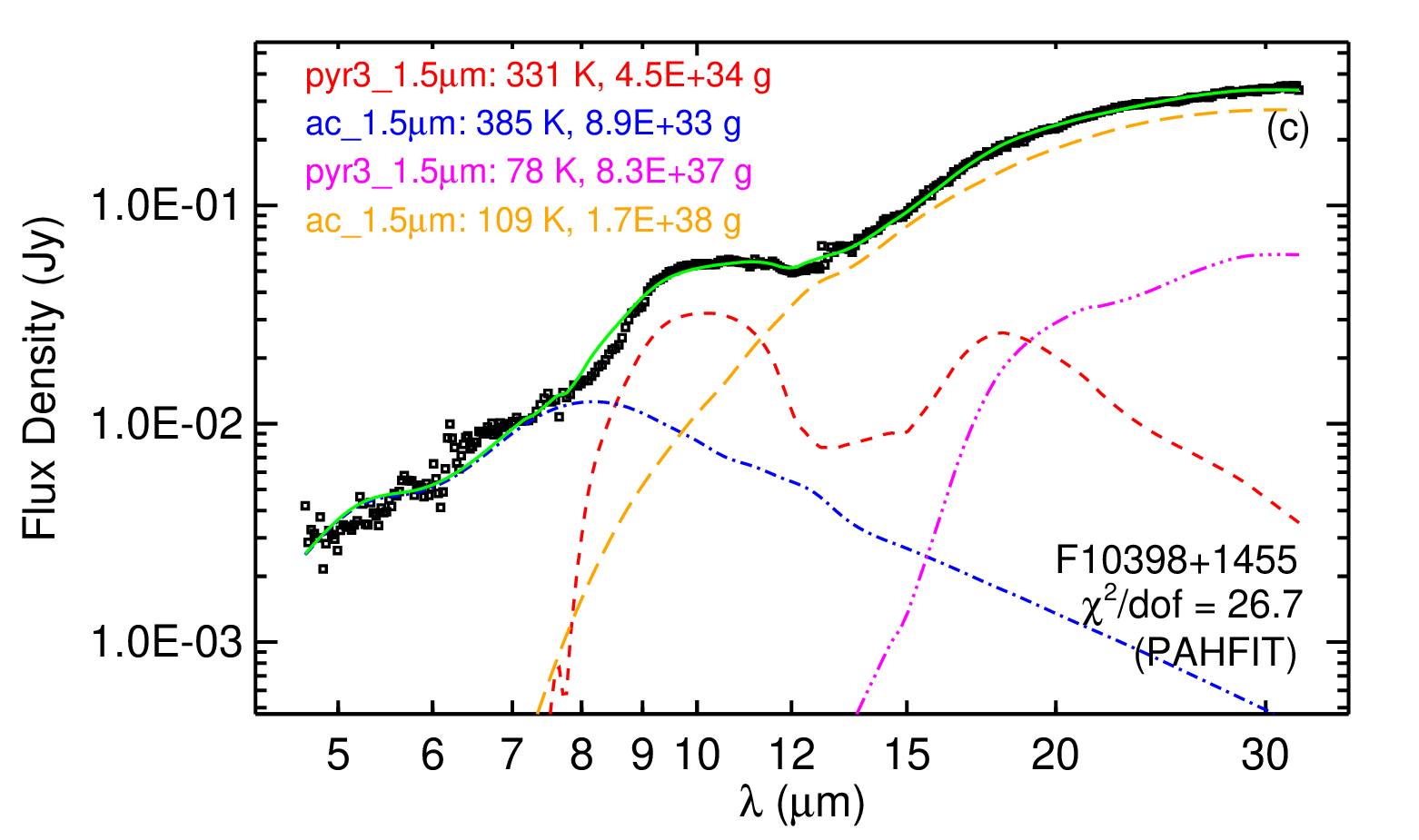}}
\end{array}
$
\caption{\footnotesize
         \label{fig:a1.5um}
         Upper panel (a): same as
         Figure~\ref{fig:dl84ac}(a)
         but with $a=1.5\mum$ for all dust components.
         Middle panel (b): same as
         Figure~\ref{fig:olivine_ac}(a)
         but with $a=1.5\mum$ for all dust components.
         Bottom panel (c): same as
         Figure~\ref{fig:pyroxene_ac}(a)
         but with $a=1.5\mum$ for all dust components.
         }
\end{center}
\end{figure*}

\begin{figure*}
\begin{center}
$
\begin{array}{c}
\resizebox{0.7\hsize}{!}{
\includegraphics{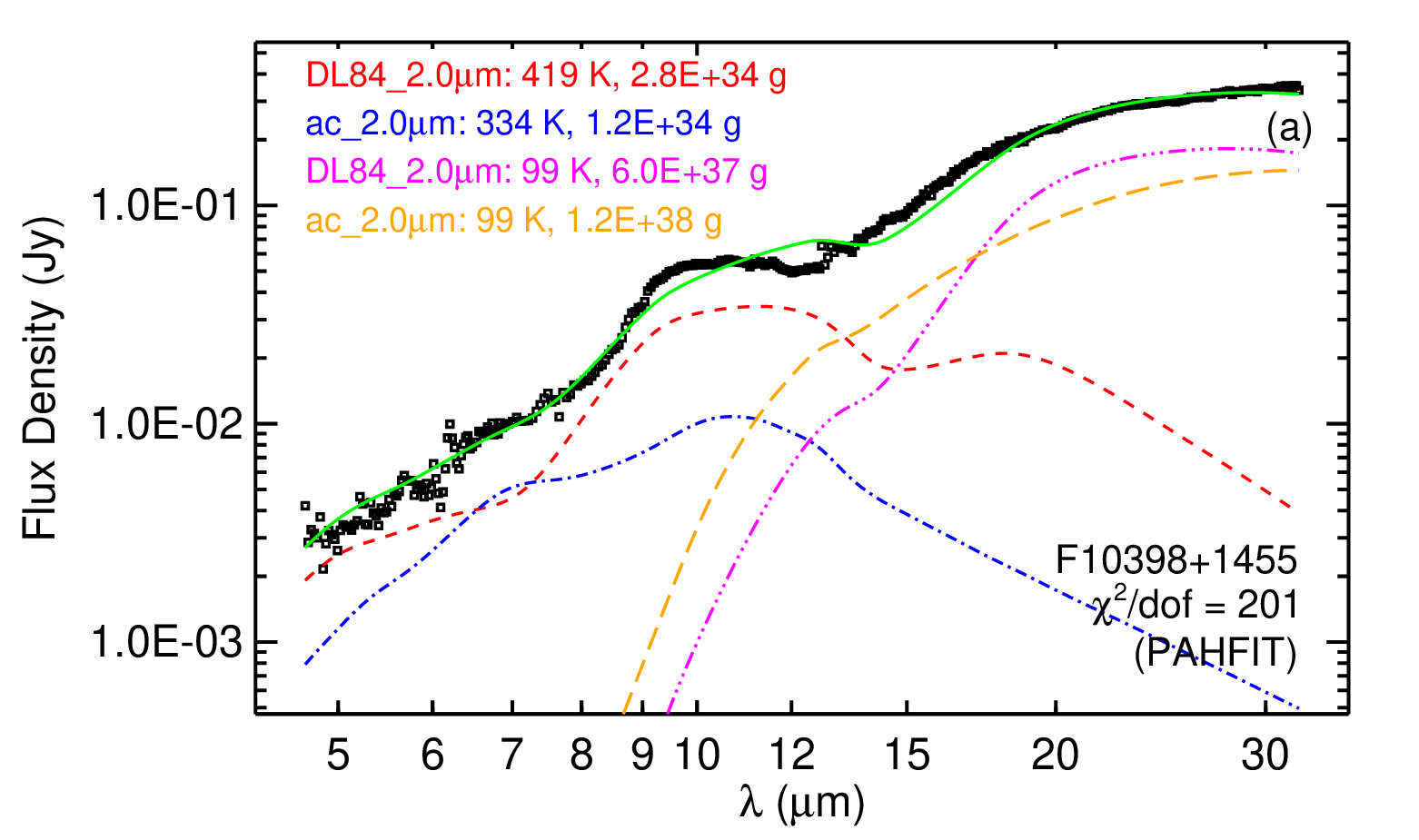}} \\
\resizebox{0.7\hsize}{!}{
\includegraphics{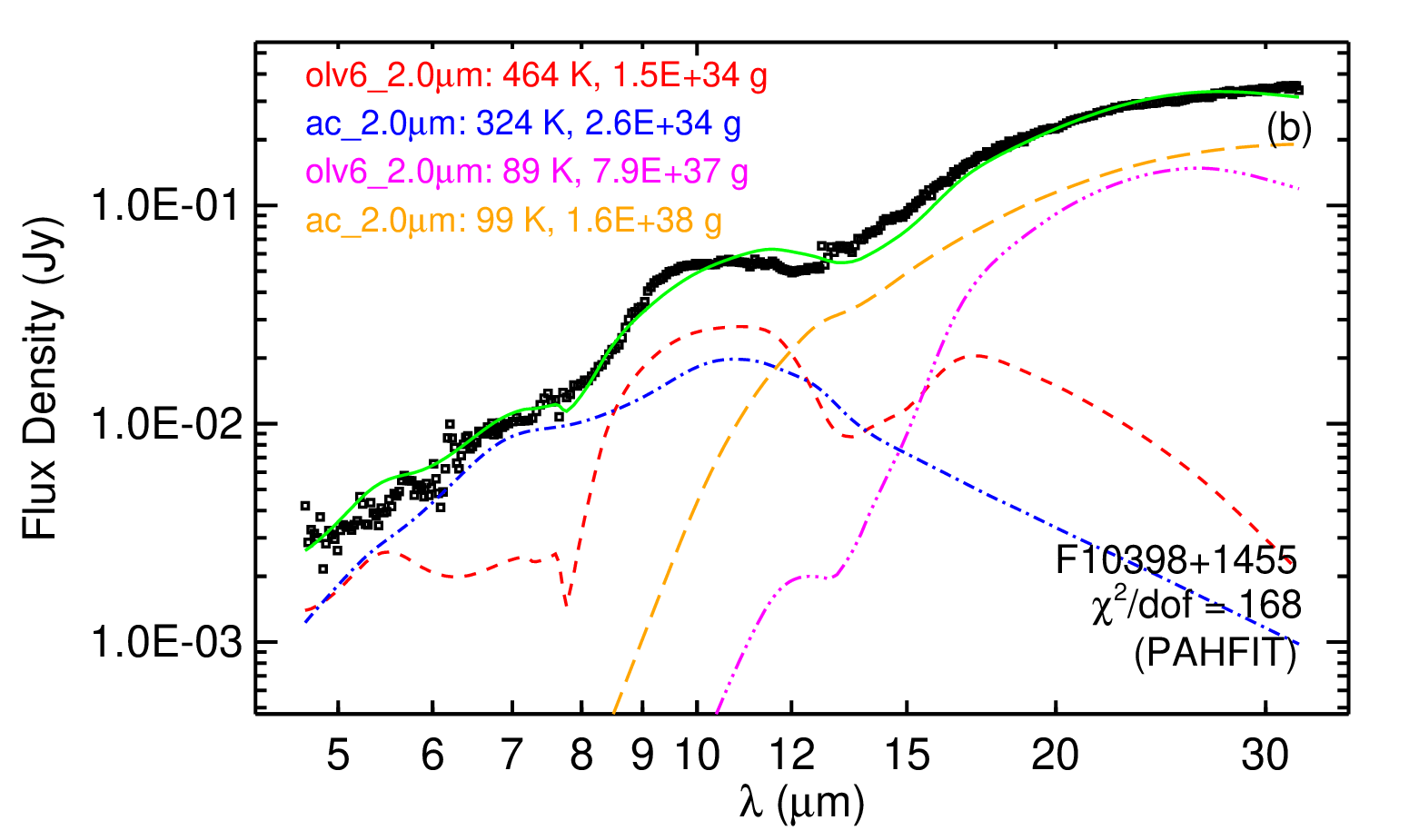}} \\
\resizebox{0.7\hsize}{!}{
\includegraphics{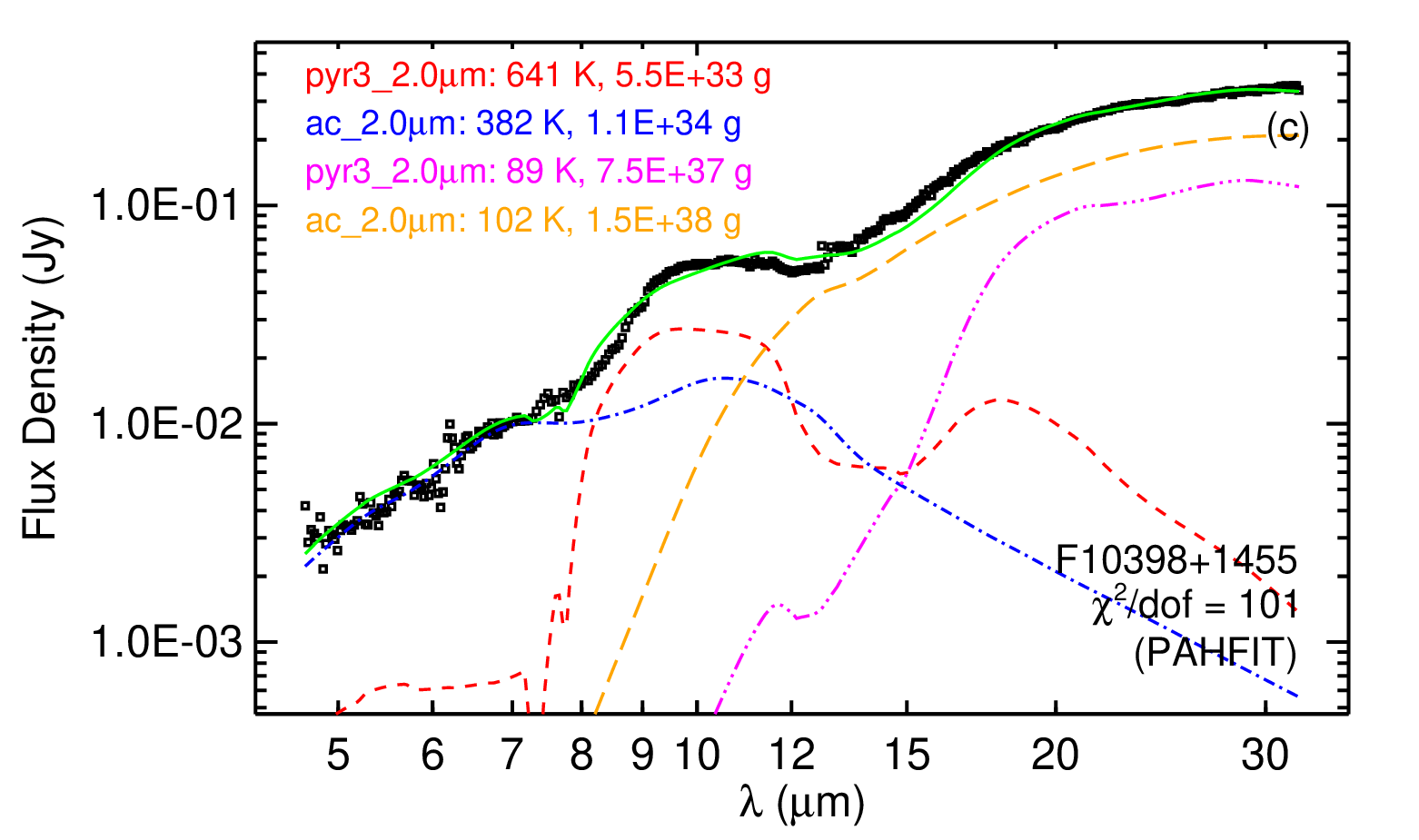}}
\end{array}
$
\caption{\footnotesize
         \label{fig:a2um} 
         Same as Figure~\ref{fig:a1.5um}
         but with $a=2.0\mum$ for all dust components.
         }
\end{center}
\end{figure*}

\subsection{Dust Sizes\label{sec:size}}

Xie et al.\ (2014) speculated that the steep 
$\sim$5--8$\mum$ emission continuum seen 
in these spectroscopically anomalous galaxies
could arise from sub-$\mu$m-sized silicate dust 
instead of $\mu$m-sized silicate dust
since the latter displays a larger opacity
in the $\sim$5--8$\mum$ wavelength range
(see Figure~\ref{fig:kappa1}).
However, as demonstrated in \S\ref{sec:composition},
the $\sim$5--8$\mum$ emission continuum 
is dominated by the carbon dust component.
Therefore, the $\sim$5--8$\mum$ continuum 
is not a powerful constraint 
on the silicate dust grain size.  
Instead, the 9.7$\mum$ feature allows 
us to place a stronger constraint.
Indeed, as shown in 
Figures~\ref{fig:a1um}--\ref{fig:a2um},
for models consisting of 
different silicate and carbon dust species
(``astronomical silicate,'' olivine, pyroxene,
graphite, and amorphous carbon),
the observed IR emission could still
be closely fitted with $a=1.0\mum$,
but not with $a=1.5\mum$ or $a=2.0\mum$. 
This is easy to understand: 
as illustrated in 
Figure~\ref{fig:kappa1},
the 9.7$\mum$ profile is
insensitive to dust size 
for $a\simlt1\mum$,
but it considerably broadens
for larger sizes.
The best-fit models with $a=1.0\mum$
lead to an appreciably higher temperature
for the warm silicate component ($\Twsil$)
compared with that of the best-fit models 
with $a=0.1\mum$.
Consequently, the mass required by 
the warm silicate component ($\Mwsil$)
drops by a factor of $\simali$3--5.
This can be understood in the sense that,
compared with silicate dust of $a=0.1\mum$, 
the ``9.7\,$\mum$'' silicate feature
of the opacity $\kappa_{\rm abs}(\nu)$
profile of silicate dust of $a=1.0\mum$ 
slightly shifts to a longer wavelength.
To compensate this ``redshift,'' 
one requires 
an increased temperature 
which blueshifts the peak wavelength
of the Planck function $B_\nu(T)$,
and therefore the resulting emission spectrum
which is the product of $B_\nu(T)$
and $\kappa_{\rm abs}(\nu)$ remains
at the observed peak wavelength.
The effects of grain size 
on the derived dust temperature 
are manifested in 
Figures~\ref{fig:a1.5um} and \ref{fig:a2um}. 


\section{Discussion\label{sec:discussion}} 
We have shown in \S\ref{sec:results} that a simple
model consisting of four dust components
(warm and cold silicate dust, and warm and cold
carbon dust) closely reproduces the observed IR
emission of all three spectroscopically anomalous
galaxies, including the steeply rising $\sim$5--8$\mum$
emission continuum and the 9.7$\mum$ emission feature.
The model-fitting results are insensitive to 
the exact silicate or carbonaceous dust composition. 
As shown in Figures~\ref{fig:dl84ac}--\ref{fig:dl84gra}, 
models consisting of ``astronomical'' silicate,
amorphous olivine, or pyroxene,
combined with amorphous carbon or graphite,
are all capable of successfully fitting 
the observed IR emission.
The $\sim$5--8$\mum$ emission continuum 
is dominated by the warm carbon dust component
with $T$ in the range of $\sim$250--400$\K$.
The 9.7$\mum$ silicate emission is predominantly
due to silicate dust of T $\sim$ 200--400$\K$.
The emission at $\lambda>14\mum$ is dominated
by the cold carbon dust component mostly
with $T$ $\sim$ 110--120$\K$.
The 9.7$\mum$ emission feature constrains
the silicate dust size to not exceed 
$\sim$1.0$\mum$ (i.e., $a\simlt1.0\mum$,
see Figures~\ref{fig:a1um}--\ref{fig:a2um}).
The mass ratio of the warm carbon dust to
the warm silicate dust ($\Mwsil/\Mwcarb$)
ranges from 
$\simali$0.2 to $\simali$2.0,
with a mean ratio of $\simali$0.98.
Similarly, 
the mass ratio of the cold carbon dust to
the cold silicate dust ($\Mcsil/\Mccarb$)
ranges from $\simali$0.34 to $\simali$2.0,
with a mean ratio of $\simali$1.5.
The total dust mass is dominated 
by the cold components, 
with the warm components only accounting for
$<$\,0.15\% of the total dust mass.

\begin{figure*}
\begin{center}
$
\begin{array}{c}
\resizebox{0.7\hsize}{!}{
\includegraphics{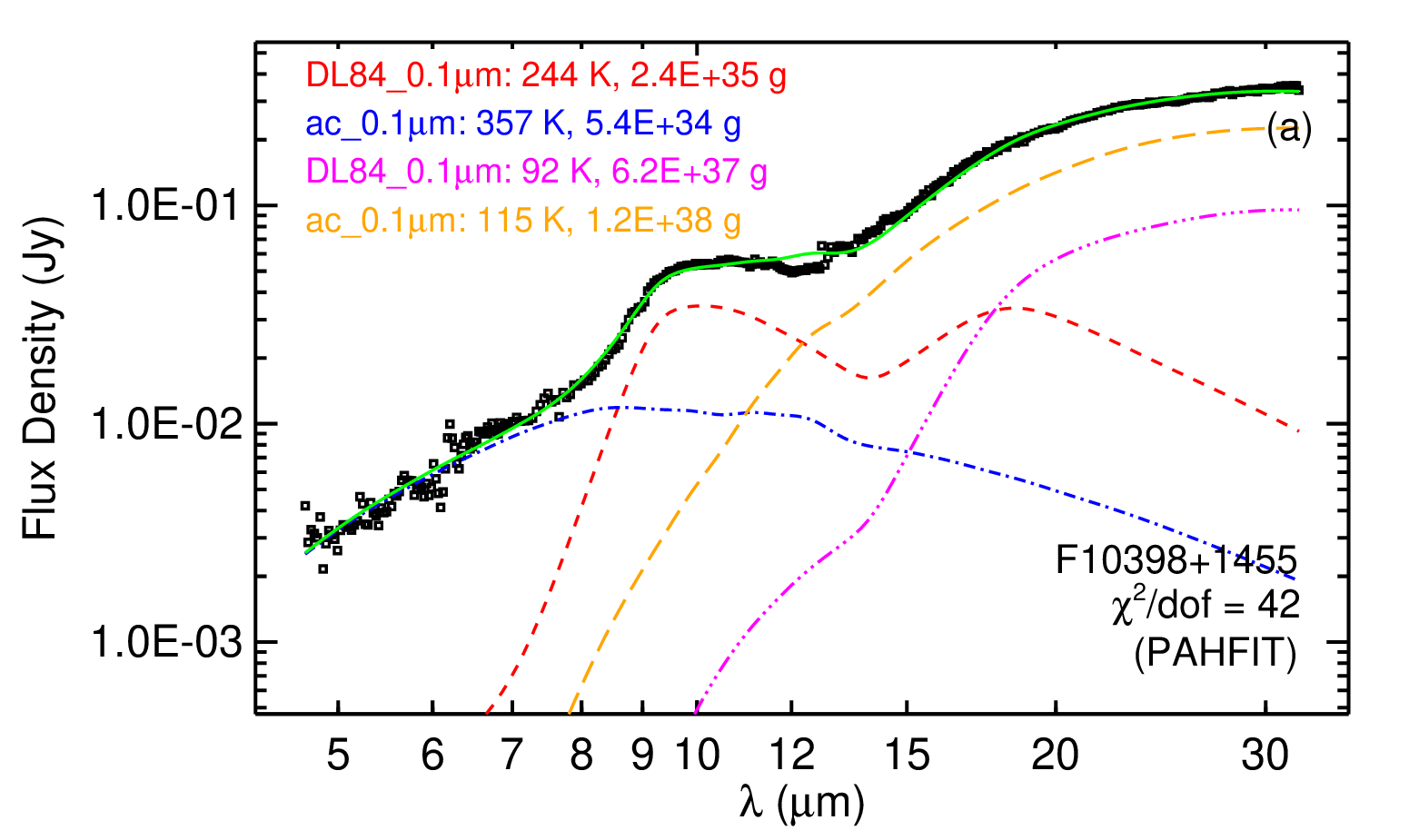}} \\
\resizebox{0.7\hsize}{!}{
\includegraphics{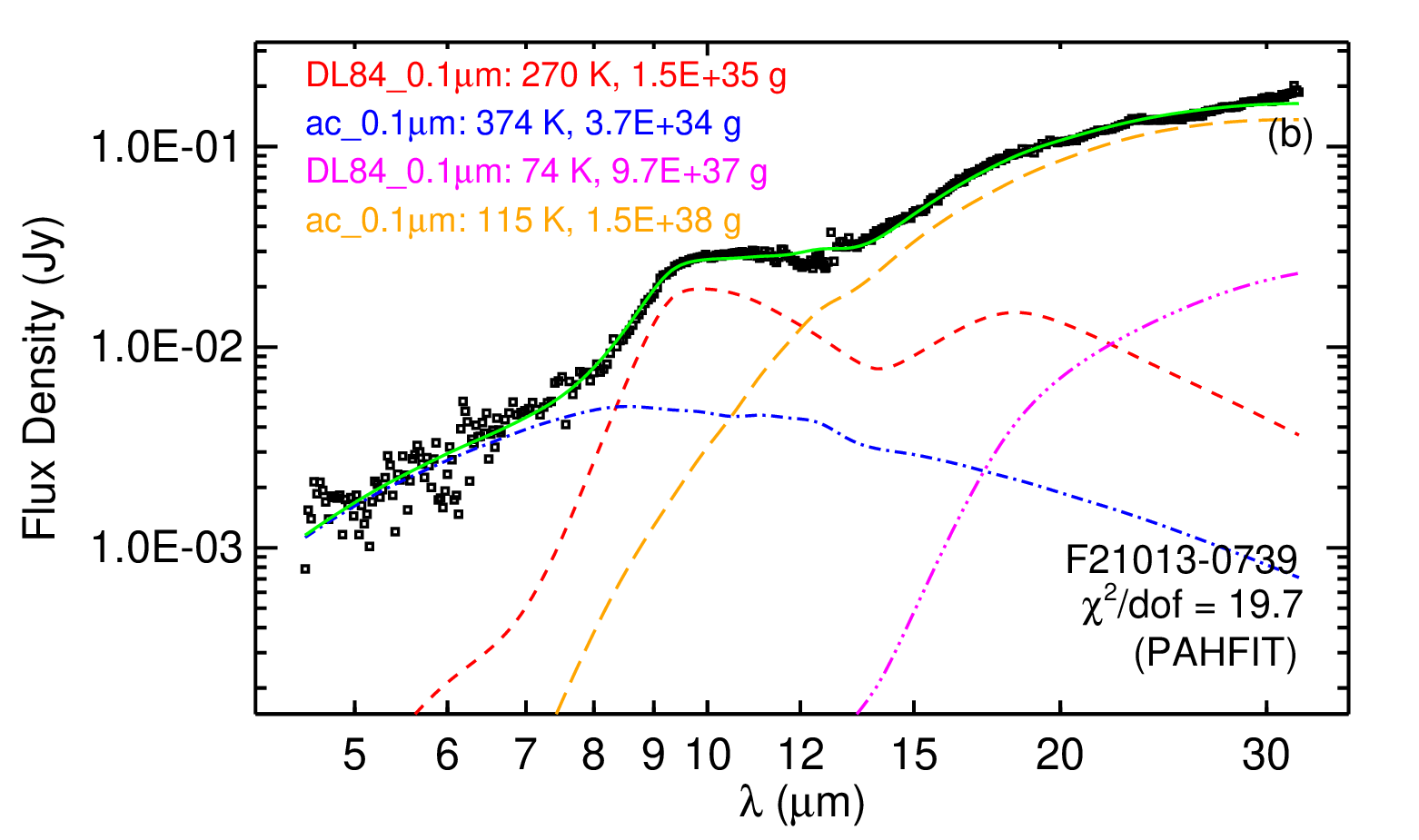}} \\
\resizebox{0.7\hsize}{!}{
\includegraphics{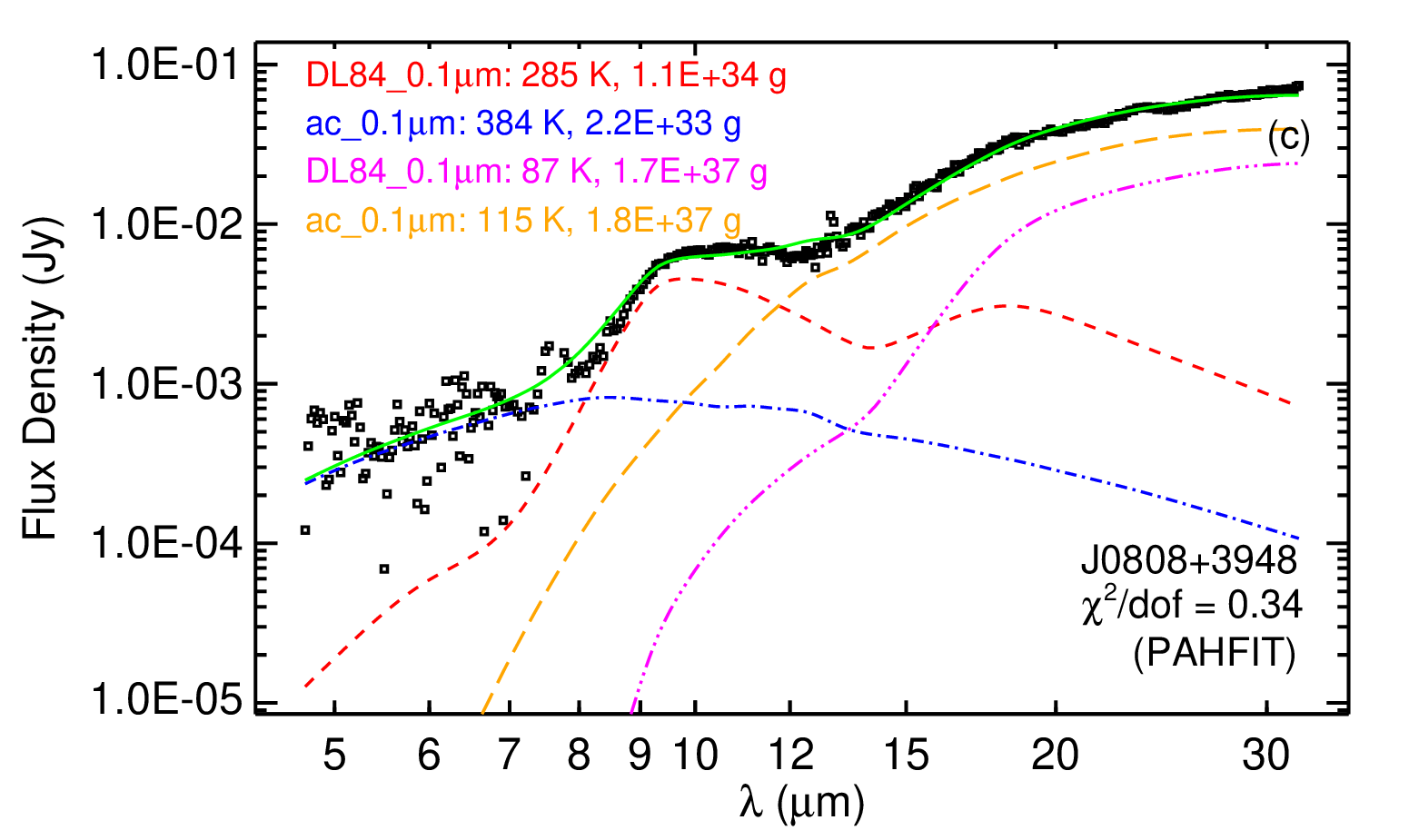}}
\end{array}
$
\caption{\footnotesize
         \label{fig:tccarbfix}
         Same as Figure~\ref{fig:dl84ac}
         but with the temperature of 
         the cold carbon component
         (orange long dashed line) 
         fixed at $\Tccarb=115\K$
         (see \S\ref{sec:discussion}).
         }
\end{center}
\end{figure*}

To investigate the significance of 
the dust temperature and mass 
yielded from our dust IR emission modeling, 
for all three sources we fix the temperature 
of the cold carbon dust component 
to be $\Tccarb = 115\K$ 
and then probe the dust temperature 
and mass variations for the other components. 
Figure~\ref{fig:tccarbfix} shows 
the model spectra for all three sources
obtained from the mixture of 
``astronomical silicate''
and amorphous carbon (``DL84\,+\,ac'')
with $a = 0.1\mum$ and $\Tccarb =115\K$. 
The model spectra are in close agreement
with the observed spectra.
Compared with the same model but with $\Tccarb$
treated as a free parameter 
(see Figure~\ref{fig:dl84ac}),
the temperatures derived for the other three
components ($\Twsil$, $\Twcarb$, and $\Tcsil$) 
vary within $<$\,10\%, while the derived masses
($\Mwsil$, $\Mwcarb$, and $\Mcsil$) 
vary within $<$\,75\%. 

To test the reasonableness of the dust mass 
derived in \S\ref{sec:results}, 
for each source we compare the dust mass
with the stellar mass. 
On average, the dust-to-stellar mass ratio 
of these three galaxies is $\simali$10$^{-5}$,
much smaller than that of the Milky Way
($\simali$10$^{-3}$; see Li 2004b).
This ratio appears reasonable 
since the mid-IR emission considered here
only probes the dust of $T>100\K$
(see Table~\ref{tab:modpara}),
while the bulk dust mass is in dust
with $T<100\K$ in the host galaxy
which emits in the far-IR
and escapes from detection by {\it Spitzer}/IRS. 
%

\begin{figure*}
\begin{center}
$
\begin{array}{c}
\resizebox{0.7\hsize}{!}{
\includegraphics{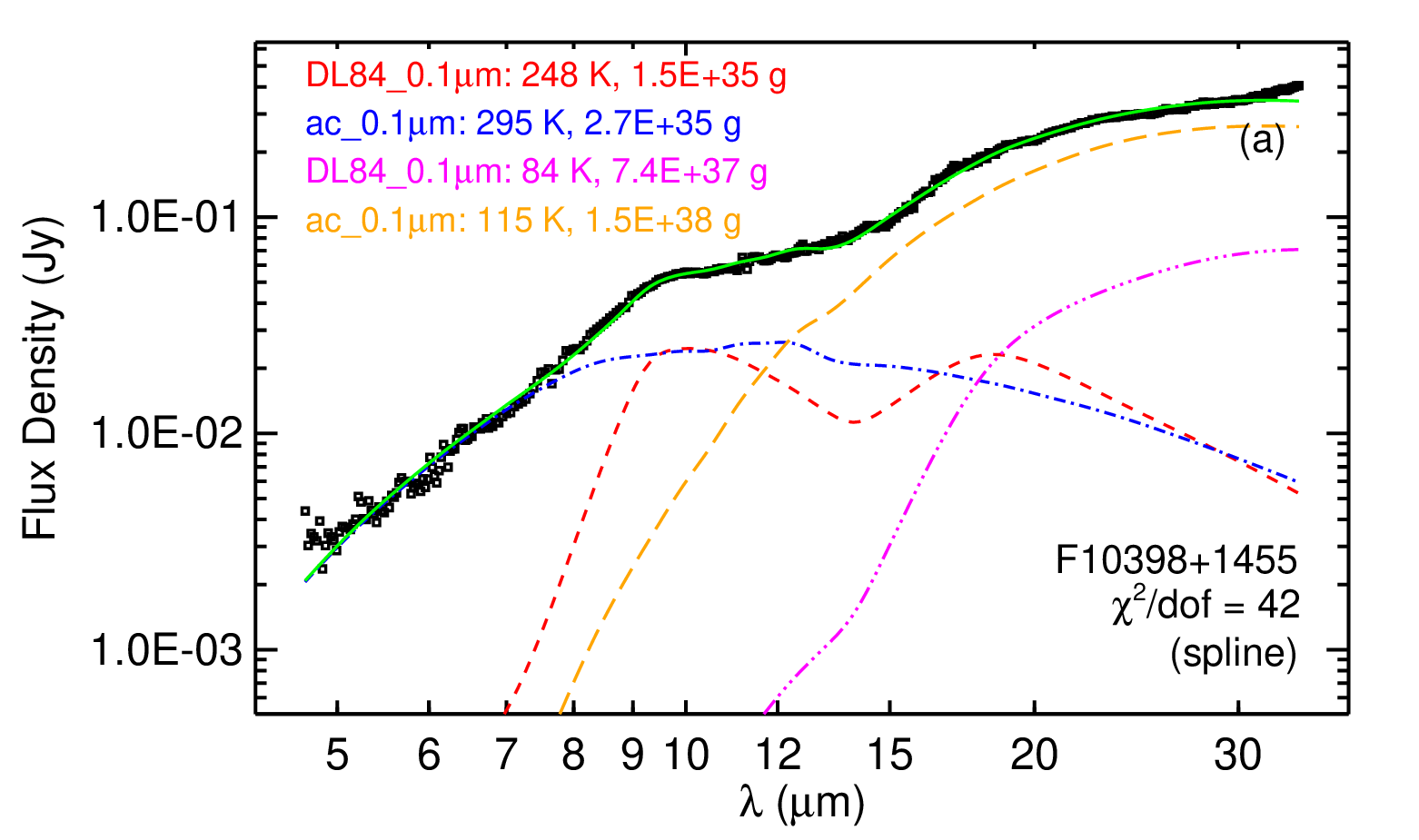}} \\
\resizebox{0.7\hsize}{!}{
\includegraphics{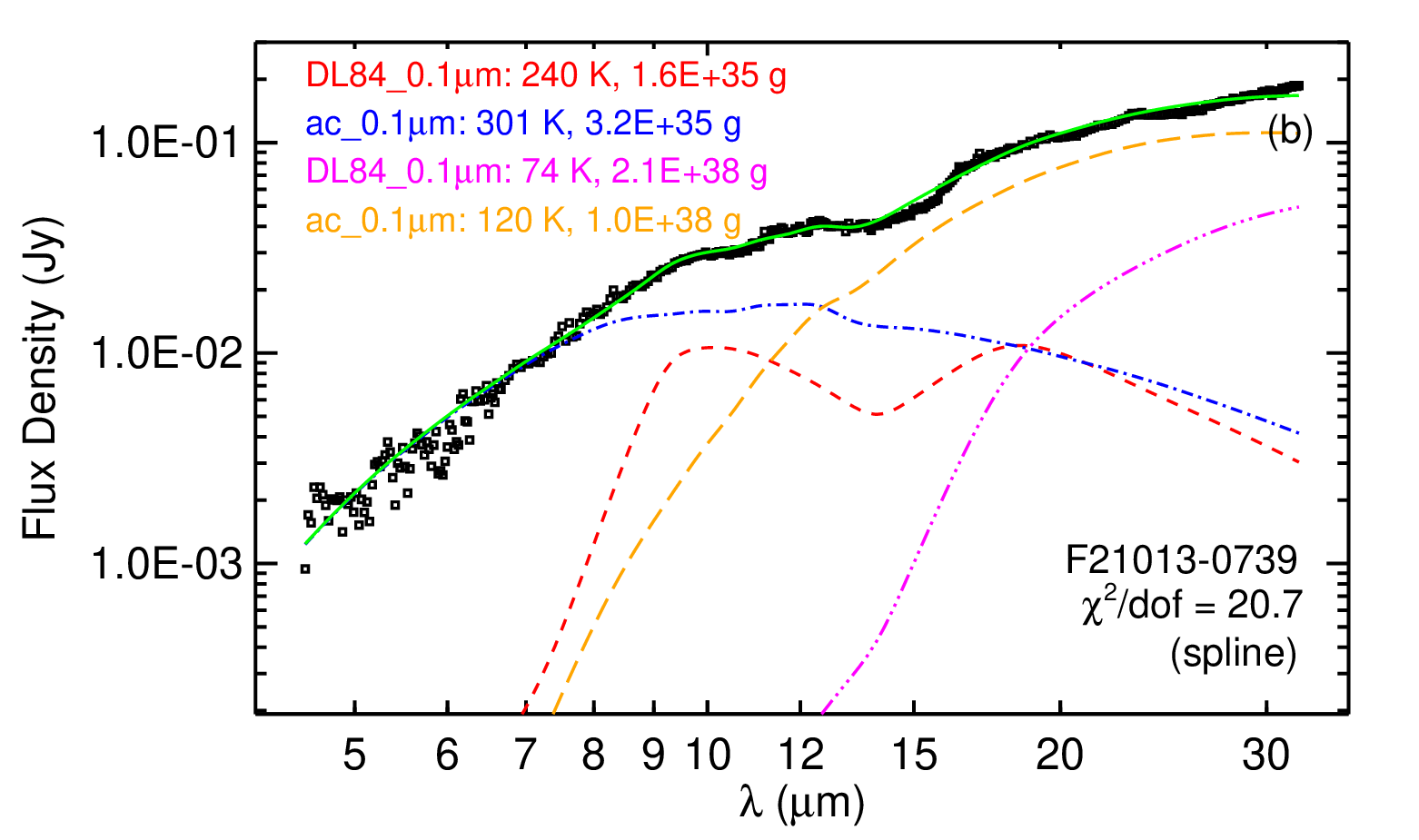}} \\
\resizebox{0.7\hsize}{!}{
\includegraphics{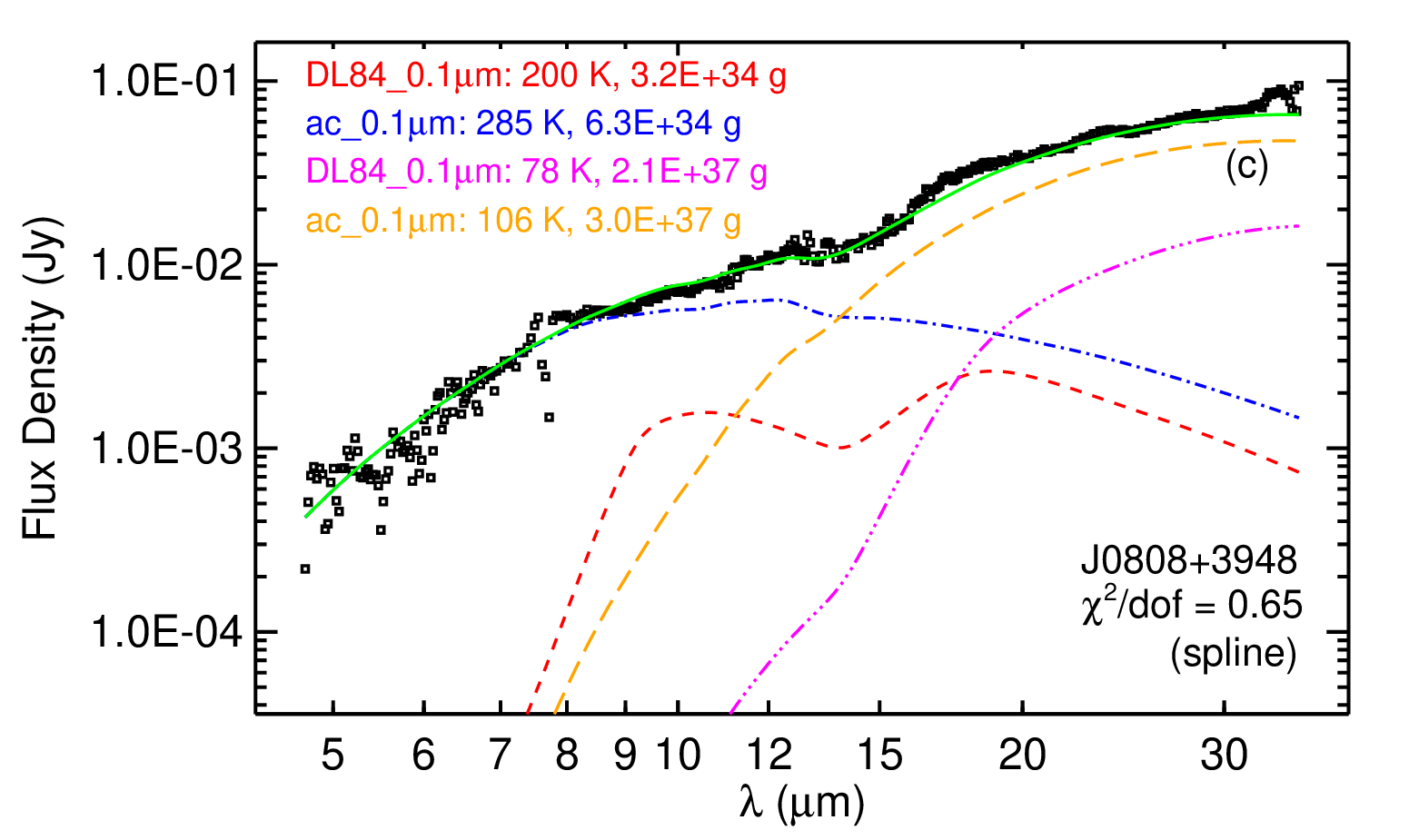}}
\end{array}
$
\caption{\footnotesize
         \label{fig:dl84ac_spline}
         Same as Figure~\ref{fig:dl84ac}
         but with the ``observed'' spectra 
         (black open squares)
         obtained by subtracting from 
         the \textit{Spitzer}/IRS spectra 
         the PAH and ionic emission lines 
         determined with the spline method
         (see \S\ref{sec:intro}).
         }
\end{center}
\end{figure*}

\subsection{PAHFIT Versus Spline\label{sec:spline}}
The results presented in \S\ref{sec:results}
are derived by fitting 
the ``observed'' IR emission
obtained by subtracting from 
the \textit{Spitzer}/IRS spectra 
the PAH and ionic emission lines 
determined with the PAHFIT method.
As illustrated in Figure~\ref{fig:sil_pahfit_spline},
the IR emission spectra derived from
the spline method show 
a relatively broad and weak 9.7$\mum$ 
silicate emission feature 
in comparison with that from PAHFIT.
Furthermore, the spline method yields 
a steeper slope for the $\simali$5--8$\mum$ 
continuum than PAHFIT
(see \S\ref{sec:data}). 
We have also modeled the spline-based 
``observed'' IR emission of all three 
spectroscopically anomalous galaxies.
As illustrated in Figure~\ref{fig:dl84ac_spline},
the simple four-component model 
consisting of warm/cold silicate dust
and warm/cold carbon dust closely
reproduces the ``observed'' IR emission
obtained from the spline method.
The temperatures and masses of 
the cold components for all sources 
do not change much 
($<$\,10\% for $T$, and $<$\,30\% for $M$;
see Table~\ref{tab:modpara})
since the PAHFIT- and spline-based 
IR emission spectra exhibit little 
difference beyond $\lambda$\,$\simali$14$\mum$ 
where the cold components dominate the radiation.
In contrast, the temperature of 
the warm silicate component derived
from the spline method drops by $\sim$10--25\%
and its mass changes by a factor of $\sim$1.1--1.9.
For the warm carbon component,
on average, its temperature 
drops by $\simali$20\%\footnote{%
  Compared with PAHFIT,
  the spline method yields a steeper
  and higher $\simali$5--8$\mum$ continuum 
  (see Figure~\ref{fig:sil_pahfit_spline})
  and therefore a lower temperature
  and larger mass
  for the warm carbon component
  which dominates the $\simali$5--8$\mum$ continuum.
  }
and the corresponding mass 
increases by a factor of 
$\simali$8.7, $\simali$6.6 and $\simali$21 for 
F10398+1455, F21013-0739 and 
SDSS~J0808+3948, respectively.
%

\begin{figure}[ht]
\begin{center}
\resizebox{\hsize}{!}{
\includegraphics{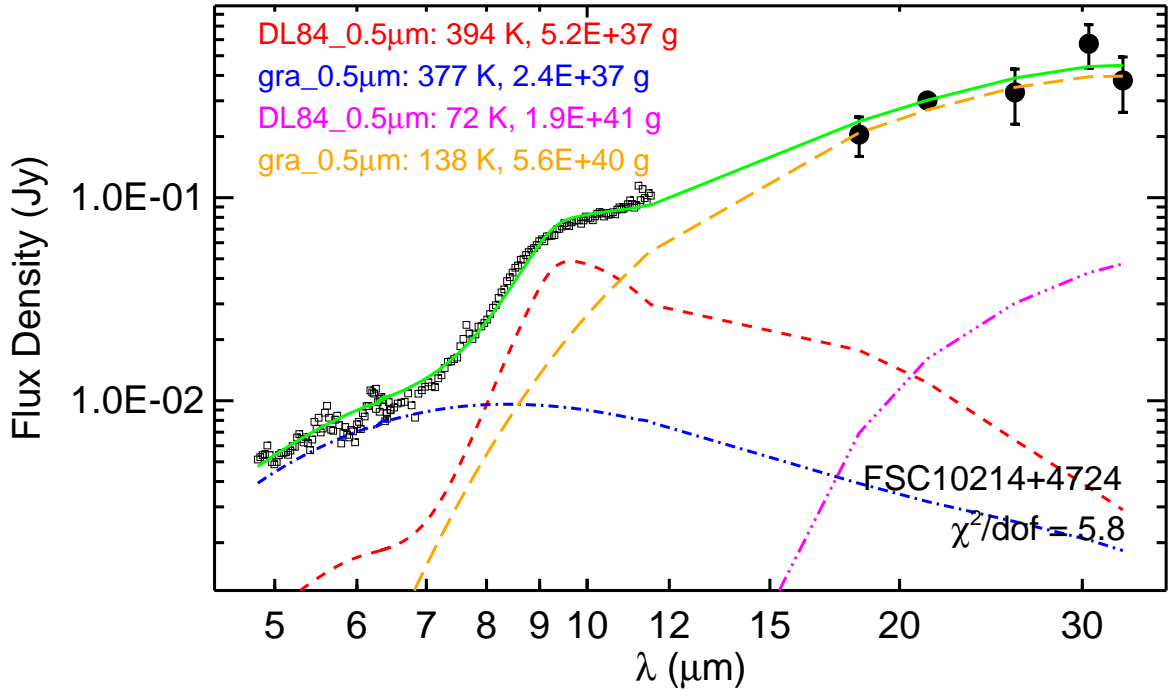}}
\caption{\footnotesize
         \label{fig:ulirg}
         Comparison of the observed IR emission
         of IRAS~FSC10214+4724 
         (black open squares; Teplitz \etal 2006) 
         with the model emission (solid green line)
         which is a sum of four components:
         warm DL84 silicate (red dashed line),
         cold DL84 silicate (magenta dot-long dashed line),
         warm graphite (blue dotted--dashed line), and
         cold graphite (orange long dashed line).
         All dust components have a size of $a=0.5\mum$.
         }
\end{center}
\end{figure} 

\begin{figure}[ht]
\begin{center}
\resizebox{\hsize}{!}{
\includegraphics{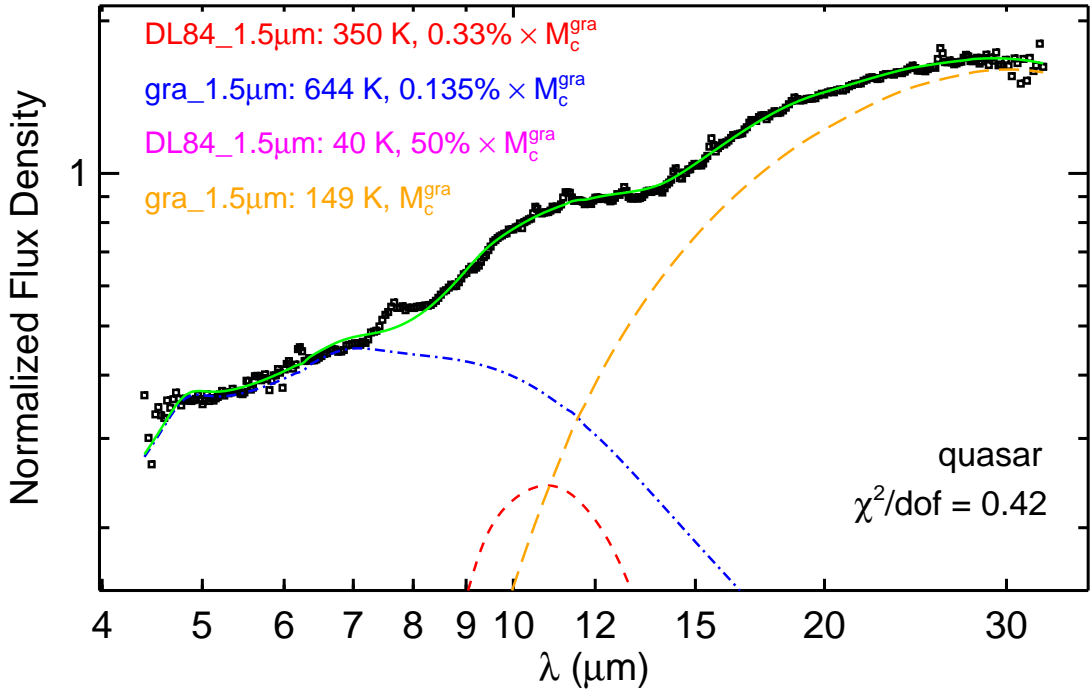}}
\caption{\footnotesize
         \label{fig:quasar}
         Same as Figure~\ref{fig:ulirg}
         but for the average spectrum of
         quasars (see Hao et al.\ 2007).
         }
\end{center}
\end{figure}

\subsection{Comparison with ULIRG 
            and Quasars\label{sec:quasar}}
The anomalous IR emission 
(i.e., a steep $\simali$5--8$\mum$ emission
continuum combined with silicate emission)
seen in IRAS~F10398+1455, IRAS~F21013-0739 
and SDSS~J0808+3948 is rarely seen
in other sources.
To the best of our knowledge, 
the only similar source 
is IRAS~FSC 10214+4724, 
a lensed, starburst-powered ULIRG 
at a redshift of $z\approx 2.29$ (Teplitz \etal 2006).
The major difference between IRAS~FSC 10214+4724
and our three galaxies is that  
IRAS FSC 10214+4724 lacks the PAH emission features
which are prominent in our three galaxies.
We have also modeled the IR emission of
IRAS~FSC 10214+4724. 
As shown in Figure~\ref{fig:ulirg},
a mixture of warm/cold ``astronomical silicate''
and warm/cold graphite of $a\approx0.5\mum$
closely fits the observed IR emission.
The derived parameters 
(see Table~\ref{tab:modpara})
are similar to that of our three galaxies. 

We have also modeled the average spectrum 
of quasars (Hao et al.\ 2007).\footnote{%
   The averaged spectrum of quasars is 
   constructed with individual quasar 
   spectra 
   normalized at $\lambda=14.5\mum$, 
   in order to avoid the most luminous quasar 
   dominating the resulting spectrum.
   }
As shown in Figure~\ref{fig:quasar},
the average quasar spectrum exhibits
a flat $\simali$5--8$\mum$ emission continuum
and a broad silicate emission feature at 9.7$\mum$.
The best-fit is achieved with dust of $a\approx1.5\mum$,
with $\Twsil\approx350\K$ for warm silicate,
$\Twcarb\approx644\K$ for warm graphite,
 $\Tcsil\approx40\K$ for cold silicate, and
$\Tccarb\approx149\K$ for cold graphite.
The flat $\simali$5--8$\mum$ emission continuum
seen in quasars is caused by warm graphite
which is about twice as hot as that in
our three galaxies. 
%
%

In principal, it would be desireable to model
the IR emission spectra of starbursts
to which the steep $\simali$5--8$\mum$ continuum
emission and the PAH emission features of our
three galaxies bear a close resemblance.
%
However, to obtain any meaningful information 
about the properties of the dust 
(e.g., size, composition, temperature) 
from modeling the IR spectra of starbursts,
one has to make assumptions concerning 
the distribution of stars and dust 
within the starburst.
We therefore intend not to model 
the IR spectra of starbursts
as we know little about how stars
and dust are spatially distributed.
%


Y.~Xie et al.\ (2015, in preparation) 
examine the multi-wavelength data 
of these three galaxies 
to investigate their powering sources. 
They find that F21013-0739 and \sd\  
have detections in hard X-rays (Jia \etal 2011) 
which indicates the presence of AGN. 
However, the diagnostics from the UV, 
optical, mid-IR, and radio 
do not show any strong evidence of AGN. 
They conclude that AGN may be present 
in these galaxies, but they are too young 
to be detected in all wavelength bands.
The lower dust temperature 
($T$ $\sim$ 250--400$\K$) 
and smaller grain size 
(with an upper limit of $a\simlt1.0\mum$)
inferred for our three galaxies 
compared with that of quasars
could be due to the fact that our galaxies 
may harbor a young/weak AGN 
which is not maturely developed yet.
Compared with quasars,
such a young AGN emits much less 
and softer in the X-ray and UV.
Therefore, sub-$\mu$m-sized grains 
could survive against sublimation
and photo-sputtering in our galaxies
while only $\mu$m-sized grains 
could survive in quasars. 
Exposed to a young AGN with 
a lower UV luminosity $L_{\rm UV}$,
the dust temperature is expected to be lower 
since $T\propto L_{\rm UV}^{1/\left(4+\beta\right)}$,
where $\beta$ is the dust opacity power-law
exponent in the far-IR 
(i.e., $\kappa_{\rm abs}\propto \lambda^{-\beta}$).
Indeed, the bolometric luminosity of 
our galaxies is in the order of
$\simali$1$0^{44}\erg\s^{-1}$ 
(Y.~Xie \etal 2015, in preparation), 
about two orders of magnitude 
lower than that of typical quasars. 

\section{Summary\label{sec:summary}}
We have modeled the \textit{Spitzer}/IRS spectra of 
three spectroscopically anomalous galaxies
(IRAS~F10398+1455, IRAS~F21013-0739 
and SDSS~J0808+3948)
in terms of a simple model consisting of 
a mixture of warm and cold silicate dust, and
warm and cold carbon dust.
The IR spectral characteristics of 
these galaxies are unique in the sense that they 
show silicate emission 
which is characteristic of AGN 
while they also show a steep $\simali$5--8$\um$ 
emission continuum and strong PAH emission features 
which are characteristic of starburst galaxies.
In contrast, AGN exhibit a {\it flat} emission 
continuum at $\simali$5--8$\mum$ and lack the PAH 
emission features. The 9.7 and 18$\mum$ silicate 
features seen in emission in these three galaxies
are seen in absorption in starbursts.
%
We find that our model with four components 
(cold and warm carbons, cold and warm silicates) 
closely reproduces the observed IR emission of 
all three galaxies, including the steep 
$\simali$5--8$\mum$ emission continuum and 
the 9.7$\mum$ silicate emission feature.
The model-fitting results are insensitive to
the exact silicate and carbon dust composition 
(e.g., ``astronomical'' silicate or laboratory analogs, 
olivine or pyroxene, iron-poor silicate 
or iron-rich silicate, amorphous carbon or graphite)
and dust size, provided the dust size 
does not exceed one micrometer
(i.e., $a\simlt1.0\mum$).
It is also found that the dust temperature 
is the primary cause in regulating the 
steep $\simali$5--8$\um$ emission continuum 
and the prominent silicate emission feature.
Compared with the average spectrum of quasars
which exhibits a flat $\simali$5--8$\mum$ continuum, 
the steep $\simali$5--8$\mum$ continuum seen
in our three galaxies predominantly arises from
warm carbon dust of temperatures 
of $T$ $\sim$ 250--400$\K$,
much lower than that of quasars
($T$ $\sim$ 640$\K$).
The 9.7$\mum$ silicate emission 
of the quasar average spectrum
is best fitted with silicate dust 
of size $a\approx1.5\mum$,
larger than the upper size limit
of $a\simlt1.0\mum$ inferred for 
our three galaxies. 
The lower dust temperature 
and smaller dust size
inferred for our three galaxies 
compared with that of quasars
could be due to the fact that our galaxies
may harbor a young/weak AGN 
which is not maturely developed yet. 
To test the simple dust model used in current work, 
we would like to present a series of physical AGN models 
(e.g Clumpy models) for these three galaxies as well as 
the quasar average spectrum and ULIRG FSC 10214+4724 
using the model parameters derived in this paper 
(R. Nikutta et al. 2015, in preparation).  

\acknowledgments
We thank V.~Charmandaris, A.~Mishra,
J. Y.~Seok, S.~Wang 
and the anonymous referee
for helpful suggestions and discussions.  
L. H. and X. Y. X. are supported by 
the National Natual Science Foundation 
of China under grants No. 11473305, 
by the Strategic Priority Research Program 
"The Emergence of Cosmological Structures" 
of Chinese Academy of Sciences, 
Grant No. XDB09030200.
A. L. and X. Y. X. are supported in part 
by NSF AST-1311804 and NASA NNX14AF68G. 
R. N. acknowledges support by FONDECYT grant No. 3140436.
The Cornell Atlas of \spitzerirs Sources (CASSIS) 
is a product of the Infrared Science Center 
at Cornell University, supported by NASA and JPL.



\begin{thebibliography}{}
\expandafter\ifx\csname natexlab\endcsname\relax\def\natexlab#1{#1}\fi

\bibitem[\protect\citeauthoryear{Allamandola, Tielens \& Barker}
{Allamandola, Tielens \& Barker}{1985}]{alla_etal_1985}
Allamandola, L.~J., Tielens, A.~G.~G.~M., \& Barker, J.~R. 1985, 
\apjl, 290, L25

\bibitem[{{Antonucci}(1993)}]{Antonucci1993}
{Antonucci}, R. 1993, \araa, 31, 473

\bibitem[\protect\citeauthoryear{{Asplund} et al.}
{{Asplund} et al.}{2009}]{aspl09}
Asplund, M., Grevesse, N., Sauval, A.~J., \& Scott, P. 2009, 
\araa, 47, 481

\bibitem[Baldwin et al.(1981)]{baldwin81} 
Baldwin, J.~A., Phillips, M.~M., \& Terlevich, R.\ 1981, \pasp, 93, 5 


\bibitem[{{Bohren} \& {Huffman}(1983)}]{BHbook}
{Bohren}, C.~F., \& {Huffman}, D.~R. 1983, 
{Absorption and scattering of light by small particles}

\bibitem[\protect\citeauthoryear{{Dorschner} et al.}
{{Dorschner} et al.}{1995}]{dors_etal_95}
{Dorschner}, J., Begemann, B., Henning, T., 
Jaeger, C., \& Mutschke, H. 1995, \aap, 300, 503

\bibitem[\protect\citeauthoryear{Draine \& Lee}
{Draine \& Lee}]{D&L1984}
Draine, B.~T., \& Lee, H.~M. 1984, \apj, 285, 89

\bibitem[\protect\citeauthoryear{Draine \& Li}
{Draine \& Li}]{D&L2001}
Draine, B.~T., \& Li, A. 2001, \apj, 551, 807

\bibitem[\protect\citeauthoryear{{Hao} et al.}
{{Hao} et al.}{2007}]{hao_etal_07}
{Hao}, L., Weedman, D.~W., 
Spoon, H.~W.~W., et al. 2007, \apjl, 655, L77

\bibitem[\protect\citeauthoryear{Henning, T.}
{Henning, T.}{2010}]{henning10}
Henning, T. 2010, \araa, 48, 21

\bibitem[\protect\citeauthoryear{{Houck} et al.}{{Houck}
  et al.}{2004}]{hou_etal_04}
{Houck}, J.~R., {Roellig}, T.~L., {van Cleve}, J.,  
et al. 2004, \apjs, 154, 18

\bibitem[\protect\citeauthoryear{{Imanishi} et al.}{{Imanishi}
 et al.}{1997}]{imanishi97}
{Imanishi}, M., {Terada}, H., {Sugiyama}, K., 
et al. 1997, \pasj, 49, 69 

\bibitem[\protect\citeauthoryear{{Jia} et al.}
{{Jia} et al.}{2011}]{jia11}
{Jia}, J., {Ptak}, A., {Heckman}, T.~M., 
et al. 2011, \apj, 731, 55

\bibitem[\protect\citeauthoryear{K$\rm \ddot{o}$hler \& Li}
{K$\rm \ddot{o}$hler \& Li}{2010}]{kholer10}
K$\rm \ddot{o}$hler, M., \& Li, A. 2010, \mnras, 406, L6

\bibitem[\protect\citeauthoryear{Laor \& Draine}
{Laor \& Draine}{1993}]{LD93}
{Laor}, A., \& {Draine}, B.~T. 1993, \apj, 402, 441

\bibitem[\protect\citeauthoryear{{Lebouteiller}et al.}{{Lebouteiller}
et al.}{2011}]{Lebou_etal_2011}
{Lebouteiller}, V., {Barry}, D.~J., {Spoon}, H.~W.~W., 
et al. 2011, \apjs, 196, 8

\bibitem[\protect\citeauthoryear{Leger \& Puget}{1984}]{Leger}
L\'eger, A., \& Puget, J.~L. 1984, \aap, 137, L5

\bibitem[]{}Li, A.\ 2004a, 
            in Astrophysics of Dust (ASP Conf. Ser. 309), 
            ed. A.N. Witt, G.C. Clayton, \& B.T. Draine 
            (San Francisco, CA: ASP), 417

\bibitem[]{}Li, A.\ 2004b, 
            in Penetrating Bars Through Masks 
            of Cosmic Dust: The Hubble Tuning Fork 
            Strikes a New Note,
            ed. D.L. Block, I. Puerari, K.C. Freeman, 
            R. Groess, \& E.K. Block 
            (Dordrecht: Kluwer) 535 

\bibitem[\protect\citeauthoryear{Li}
{Li}{2007}]{LiAG07}
Li, A. 2007, in The Central Engine of Active Galactic Nuclei (ASP Conf. Ser. 373), ed. L. C. Ho \& J.-M. Wang (San Francisco, CA: ASP), 561

\bibitem[\protect\citeauthoryear{Li}
{Li}{2009}]{LiAG09}
Li, A. 2009, in Small Bodies in Planetary Sciences (Lecture Notes in Physics vol. 758), ed. I. Mann, A. Nakamura, \& T. Mukai , Springer, Chapter 6, 167

\bibitem[\protect\citeauthoryear{Li, Shi \& Li}
{Li, Shi \& Li}{2008}]{limoping08}
Li, M.~P., Shi, Q.~J., \& Li, A. 2008, \mnras, 391, L49

\bibitem[\protect\citeauthoryear{Lyu, Hao \& Li}
{Lyu, Hao \& Li}{2014}]{lyu14}
{Lyu}, J., {Hao}, L., \& {Li}, A. 2014, \apjl, 792, L9

\bibitem[\protect\citeauthoryear{{Maiolino} et al.}
{{Maiolino} et al.}{2001}]{maiolino01}
{Maiolino}, R., {Marconi}, A., \& {Oliva}, E. 2001, 
\aap, 365, 37

\bibitem[\protect\citeauthoryear{Markwardt}
{Markwardt}{2009}]{mpfit}
{Markwardt}, C.~B. 2009,
in Astronomical Data Analysis Software and Systems XVIII 
(ASP Conf. Ser. 411), ed. D. A. Bohlender, D. Durand, 
\& P. Dowler (San Francisco, CA: ASP), 251 

\bibitem[\protect\citeauthoryear{{Kemper} et al.}
{{Kemper} et al.}{2007}]{kemper07}
Markwick-Kemper, F., {Gallagher}, S.~C., 
{Hines}, D.~C., \& {Bouwman}, J. 2007, \apjl, 668, L107

\bibitem[\protect\citeauthoryear{{Mason} et al.}
{{Mason} et al.}{2004}]{mason04}
{Mason}, R.~E., {Wright}, G., {Pendleton}, Y., \& {Adamson}, A. 
2004, \apj, 613, 770

\bibitem[\protect\citeauthoryear{{Mason} et~al.}{{Mason}
et~al.}{2009}] {mas_etal_09}
Mason, R.~E., {Levenson}, N.~A., {Shi}, Y., 
et al. 2009, \apjl, 693, L136

\bibitem[\protect\citeauthoryear{{Mason} et~al.}{{Mason}
et~al.}{2015}] {mas_etal_15}
Mason, R.~E., {Rodriguez-Ardila}, A., {Martins}, L., 
et al. 2015, \apjs, 217, 13 



\bibitem[\protect\citeauthoryear{{Nikutta} et al.}
{{Nikutta} et al.}{2009}]{nikutta_etal_09}
Nikutta, R., {Elitzur}, M., \& {Lacy}, M. 
2009, \apj, 707, 1550

\bibitem[\protect\citeauthoryear{Pendleton \& Allamandola}
{Pendleton \& Allamandola}{2002}]{PA02}
Pendleton, Y.~J., \& Allamandola, L.~J. 2002, \apjs, 138, 75

\bibitem[\protect\citeauthoryear{{Roche}, P.~F.}
{{Roche}, P.~F.}{1991}]{roche91} 
Roche, P.~F., {Aitken}, D.~K., {Smith}, C.~H., \& {Ward}, M.~J. 
1991, \mnras, 606, 629

\bibitem[\protect\citeauthoryear{Rouleau \& Martin}
{Rouleau \& Martin}{1991}]{RM1991}
Rouleau, F., \& Martin, P.~G. 1991, \apj, 377 526

\bibitem[\protect\citeauthoryear{{Siebenmorgen}, R.} 
{{Siebenmorgen}, R.}{2004}]{sieben04} 
{Siebenmorgen}, R., Kr{\"u}gel, E., \& {Spoon} H.~W.~W.
2004, \aap, 414, 123

\bibitem[\protect\citeauthoryear{{Smith} et al.}
{{Smith} et al.}{2007}]{smith07}
Smith, J.~D.~T., {Draine}, B.~T., {Dale}, D.~A., 
et al. 2007, \apj, 656, 770

\bibitem[\protect\citeauthoryear{{Smith} et al.}
{{Smith}et al.}{2010}]{smith.h10}
{Smith}, H.~A., {Li}, A., {Li}, M.~P., 
et al. 2010, \apj, 716, 490

\bibitem[\protect\citeauthoryear{{Sofia} et al.}
{{Sofia}et al.}{2010}]{sofia11}
{Sofia}, U.~J., {Parvathi}, V.~S., {Babu}, B.~R.~S., 
\& {Murthy}, J. 2011, \aj, 141, 22

\bibitem[\protect\citeauthoryear{{spoon} et~al.}
{{spoon} et~al.}{2004}]{spoon_etal_04}
{Spoon}, H.~W.~W., {Armus}, L., {Cami}, J., 
et al. 2004, \apjs, 154, 184

\bibitem[\protect\citeauthoryear{{sturm} et~al.}{{sturm}
et~al.}{2006}]{sturm_etal_06}
{Sturm}, E., {Hasinger}, G., {Lehmann}, I., 
et~al. 2006, \apj, 642, 81

\bibitem[\protect\citeauthoryear{{Teplitz} et al.}
{{Teplitz} et al.}{2006}]{tepl_etal_06}
Teplitz, H.~I., {Armus}, L., {Soifer}, B.~T., 
et al. 2006, \apjl, 638, L1

\bibitem[\protect\citeauthoryear{Urry \& Padovani}
{Urry \& Padovani}{1995}]{urry95}
 Urry, C.~M., \& Padovani, P. 1995, \pasp, 107, 803

\bibitem[\protect\citeauthoryear{{Voit}, G.~M.}{{Voit}, G.~M.}
{1991}]{voit91} Voit, G.~M. 1991, \apj, 379, 122

\bibitem[\protect\citeauthoryear{{Voit}, G.~M.}
{{Voit}, G.~M.}{1992}]{voit92} 
Voit, G.~M. 1992, \mnras, 258, 841

\bibitem[\protect\citeauthoryear{Xie, Hao \& Li}
{Xie, Hao \& Li}{2014}]{xieyx14}
{Xie}, Y., {Hao}, L., \& {Li}, A. 2014, \apjl, 794, L19

\end{thebibliography}
\end{document}